\documentclass[%
 reprint,
 superscriptaddress,
 amsmath,amssymb,
 aps,
prl,
]{revtex4-2}

\usepackage{graphicx}
\usepackage{overpic}
\usepackage{caption}
\usepackage{float}
\usepackage{dblfloatfix}
\usepackage{dcolumn}
\usepackage{bm}
\usepackage{bbold}
\usepackage{mathtools}
\usepackage{hyperref}  
\hypersetup{
    colorlinks=true,
    linkcolor=blue,
    filecolor=blue,      
    urlcolor=blue,
    citecolor=blue,
    }
\usepackage{xcolor}
\usepackage{caption}
\usepackage{multirow}
\usepackage[export]{adjustbox}
\usepackage{empheq}
\usepackage{url}

\renewcommand{\k}{\boldsymbol{k}}
\newcommand{\g}{\boldsymbol{g}}
\newcommand{\q}{\boldsymbol{q}}

\newcommand{\K}{{\boldsymbol{K}}}
\newcommand{\rr}{{\boldsymbol{r}}}
\newcommand{\R}{\boldsymbol{R}}
\newcommand{\rp}{\boldsymbol{r'}}
\newcommand{\Rp}{\boldsymbol{R'}}
\newcommand{\db}{\boldsymbol{\delta}}
\newcommand{\dbp}{\boldsymbol{\delta '}}

\makeatletter
\def\maketitle{
\@author@finish
\title@column\titleblock@produce
\suppressfloats[t]}
\makeatother

\begin{document}


\title{Fermi velocity, interlayer couplings, and magic angle renormalization in twisted bilayer graphene.}



\author{Miguel Sánchez Sánchez}
\email[]{miguel.sanchez@csic.es}
\affiliation{Instituto de Ciencia de Materiales de Madrid ICMM-CSIC. Madrid (Spain)}

\author{José González}
\affiliation{Instituto de Estructura de la Materia IEM-CSIC. Madrid (Spain)}

\author{Tobias Stauber}
\affiliation{Instituto de Ciencia de Materiales de Madrid ICMM-CSIC. Madrid (Spain)}


\begin{abstract}

Through extensive self-consistent Hartree-Fock calculations in a tight-binding model of twisted bilayer graphene (TBG), we show that many-body effects lead to a considerable increase of the bandwidth of the flat bands and, concomitantly, to a shift of the magic angle (defined by the condition of minimum bandwidth). Specifically, we predict a shift from the $\textit{ab initio}$ magic angle of $0.99^\circ$ to a renormalized value of $0.88^\circ$ for a TBG sample suspended between metallic gates with a gate-to-gate distance of $10 \text{ nm}$. We derive analytical expressions for the renormalized Fermi velocity and interlayer couplings, finding good agreement with the numerical results, and investigate the convergence toward the numerical solutions with respect to the number of renormalized couplings of a generalized Bistritzer-MacDonald (BM) model. Using the derived analytical formulas, we demonstrate the possibility of tuning the flat bands via different dielectric environments and gate geometries in the experiments. Furthermore, we predict a significant enhancement of the flat-band Fermi velocity at intermediate twist angles relative to the bare value, and propose measurements in this range as a probe of the effective couplings of TBG. Our results imply a change of paradigm whereby the maximum $T_c$ for superconductivity would correspond to a condition of small but not minimum bandwidth.

\end{abstract}

\maketitle

\paragraph{\underline{Introduction.}}




In graphene, the velocity of the Dirac cones is heavily renormalized by many-body effects, a phenomenon that has been extensively studied \cite{Gonzalez1994,gonzalez99,hofmann14,barnes14,sodemann12,jung11,dejuan10,frassdorf17,bauer15,stauber17,guandalini24,Tang2018,Polini2007,kotov12,Elias2011,Chae12,Siegel2011,Ryu2017,Yu2013,knox11,Hwang2012,Lee2026}. This naturally raises the question of the magnitude of analogous effects in twisted bilayer graphene (TBG). A seminal paper by Kang and Vafek \cite{kang20} addressed this question using a renormalization group approach, and since then a number of works in this line have followed \cite{sanchez25,sanchez26,huang25,lu2025numericalperspectivemoiresuperlattices,lu2026generalmanybodyperturbationframework,guo24}. Here we perform tight-binding calculations with up to $\sim 18,000$ atoms per unit cell using the recently developed FORGE package \cite{forgezenodo}. We obtain the renormalized symmetry-preserving state of TBG in self-consistent Hartree-Fock theory at the charge neutrality point. The results show a widening of the flat bands and a concomitant renormalization of the magic angle, showcasing many-body effects of comparable magnitude to those from lattice relaxation or strain \cite{carr18,carr19,koshino20_1,koshino20_2,kang23, bi19,arbeitman25}.

We trace these outcomes back to the renormalization of the various parameters of the Bistritzer-MacDonald (BM) model \cite{Bistritzer2011}. We analytically derive the corrections of the Fermi velocity $v_F$ and the interlayer couplings $w_0$ and $w_1$ to first order in the Coulomb interaction and the interlayer couplings, finding remarkably good agreement with the numerical values. Moreover, we demonstrate that the band structures can be conveniently reproduced by way of renormalizing the Fermi velocity, the interlayer potential $T(\rr)$ and the intralayer pseudo-gauge field $A^\ell (\rr)$, without introducing additional parameters. We also consider the dependence of the band renormalization on different experimental setups, and discover a degree of tunability of the flat bands that can be explored experimentally and would provide a test of our results.

Finally, we discuss possible extensions of our work and draw implications for the interpretation of experiments. In particular, we describe how superconductivity could emerge at twist angles $\sim 1.1^\circ$ while it is phase fluctuations-limited at the renormalized magic angle of $\lesssim1.0^\circ$. 

\begin{figure}[!t]
    \centering
    \includegraphics[width=.97\linewidth]{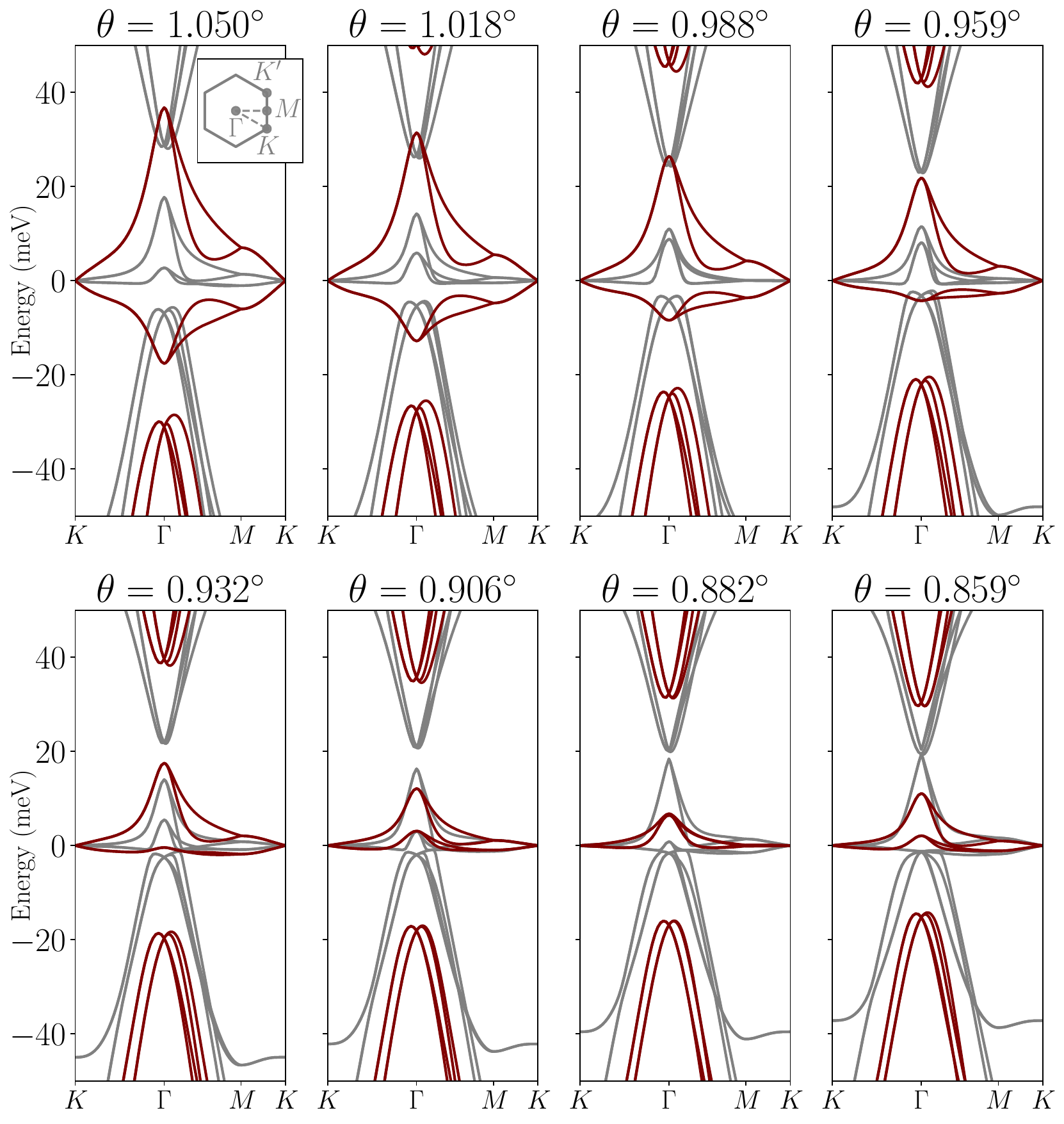}
    \caption{Band structures of renormalized TBG, in red, and of non-interacting TBG, in gray, for twist angles between $1.050^\circ$ and $0.859^\circ$. In the renormalized state the magic angle is shifted towards a lower value, from $0.988^\circ$ to $0.882^\circ$, and the gap between the flat and remote bands increases. The Dirac cones are set to zero energy. We set $\epsilon=10$ and gate distance $\xi=10$ nm.}
    \label{renormbands}
\end{figure}

\paragraph{\underline{Symmetric state.}}

We consider commensurate bilayer structures with twist angles $\theta$ given by $\cos(\theta) = 1 - 1/(6n_\theta^2 + 6n_\theta + 1)$ for integer $n_\theta$. There are $4(3n_\theta^2 + 3n_\theta + 1)$ atoms per moiré unit cell and the moiré lattice constant is $L_M = a/(2\sin(\theta/2))$ with $a=2.46 \text{ \AA}$ the lattice constant of graphene. We adopt the tight-binding functions from Refs. \cite{fang16,kang23}, 
and the in-plane relaxation of the lattice is obtained by solving the elastic theory from Ref. \cite{carr18}, which provides quantitative agreement with experimental data \cite{kang25,Kazmierczak2021}.

The electron-electron interactions are described by the double-gated Coulomb potential, $V(\rr) = \frac{e^2}{\epsilon}\sum_{n=-\infty}^\infty \frac{(-1)^n}{||\rr + n\xi \boldsymbol{\hat{z}}||}$, where the sample is placed between two metallic gates at a distance $\xi$. We fix $\xi=10$ nm throughout this work. We also include an on-site Hubbard term of $U=4$ eV, within the range $3-9$ eV \cite{wehling11,schuler13}.

Lacking a quantitative treatment of the dielectric function, we set the dielectric constant to the static random phase approximation (RPA) value for $8$ ($2$ spins $\times$ $2$ valleys $\times$ $2$ layers) Dirac fermions \cite{gonzalez99,barnes14,kotov12},
\begin{align}
    \epsilon = \epsilon_{\text{env}} + \frac{\pi e^2}{\hbar v_F} \approx \epsilon_{\text{env}} + 8.6,  
\end{align}
with $\epsilon_{\text{env}}$ the dielectric constant of the environment and $v_F$ the Fermi velocity of graphene. In, graphene, the Hartree-Fock + static RPA approximation is broadly consistent with experiments \cite{barnes14,stauber17,Yu2013,guandalini24,Lee2026}. A suspended sample with $\epsilon_{\text{env}} = 1$ gives $\epsilon \approx 10 $ and a hBN encapsulated sample with $\epsilon_{\text{env}} = 6$ gives $\epsilon \approx 15$. We fix $\epsilon=10$ throughout this work.

We solve the interacting model at charge neutrality in self-consistent Hartree-Fock theory. By explicitly enforcing all the symmetries of the problem: lattice translations, the crystallographic symmetries, spin rotations, time-reversal symmetry and the emergent valley charge conservation, we converge to a renormalized symmetry-preserving state. Details of the Hartree-Fock tight-binding approach are given in the Supplementary Materials \cite{supplement}. An effective low-energy theory can be constructed via the (many-body) projection onto the desired degrees of freedom of the renormalized state \cite{sanchez25}, a process similar to Wilsonian renormalization \cite{huang25,kang20}. We argue that this effective theory is a better starting point than the projection onto the non interacting model, as it incorporates the non-local exchange contributions from the integrated-out modes \cite{sanchez26}.  

In Fig. \ref{renormbands} we show the bands of the the converged solutions as well as the non interacting bands for $\theta$ between $1.050^\circ$ $(n_\theta=31)$ and $0.859^\circ$ $(n_\theta=38)$. The magic angle shall be defined as the angle at which the band inversion at $\Gamma$ takes place \cite{carr19,fang2019angledependentitabinitio} --- this  is also approximately the angle at which the full bandwidth of the flat bands is minimal. We observe a widening of the bands in the renormalized solutions, and a shift of the magic angle from $\approx 0.988 ^\circ$ in the non-interacting model towards a lower value of $\approx 0.882^\circ$. An increase of the gap between the flat and remote bands is also noticeable, in line with Ref. \cite{huang25}. In summary, Fig. \ref{renormbands} underscores nontrivial many-body corrections to the flat bands of TBG, which can be of the same scale as those from lattice relaxation or strain necessary for comparison to experiments \cite{carr18,carr19,koshino20_1,koshino20_2,kang23, bi19,arbeitman25,Kong2025, kwan21, crippa2026interplaymanybodycorrelationsstrain, Kazmierczak2021,Nuckolls2023, Yoo2019, Pierce2021, Xie2019,zhang2025heavyfermionsmassrenormalization}. In the following Sections we try to understand analytically the origin of these new features. 

\begin{figure}
    \raggedright
    \hspace{0.1cm} (a) \hspace{3.0cm} (b) \\
    \centering
    \includegraphics[width=0.4\linewidth]{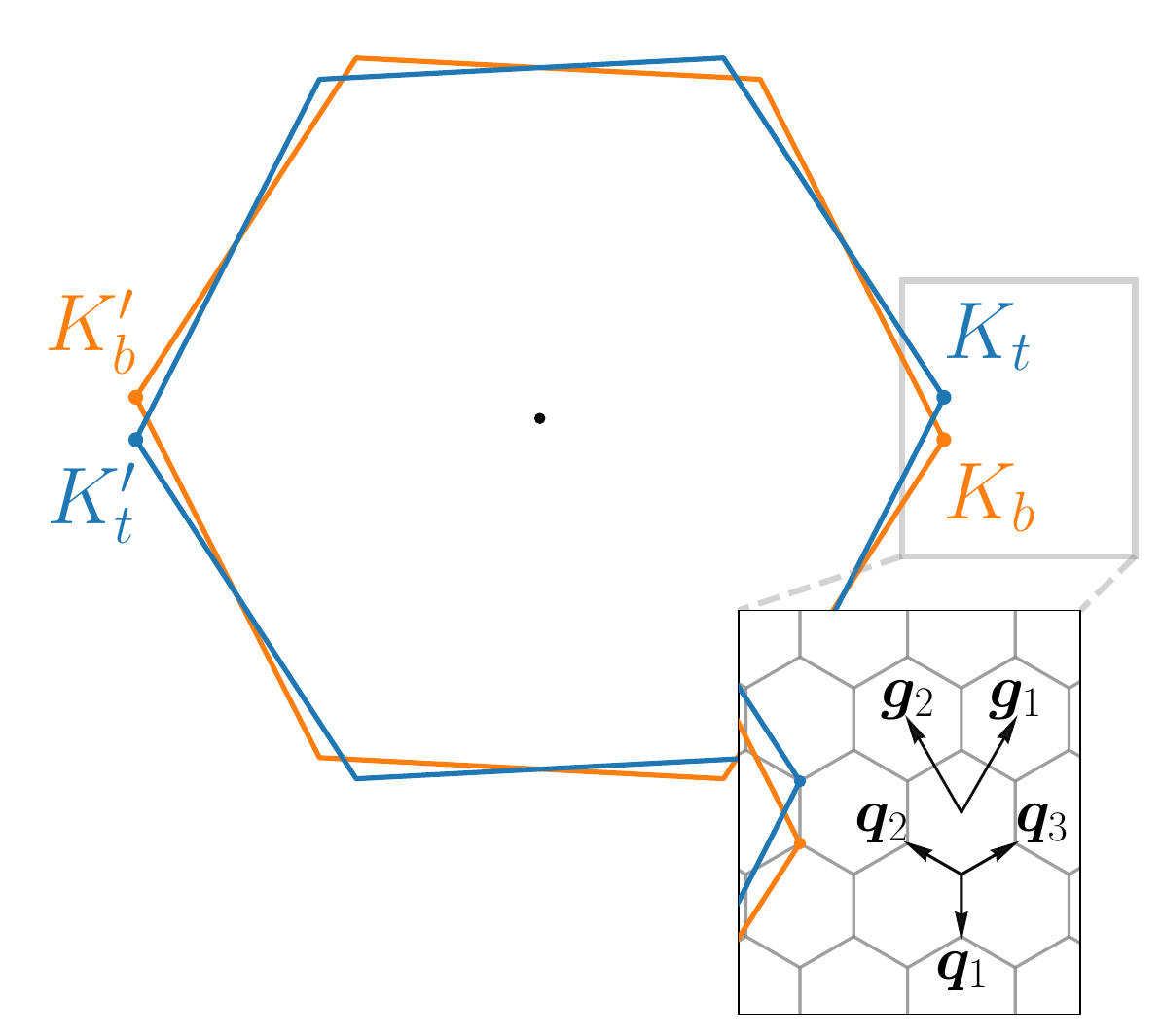}
    \includegraphics[width=0.55\linewidth]{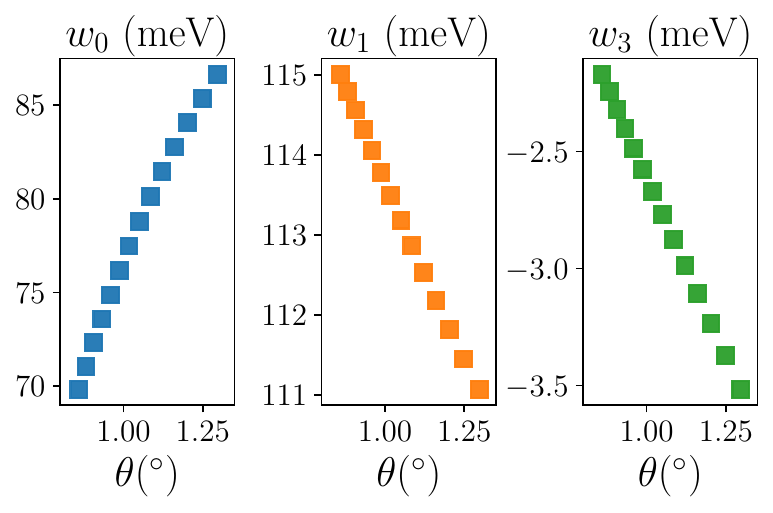}
    \caption{(a) Brillouin zones of the top (blue) and bottom (orange) layers, with their respective $K(K')$ points labeled. Inset: folded moiré Brillouin zones, the reciprocal lattice vectors $\g_{1,2}$ and the momenta $\q_{1,2,3}$. (b) BM model parameters $w_0$, $w_1$ and $w_3$ as a function of the twist angle.}
    \label{bmsetup}
\end{figure}
\paragraph{\underline{Generalized BM model.}}

Let us consider the plane waves \cite{miao23,fang2019angledependentitabinitio,guinea19},
\begin{align}
    |\k,\ell,\sigma, \eta \rangle = \frac{2}{\sqrt{N_{at}}}\sum_{\rr \in \sigma \ell} e^{i (\eta \K_\ell + \k)\cdot \rr}| \rr \rangle,
    \label{bwaves}
\end{align}
with $\k =(k_x,k_y)$, $\ell =$ top($t,+$), bottom($b,-$) denotes the layer, $\sigma=A,B$ the graphene sublattice and $\eta=K(+1), K'(-1)$ the valley. $+(-) \K_\ell$ is the $K(K')$ point of layer $\ell$, "$\rr \in \sigma \ell$" is short for the sum over atomic sites on sublattice $\sigma$ and layer $\ell$, and $N_{at}$ is the number of atoms in the system. 

The low-energy states are supported near the Dirac cones of the monolayers, i.e. around $\k=0$. It is then natural to project the tight-binding hamiltonian $H_{\text{TB}}$ onto the basis of plane waves. From now on, we focus on the $K$ valley and omit the $\eta$ label (the two valleys are related by time-reversal symmetry). We keep the description concise and provide all details in the Supplementary Materials \cite{supplement}.

The 'graphene blocks' feature the Dirac ones of the parent graphene layers,
\begin{align}
    \langle \k,\ell| H_{\text{TB}}|\k,\ell \rangle =& v_F \bar{\boldsymbol{\sigma}}_{\ell\theta/2} \cdot \k
    \label{gblock}
\end{align}
with $\bar{\boldsymbol{\sigma}} = (\sigma_x, -
\sigma_y)$,  $\bar{\boldsymbol{\sigma}}_{\ell\theta/2} = e^{i \ell \theta \sigma_z /4
} \bar{\boldsymbol{\sigma}} e^{-i \ell \theta  \sigma_z /4}$, $\sigma_{x,y,z}$ are Pauli matrices and $v_F=2.179  \text{ eV} \cdot a$ (in units where $\hbar=1$) is the graphene Fermi velocity. 
Above and in the following we omit the sublattice labels; the matrix element is then to be read as a $2\times2$ block in the sublattice space.

On the other hand, consider the interlayer block
\begin{align}
    \langle \k + \q_1,t | H_{\text{TB}} | \k,b \rangle =& w_0^{\q_1} + w_1^{\q_1} \sigma_x + w_2^{\q_1} \sigma_y + iw_3^{\q_1} \sigma_z,
\end{align}
with $\q_1 = -(\g_1 + \g_2)/3 = (0,-4\pi/(3L_M))$, see Fig. \ref{bmsetup}(a) \footnote{Note that $\q_{1}$ is not a reciprocal lattice vector. This is because the origin of momentum in Eq. (\ref{bwaves}) is different for different layers.}. From symmetry constraints $w_2^{\q_1}=0$, and $w_0^{\q_1}$ and $w_1^{\q_1}$ correspond to the parameters of the BM model \cite{Bistritzer2011},
\begin{align}
    w_0 \equiv w_0^{\q_1}, \hspace{1.5cm} w_1\equiv w_1^{\q_2},
\end{align}
with $w_0 \approx 75$ meV, $w_1 \approx 114$ meV near the magic angle, see Fig. \ref{bmsetup}(b). The coupling $w_3 \equiv w_3^{\q_1} \approx -2.5$ meV is also allowed in general. Analogous terms for momenta $\q_2$ and  $\q_3$ (Fig. \ref{bmsetup}(a)) are related by three-fold rotational symmetry. 

In full generality, interlayer couplings $w_i^{\q}$ with generic momentum exchange $\q = \q_1 + \g$, with $\g$ a reciprocal lattice vector, are non-zero, and $w_i^{\q} = w_i^{\q}(\k)$ are functions of $\k$  \cite{carr19,fang2019angledependentitabinitio,koshino20,kang23,guinea19}. We define $w_i^{\q} \equiv w_i^{\q}(-\q/2)$ (this choice dictated by symmetry constraints); the residuals $w_i^{\q}(\k) - w_i^{\q}$ are denoted 'non-local interlayer couplings' \cite{arbeitman25}. The approximation of constant matrix elements amounts to a local coupling between the layers via the potential \footnote{Keeping momenta $\q$ with $q \leq \sqrt{19}q_1$ is sufficient for convergence.}
\begin{align}
    T(\rr) = \sum_{\q} \big( w_0^{\q}+w_1^{\q} \sigma_x+w_2^{\q} \sigma_y+ iw_3^{\q} \sigma_z \big) e^{i\q \cdot \rr},
    \label{Tfield}
\end{align}

\begin{figure}
    \raggedright
    (a) \hspace{4cm} (b) \\
    \centering
    \includegraphics[width=0.99\linewidth]{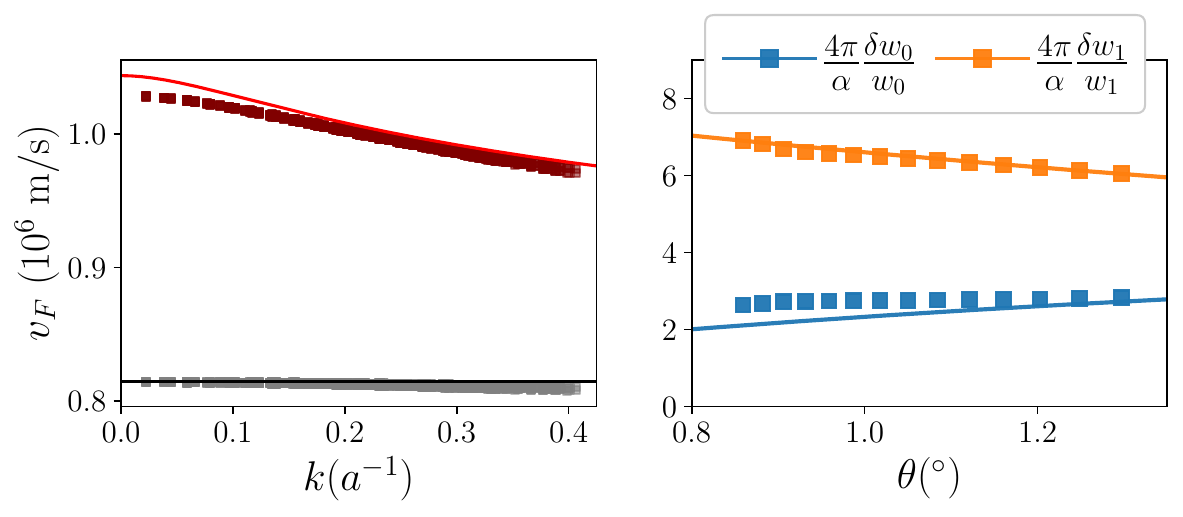}
    \raggedright
    (c) \\
    \centering
    \includegraphics[width=0.99\linewidth]{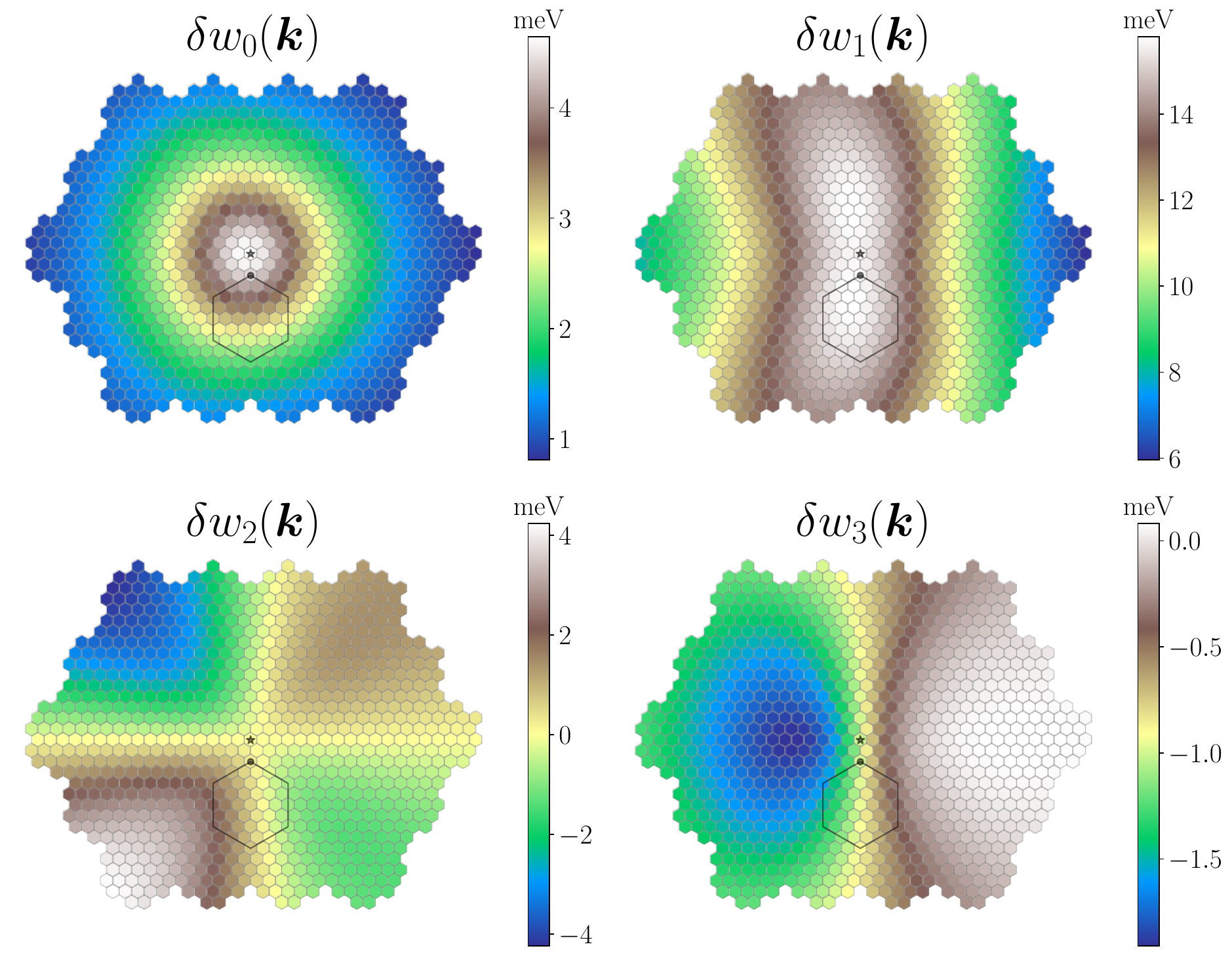}
    \caption{(a) Fermi velocity renormalization in TBG. Red and gray colors correspond to non-interacting and renormalized quantities, respectively. (b) Renormalization of the interlayer couplings, $w_0$, $w_1$, with $\alpha = e^2/(v_F \epsilon)$. The squares correspond to the numerical results and the curves to the analytical predictions. (c) Momentum dependence of the renormalizations of the interlayer couplings. For reference, one moiré Brillouin zone, $\k = \boldsymbol{0}$ (dot) and $\k=-\q_1/2$ (star) are depicted. We set $\epsilon=10$ and $\xi=10$ nm in all panels and $\theta=1.050^\circ$ in (a) and (c).}
    \label{renormbm}
\end{figure}

\paragraph{\underline{Fermi velocity and interlayer couplings renormalization.}}

To first order in the Coulomb interaction, the renormalization of the Fermi velocity $\delta v_F$ reads \cite{Gonzalez1994,Lee2026,barnes14,hofmann14},
\begin{align}
    \frac{\delta v_F(k)}{v_F} = \frac{\alpha}{4}\log\bigg( \frac{4\Lambda/\sqrt{e}}{k}\bigg), 
\end{align}
where $k=|\k|$, $\Lambda$ is a momentum cutoff and 
\begin{align}
    \alpha=\frac{e^2}{\epsilon v_F} \approx \frac{2.7}{\epsilon}
\end{align}
is the graphene fine structure constant.
In the proximity of the metallic gates
the Fermi velocity correction closely follows the analytical formula 
\begin{align}
    \frac{\delta v_F(k)}{v_F} = \frac{\alpha}{4}\log\bigg( \frac{4 \Lambda /\sqrt{e}}{k + p(k\xi)\xi^{-1}}\bigg), \label{vfrenorm}
\end{align}
with $p(k\xi)$ an $O(1)$ function that smoothly interpolates between the $k\xi \to 0$ and $k\xi \to \infty$ limits \cite{supplement}. Fitting to monolayer graphene data yields $\Lambda=3.92 a^{-1}$.

The renormalizations of the interlayer couplings $\delta w_0$, $\delta w_1$ are given to first order in $\alpha$ and $w_{0}$, $w_1$ by
\begin{align}
\delta &w_0 = \frac{\alpha w_0}{4\pi} \int {d\q} \frac{1}{|\q+\frac{\q_1}{2}| + |\q-\frac{\q_1}{2}|} \nonumber \\ 
&\times \Bigg\{ 1 - \frac{q^2-\frac{q_1^2}{4}}{|\q+ \frac{\q_1}{2}||\q - \frac{\q_1}{2}|}\Bigg\} \Bigg\{\frac{e^{-qd}}{q} -\frac{1 - \tanh \big(\frac{q\xi}{2} \big)}{q}  \Bigg\}, \label{dw0main}
\end{align}  
\begin{align}  
\delta &w_1 = \frac{\alpha w_1}{4\pi} \int {d\q} \frac{1}{|\q+\frac{\q_1}{2}| + |\q-\frac{\q_1}{2}|} \nonumber \\
& \times \Bigg\{ 1 - \frac{q_x^2 - q_y^2 + \frac{q_{1}^2}{4}}{|\q+\frac{\q_1}{2}||\q-\frac{\q_1}{2}|}\Bigg\} \Bigg\{\frac{e^{-qd}}{q} -\frac{1 - \tanh \big( \frac{q\xi}{2} \big)}{ q}  \Bigg\},  \label{dw1main} 
\end{align}  
and $\delta w_2 = 0$ as required by symmetry. $d=3.35 \text{ \AA}$ is the interlayer distance. The leading terms of $\delta w_0$, $\delta w_1$ can be computed explicitly in the $q_1d \ll 1$ limit, applicable near the magic angle, for $\xi=\infty$:
\begin{align}
    &\frac{4\pi}{\alpha} \frac{\delta w_0}{w_0} \approx 2\pi - \frac{\pi^2}{4} q_1d,\\
    &\frac{4\pi}{\alpha} \frac{\delta w_1}{w_1} \approx \pi \log \bigg(\frac{8 e^{3/2-\gamma - \pi}}{q_1d}\bigg) + \frac{\pi^2}{4} q_1d, 
\end{align}
with $\gamma$ the Euler-Mascheroni constant. The logarithmic dependence on the UV cutoff ($d^{-1}$) was first derived by Kang and Vafek \cite{kang20}. We identify the IR cutoff $q_1$ and compute explicitly the constant and linear coefficients of $\delta w_1$ and $\delta w_0$ \cite{supplement}. 

An analogous formula for $\delta w_3$ can be readily derived. However, due to its smallness, the first nontrivial correction to $w_3$ must also incorporate contributions of higher order in parametrically larger couplings. We postpone such higher-order calculations for future works.

By projecting the mean-field hamiltonian to the plane-wave basis, we extract the numerical values of $\delta v_F(k)$, $\delta w_0$ and $\delta w_1$. We compare the numerical data with the predictions from Eqs. (\ref{vfrenorm}), (\ref{dw0main}), (\ref{dw1main}) in Fig. \ref{renormbm}(a) and (b), finding a remarkably good agreement (in Fig. \ref{renormbm}(a) $v_F(k) = v_F + \delta v_F(k)$ is reported in physical units). Our results might be viewed the starting point of a systematic perturbative treatment of single-particle terms in interacting TBG \cite{escudero24}.

\paragraph{\underline{Reproducing the numerical solution.}}

In Fig. \ref{convergebands} we consider the band structure at the renormalized magic angle. In Fig. \ref{convergebands}(a) we include $\delta v_F(k)$, $\delta w_0$ and $\delta w_1$ according to the derived analytical expressions as well as directly extracted from the numerics. All remaining couplings are kept at their bare values. The phenomenology of the magic angle renormalization and the increased gap to the remote bands is well captured by this minimal renormalization scheme. 

More renormalized couplings are however necessary to achieve quantitative agreement with the converged bands. At this point, let us introduce the notation for the intralayer blocks (in the approximation of constant matrix elements),
\begin{align}
    \langle \k + \g, \ell | H_{\text{TB}} | \k, \ell \rangle =&  A_0^{\ell,\g} + A_1^{\ell,\g} \sigma_x + A_2^{\ell,\g} \sigma_y + i A_3^{\ell,\g} \sigma_z,
\end{align}
or, equivalently, introduce the intralayer potentials \footnote{Keeping momenta $\g$ with $g \leq 3g_1$ is sufficient for convergence.}
\begin{align}
    A^{\ell}(\rr) \equiv \sum_{\g \neq \boldsymbol{0}} &\big( A_0^{\ell,\g}+ A_1^{\ell,\g} \sigma_x + A_2^{\ell,\g} \sigma_y+ i A_3^{\ell,\g} \sigma_z \big) e^{i\g \cdot \rr},
    \label{Afield}
\end{align}
with $\g$ a reciprocal lattice vector. This term, with $A^{\ell,\g}_{0}$ and $A^{\ell,\g}_3$ negligible, is akin to a pseudo-gauge field coupled to the graphene Dirac cones \cite{kang23,koshino20,Ceferino2024}. We note in passing that, similar to $\delta w_3$, the first nontrivial corrections to $A^\ell(\rr)$ must account for higher-order perturbations. For instance, for $g=g_1$ an order-of-magnitude estimate gives $ \big | \delta A_{1,2}^{\ell,\g} \big| \sim \alpha \big| A_{1,2}^{\ell,\g} \big| \sim 17\alpha \text{ meV}$ to first order, whereas second-order interlayer processes give contributions $\big| \delta A^{\ell,\g}_i\big| \sim \frac{\alpha w_{0,1}^2}{v_F q_1} \sim 60\alpha \text{ meV}$ \footnote{As a remark, even if $A_{0,3}^{\ell,\g}$ are negligible, the renormalizations $\delta A_{0,3}^{\ell,\g}$ can be nonzero in general}.

In Fig. \ref{convergebands}(b) we consider the renormalized the Fermi velocity and the local fields $T(\rr)$ and $A^\ell(\rr)$, with all remaining terms equal to their bare values, and find overall good agreement with the converged bands. In summary, Fig. \ref{convergebands} demonstrates the possibility of reproducing the numerical solution with only a handful of renormalized parameters, potentially derived analytically. 

Further including the momentum dependence of the renormalized interlayer couplings in Fig. \ref{convergebands}(b) matches the flat bands almost exactly, with discrepancies always less than $1$ meV. In Fig. \ref{renormbm}(c) we show the renormalizations of the interlayer couplings with momentum exchange $\q_1$ as a function of $\k$, $\delta w_i(\k)$, which have the predominant effect on the band structure. An analysis of their features can be found in the Supplementary Materials \cite{supplement}.

\begin{figure}
    \centering
    \includegraphics[width=0.99\linewidth]{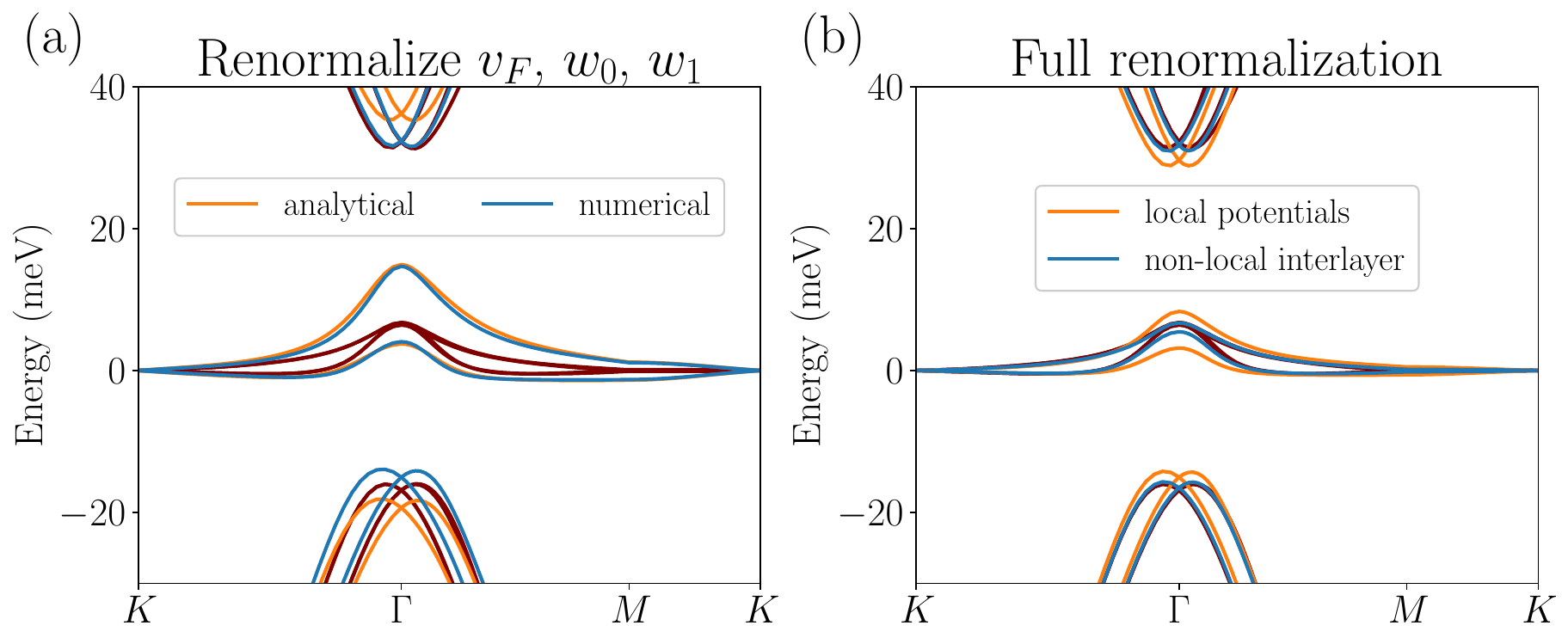}
    \caption{(a) Band structure at $0.882^\circ$ with the renormalized $v_F$, $w_0$ and $w_1$ obtained from the analytical expressions (orange) and extracted from the numerical solution (blue). (b) With renormalized $v_F$ and the interlayer and intralayer potentials $T(\rr)$ and $A^\ell(\rr)$ (orange), and further including the full $\k$-dependent (non-local) interlayer couplings (blue). In both panels the red bands are the full numerical solution, and we only show the valley $K$ bands of the various approximations.}
    \label{convergebands}
\end{figure}

\paragraph{\underline{Tuning the flat bands.}}

The strength of the Coulomb interactions depends on the dielectric and metallic environment around the material. For example, interactions are stronger in a suspended sample, with $\epsilon \approx 10$ and $\xi=60$ nm, than in an hBN encapsulated sample with $\epsilon \approx 15$ and $\xi= 10$ nm. We expect then differences in the flat bands and the renormalized magic angles in both cases. In Fig \ref{bmbands}(a) we plot the bandwidth at the $\Gamma$ point of the flat bands of the BM model with renormalized $v_F$, $w_0$ and $w_1$ according to Eqs. (\ref{vfrenorm}), (\ref{dw0main}), (\ref{dw1main}) for both setups, as well as of the bare BM model. We use the bare values $v_F = 2.416\text{ eV}\cdot a$, $w_1=110$ meV and $w_0/w_1=0.8$ \cite{arbeitman25}\footnote{Note that $v_F = 2.416\text{ eV}\cdot a$ is $10 \%$ larger than the $\textit{ab initio}$ value. This choice effectively includes Fermi velocity renormalization and places the magic angle of the BM model at $1.08^\circ$. We stress that, when comparing the bare and renormalized results for, say, the flat Fermi velocity at intermediate twist angles, the $\textit{ab initio}$ $v_F$ must be used for consistency.}, and set $\Lambda = 1.8 a^{-1}$ (this choice is clarified in the Discussion). As anticipated, the magic angle shifts towards lower values, and the shift is larger in the suspended sample. In Fig. \ref{bmbands}(b) we plot the gap between the flat and remote bands at the $\Gamma$ point on the electron-doped side. The gap increases consistently in both experimental setups. Notice that past the magic angle it reaches $0$, 
at which point the flat bandwidth starts decreasing again. In Fig. \ref{bmbands}(c) we show the Fermi velocity of the flat bands for intermediate twist angles in units of $v_F$. The data follows the approximate formula \cite{esparza25}
\begin{align}
    \frac{v_{\text{flat band}}}{v_F} \approx \frac{v_F(0) - 3\alpha_1^2 v_F(q_1)}{ v_F(1 + 3\alpha_0^2 + 3\alpha_1^2) }, 
    \label{vfb}
\end{align}
with $\alpha_i=w_i/\big(v_F(q_1)q_1\big)$.

In summary, Fig. \ref{bmbands} illustrates the possibility of engineering the flat bands by the dielectric and metallic environments of the experiment \cite{Stepanov2020}. At intermediate twist angles, the flat-band velocity is sensitive to $v_F$, $w_0$ and $w_1$. Experiments in this range could provide their effective values and test our results \cite{luican11,cao16}.

\begin{figure}
    \centering
    \includegraphics[width=0.99\linewidth]{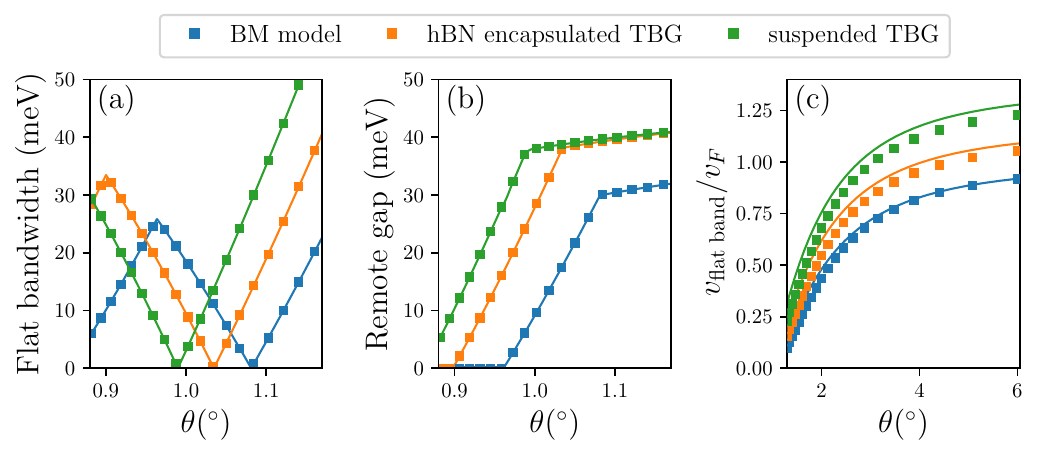}
    \caption{Renormalization in the BM model. (a) Bandwidth of the flat bands at $\Gamma$. (b) Gap between the flat and remote bands on the electron side at $\Gamma$. (c) Fermi velocity of the flat bands in units of the bare Fermi velocity. In (a) and (b) the lines are guides to the eye and in (c) they are given by Eq. (\ref{vfb}) . Blue, orange and green colors correspond to the non-interacting system, a sample encapsulated in hBN ($\epsilon=15$) with $\xi=10$ nm and a suspended sample ($\epsilon=10$) with $\xi=60$ nm, respectively.}
    \label{bmbands}
\end{figure}

\paragraph{\underline{Discussion.}}

The results of this work highlight the importance of non-local exchange and correlations beyond \textit{ab initio} methods in TBG, not only within the low-energy subspace but also associated to the bulk of remote bands \cite{kang20,huang25,lu2025numericalperspectivemoiresuperlattices,lu2026generalmanybodyperturbationframework,guo24}, a direction that merits further consideration. 

A natural next step is the quantitative treatment of screening. For instance, near the magic angle the flattening of the dispersion induces stronger screening relative to the limit of decoupled layers for $q \lesssim 1$ nm$^{-1}$ \cite{pizarro19,vanhala20,goodwin19,zhang22}. Assuming an uncorrelated flat manifold, the RPA dielectric function can grow very large at small momentum transfers due to flat-band transitions \cite{pizarro19,goodwin19}. Besides, theoretical efforts past static screening in graphene include various implementations of dynamical RPA \cite{gonzalez99,guandalini24,Polini2007,hofmann14}
among other methods \cite{bauer15,frassdorf17,Tang2018,sodemann12,dejuan10,barnes14}.

On another note, in this work we considered structurally relaxed, unstrained TBG samples. Strain can be encoded as a $\sim 10$ meV perturbation in the BM model \cite{arbeitman25}, so our main conclusions are robust against the presence of strain. Finally, let us note that we have considered carbon orbitals with effectively zero extent, and extracted a momentum cutoff $\Lambda =3.92 a^{-1}$. Accounting for the finite extent of the carbon orbitals the cutoff is reduced to $\Lambda \sim 1.8a^{-1}$ \cite{stauber17}, in good agreement with experiments \cite{Lee2026,Hwang2012,Siegel2011,Ryu2017}. 

An important observation is in order at this point: in TBG experiments, the interacting phenomena such as superconductivity \cite{Cao2018,Cao2021,Lu2019,Stepanov2020,Yankowitz2019}, the correlated insulators \cite{Cao2018_2,Lu2019, Nuckolls2023, Stepanov2020,Yankowitz2019} and the electronic 'cascades' \cite{Wong2020,Zondiner2020,Saito2021,Rozen2021,zhang2025heavyfermionsmassrenormalization,xiao2025interactingenergybandsmagic} emerge around the magic angle of $1.1^\circ$, whereas most $\textit{ab initio}$ calculations place the 'magic range' around the twist angle of $1.0^\circ$ \cite{fang2019angledependentitabinitio,zhu2026twistedbilayergraphenefirstprinciples,carr19,carr19_2}. Non-local exchange tends to further decrease this angle as we have shown --- with the precise magnitude depending subtly on the screening. The origin and interpretation of this apparent discrepancy remain to be clarified.

This tension leads us to speculate about the role of a finite dispersion in the correlated phenomena. For instance, in the context of superconductivity, the Berezinskii-Kosterlitz-Thouless (BKT) transition temperature, $k_B T_{BKT} =(\pi/2) \rho_s(T_{BKT})$ \cite{Berezinskii72, Kosterlitz1973, Kosterlitz1974, Nelson77}, is controlled by the superfluid weight $\rho_s$, which in turn can be related to the bandwidth \cite{Peotta2015,Scalapino93} (assuming not-too-large quantum geometric contributions \cite{Hu19,Julku20}). Moreover, in BCS theory the critical temperature, $k_BT_c=\Lambda_B\exp(-1/(g\rho(\varepsilon_F)))$, with energy cutoff $\Lambda_B$ and pairing coupling $g$, is enhanced by the density of states at the Fermi level, $\rho(\varepsilon_F)$. In the weak-coupling regime superconductivity could emerge with the onset of flat bands (and/or related splitting of van Hove points \cite{Gonzalez19}), but be suppressed at the magic angle by phase fluctuations according to the BKT scenario. The expected end result is a superconducting dome suppressed at the magic angle $\lesssim 1.0^\circ$, and robust superconductivity at a slightly larger twist angle $\sim 1.1^\circ$ \cite{Cao2021}. More broadly, studies of low-energy theories with a dispersive parent state \cite{sanchez26} could help elucidate the mechanism of superconductivity in TBG.

\paragraph{\underline{Data Availability.}}
All code used for raw data generation \cite{forgezenodo} and figures production \cite{figureszenodo} is openly available at Zenodo. 

\paragraph{\underline{Acknowledgments.}}

We thank Patrick Ledwith, Eslam Khalaf, Pablo Jarillo-Herrero and Suheng Xu for useful discussions. The work was supported by grants PID2023-146461NB-I00 funded by Ministerio de Ciencia, Innovaci\'on y  Universidades, and PRE2021-097070, funded by Ministerio de Ciencia, Innovaci\'on y Universidades through Agencia Estatal de Investigaci\'on, as well as by the CSIC Research Platform on Quantum Technologies PTI-001 and the Severo Ochoa Centres of Excellence program through Grant CEX2024-001445-S. The access to computational resources of CESGA (Centro de Supercomputaci\'on de Galicia) is also gratefully acknowledged.

\bibliography{biblio}

\providecommand{\noopsort}[1]{}\providecommand{\singleletter}[1]{#1}%
\begin{thebibliography}{100}%
\makeatletter
\providecommand \@ifxundefined [1]{%
 \@ifx{#1\undefined}
}%
\providecommand \@ifnum [1]{%
 \ifnum #1\expandafter \@firstoftwo
 \else \expandafter \@secondoftwo
 \fi
}%
\providecommand \@ifx [1]{%
 \ifx #1\expandafter \@firstoftwo
 \else \expandafter \@secondoftwo
 \fi
}%
\providecommand \natexlab [1]{#1}%
\providecommand \enquote  [1]{``#1''}%
\providecommand \bibnamefont  [1]{#1}%
\providecommand \bibfnamefont [1]{#1}%
\providecommand \citenamefont [1]{#1}%
\providecommand \href@noop [0]{\@secondoftwo}%
\providecommand \href [0]{\begingroup \@sanitize@url \@href}%
\providecommand \@href[1]{\@@startlink{#1}\@@href}%
\providecommand \@@href[1]{\endgroup#1\@@endlink}%
\providecommand \@sanitize@url [0]{\catcode `\\12\catcode `\$12\catcode
  `\&12\catcode `\#12\catcode `\^12\catcode `\_12\catcode `\%12\relax}%
\providecommand \@@startlink[1]{}%
\providecommand \@@endlink[0]{}%
\providecommand \url  [0]{\begingroup\@sanitize@url \@url }%
\providecommand \@url [1]{\endgroup\@href {#1}{\urlprefix }}%
\providecommand \urlprefix  [0]{URL }%
\providecommand \Eprint [0]{\href }%
\providecommand \doibase [0]{https://doi.org/}%
\providecommand \selectlanguage [0]{\@gobble}%
\providecommand \bibinfo  [0]{\@secondoftwo}%
\providecommand \bibfield  [0]{\@secondoftwo}%
\providecommand \translation [1]{[#1]}%
\providecommand \BibitemOpen [0]{}%
\providecommand \bibitemStop [0]{}%
\providecommand \bibitemNoStop [0]{.\EOS\space}%
\providecommand \EOS [0]{\spacefactor3000\relax}%
\providecommand \BibitemShut  [1]{\csname bibitem#1\endcsname}%
\let\auto@bib@innerbib\@empty
\bibitem [{\citenamefont {González}\ \emph {et~al.}(1994)\citenamefont
  {González}, \citenamefont {Guinea},\ and\ \citenamefont
  {Vozmediano}}]{Gonzalez1994}%
  \BibitemOpen
  \bibfield  {author} {\bibinfo {author} {\bibfnamefont {J.}~\bibnamefont
  {González}}, \bibinfo {author} {\bibfnamefont {F.}~\bibnamefont {Guinea}},\
  and\ \bibinfo {author} {\bibfnamefont {M.}~\bibnamefont {Vozmediano}},\
  }\href {https://doi.org/10.1016/0550-3213(94)90410-3} {\bibfield  {journal}
  {\bibinfo  {journal} {Nuclear Physics B}\ }\textbf {\bibinfo {volume}
  {424}},\ \bibinfo {pages} {595–618} (\bibinfo {year} {1994})}\BibitemShut
  {NoStop}%
\bibitem [{\citenamefont {Gonz\'alez}\ \emph {et~al.}(1999)\citenamefont
  {Gonz\'alez}, \citenamefont {Guinea},\ and\ \citenamefont
  {Vozmediano}}]{gonzalez99}%
  \BibitemOpen
  \bibfield  {author} {\bibinfo {author} {\bibfnamefont {J.}~\bibnamefont
  {Gonz\'alez}}, \bibinfo {author} {\bibfnamefont {F.}~\bibnamefont {Guinea}},\
  and\ \bibinfo {author} {\bibfnamefont {M.~A.~H.}\ \bibnamefont
  {Vozmediano}},\ }\href {https://doi.org/10.1103/PhysRevB.59.R2474} {\bibfield
   {journal} {\bibinfo  {journal} {Phys. Rev. B}\ }\textbf {\bibinfo {volume}
  {59}},\ \bibinfo {pages} {R2474} (\bibinfo {year} {1999})}\BibitemShut
  {NoStop}%
\bibitem [{\citenamefont {Hofmann}\ \emph {et~al.}(2014)\citenamefont
  {Hofmann}, \citenamefont {Barnes},\ and\ \citenamefont
  {Das~Sarma}}]{hofmann14}%
  \BibitemOpen
  \bibfield  {author} {\bibinfo {author} {\bibfnamefont {J.}~\bibnamefont
  {Hofmann}}, \bibinfo {author} {\bibfnamefont {E.}~\bibnamefont {Barnes}},\
  and\ \bibinfo {author} {\bibfnamefont {S.}~\bibnamefont {Das~Sarma}},\ }\href
  {https://doi.org/10.1103/PhysRevLett.113.105502} {\bibfield  {journal}
  {\bibinfo  {journal} {Phys. Rev. Lett.}\ }\textbf {\bibinfo {volume} {113}},\
  \bibinfo {pages} {105502} (\bibinfo {year} {2014})}\BibitemShut {NoStop}%
\bibitem [{\citenamefont {Barnes}\ \emph {et~al.}(2014)\citenamefont {Barnes},
  \citenamefont {Hwang}, \citenamefont {Throckmorton},\ and\ \citenamefont
  {Das~Sarma}}]{barnes14}%
  \BibitemOpen
  \bibfield  {author} {\bibinfo {author} {\bibfnamefont {E.}~\bibnamefont
  {Barnes}}, \bibinfo {author} {\bibfnamefont {E.~H.}\ \bibnamefont {Hwang}},
  \bibinfo {author} {\bibfnamefont {R.~E.}\ \bibnamefont {Throckmorton}},\ and\
  \bibinfo {author} {\bibfnamefont {S.}~\bibnamefont {Das~Sarma}},\ }\href
  {https://doi.org/10.1103/PhysRevB.89.235431} {\bibfield  {journal} {\bibinfo
  {journal} {Phys. Rev. B}\ }\textbf {\bibinfo {volume} {89}},\ \bibinfo
  {pages} {235431} (\bibinfo {year} {2014})}\BibitemShut {NoStop}%
\bibitem [{\citenamefont {Sodemann}\ and\ \citenamefont
  {Fogler}(2012)}]{sodemann12}%
  \BibitemOpen
  \bibfield  {author} {\bibinfo {author} {\bibfnamefont {I.}~\bibnamefont
  {Sodemann}}\ and\ \bibinfo {author} {\bibfnamefont {M.~M.}\ \bibnamefont
  {Fogler}},\ }\href {https://doi.org/10.1103/PhysRevB.86.115408} {\bibfield
  {journal} {\bibinfo  {journal} {Phys. Rev. B}\ }\textbf {\bibinfo {volume}
  {86}},\ \bibinfo {pages} {115408} (\bibinfo {year} {2012})}\BibitemShut
  {NoStop}%
\bibitem [{\citenamefont {Jung}\ and\ \citenamefont
  {MacDonald}(2011)}]{jung11}%
  \BibitemOpen
  \bibfield  {author} {\bibinfo {author} {\bibfnamefont {J.}~\bibnamefont
  {Jung}}\ and\ \bibinfo {author} {\bibfnamefont {A.~H.}\ \bibnamefont
  {MacDonald}},\ }\href {https://doi.org/10.1103/PhysRevB.84.085446} {\bibfield
   {journal} {\bibinfo  {journal} {Phys. Rev. B}\ }\textbf {\bibinfo {volume}
  {84}},\ \bibinfo {pages} {085446} (\bibinfo {year} {2011})}\BibitemShut
  {NoStop}%
\bibitem [{\citenamefont {de~Juan}\ \emph {et~al.}(2010)\citenamefont
  {de~Juan}, \citenamefont {Grushin},\ and\ \citenamefont
  {Vozmediano}}]{dejuan10}%
  \BibitemOpen
  \bibfield  {author} {\bibinfo {author} {\bibfnamefont {F.}~\bibnamefont
  {de~Juan}}, \bibinfo {author} {\bibfnamefont {A.~G.}\ \bibnamefont
  {Grushin}},\ and\ \bibinfo {author} {\bibfnamefont {M.~A.~H.}\ \bibnamefont
  {Vozmediano}},\ }\href {https://doi.org/10.1103/PhysRevB.82.125409}
  {\bibfield  {journal} {\bibinfo  {journal} {Phys. Rev. B}\ }\textbf {\bibinfo
  {volume} {82}},\ \bibinfo {pages} {125409} (\bibinfo {year}
  {2010})}\BibitemShut {NoStop}%
\bibitem [{\citenamefont {Fr\"a\ss{}dorf}\ and\ \citenamefont
  {Mosig}(2017)}]{frassdorf17}%
  \BibitemOpen
  \bibfield  {author} {\bibinfo {author} {\bibfnamefont {C.}~\bibnamefont
  {Fr\"a\ss{}dorf}}\ and\ \bibinfo {author} {\bibfnamefont {J.~E.~M.}\
  \bibnamefont {Mosig}},\ }\href {https://doi.org/10.1103/PhysRevB.95.125412}
  {\bibfield  {journal} {\bibinfo  {journal} {Phys. Rev. B}\ }\textbf {\bibinfo
  {volume} {95}},\ \bibinfo {pages} {125412} (\bibinfo {year}
  {2017})}\BibitemShut {NoStop}%
\bibitem [{\citenamefont {Bauer}\ \emph {et~al.}(2015)\citenamefont {Bauer},
  \citenamefont {R\"uckriegel}, \citenamefont {Sharma},\ and\ \citenamefont
  {Kopietz}}]{bauer15}%
  \BibitemOpen
  \bibfield  {author} {\bibinfo {author} {\bibfnamefont {C.}~\bibnamefont
  {Bauer}}, \bibinfo {author} {\bibfnamefont {A.}~\bibnamefont {R\"uckriegel}},
  \bibinfo {author} {\bibfnamefont {A.}~\bibnamefont {Sharma}},\ and\ \bibinfo
  {author} {\bibfnamefont {P.}~\bibnamefont {Kopietz}},\ }\href
  {https://doi.org/10.1103/PhysRevB.92.121409} {\bibfield  {journal} {\bibinfo
  {journal} {Phys. Rev. B}\ }\textbf {\bibinfo {volume} {92}},\ \bibinfo
  {pages} {121409} (\bibinfo {year} {2015})}\BibitemShut {NoStop}%
\bibitem [{\citenamefont {Stauber}\ \emph {et~al.}(2017)\citenamefont
  {Stauber}, \citenamefont {Parida}, \citenamefont {Trushin}, \citenamefont
  {Ulybyshev}, \citenamefont {Boyda},\ and\ \citenamefont
  {Schliemann}}]{stauber17}%
  \BibitemOpen
  \bibfield  {author} {\bibinfo {author} {\bibfnamefont {T.}~\bibnamefont
  {Stauber}}, \bibinfo {author} {\bibfnamefont {P.}~\bibnamefont {Parida}},
  \bibinfo {author} {\bibfnamefont {M.}~\bibnamefont {Trushin}}, \bibinfo
  {author} {\bibfnamefont {M.~V.}\ \bibnamefont {Ulybyshev}}, \bibinfo {author}
  {\bibfnamefont {D.~L.}\ \bibnamefont {Boyda}},\ and\ \bibinfo {author}
  {\bibfnamefont {J.}~\bibnamefont {Schliemann}},\ }\href
  {https://doi.org/10.1103/PhysRevLett.118.266801} {\bibfield  {journal}
  {\bibinfo  {journal} {Phys. Rev. Lett.}\ }\textbf {\bibinfo {volume} {118}},\
  \bibinfo {pages} {266801} (\bibinfo {year} {2017})}\BibitemShut {NoStop}%
\bibitem [{\citenamefont {Guandalini}\ \emph {et~al.}(2024)\citenamefont
  {Guandalini}, \citenamefont {Leon}, \citenamefont {D'Amico}, \citenamefont
  {Cardoso}, \citenamefont {Ferretti}, \citenamefont {Rontani},\ and\
  \citenamefont {Varsano}}]{guandalini24}%
  \BibitemOpen
  \bibfield  {author} {\bibinfo {author} {\bibfnamefont {A.}~\bibnamefont
  {Guandalini}}, \bibinfo {author} {\bibfnamefont {D.~A.}\ \bibnamefont
  {Leon}}, \bibinfo {author} {\bibfnamefont {P.}~\bibnamefont {D'Amico}},
  \bibinfo {author} {\bibfnamefont {C.}~\bibnamefont {Cardoso}}, \bibinfo
  {author} {\bibfnamefont {A.}~\bibnamefont {Ferretti}}, \bibinfo {author}
  {\bibfnamefont {M.}~\bibnamefont {Rontani}},\ and\ \bibinfo {author}
  {\bibfnamefont {D.}~\bibnamefont {Varsano}},\ }\href
  {https://doi.org/10.1103/PhysRevB.109.075120} {\bibfield  {journal} {\bibinfo
   {journal} {Phys. Rev. B}\ }\textbf {\bibinfo {volume} {109}},\ \bibinfo
  {pages} {075120} (\bibinfo {year} {2024})}\BibitemShut {NoStop}%
\bibitem [{\citenamefont {Tang}\ \emph {et~al.}(2018)\citenamefont {Tang},
  \citenamefont {Leaw}, \citenamefont {Rodrigues}, \citenamefont {Herbut},
  \citenamefont {Sengupta}, \citenamefont {Assaad},\ and\ \citenamefont
  {Adam}}]{Tang2018}%
  \BibitemOpen
  \bibfield  {author} {\bibinfo {author} {\bibfnamefont {H.-K.}\ \bibnamefont
  {Tang}}, \bibinfo {author} {\bibfnamefont {J.~N.}\ \bibnamefont {Leaw}},
  \bibinfo {author} {\bibfnamefont {J.~N.~B.}\ \bibnamefont {Rodrigues}},
  \bibinfo {author} {\bibfnamefont {I.~F.}\ \bibnamefont {Herbut}}, \bibinfo
  {author} {\bibfnamefont {P.}~\bibnamefont {Sengupta}}, \bibinfo {author}
  {\bibfnamefont {F.~F.}\ \bibnamefont {Assaad}},\ and\ \bibinfo {author}
  {\bibfnamefont {S.}~\bibnamefont {Adam}},\ }\href
  {https://doi.org/10.1126/science.aao2934} {\bibfield  {journal} {\bibinfo
  {journal} {Science}\ }\textbf {\bibinfo {volume} {361}},\ \bibinfo {pages}
  {570–574} (\bibinfo {year} {2018})}\BibitemShut {NoStop}%
\bibitem [{\citenamefont {Polini}\ \emph {et~al.}(2007)\citenamefont {Polini},
  \citenamefont {Asgari}, \citenamefont {Barlas}, \citenamefont
  {Pereg-Barnea},\ and\ \citenamefont {MacDonald}}]{Polini2007}%
  \BibitemOpen
  \bibfield  {author} {\bibinfo {author} {\bibfnamefont {M.}~\bibnamefont
  {Polini}}, \bibinfo {author} {\bibfnamefont {R.}~\bibnamefont {Asgari}},
  \bibinfo {author} {\bibfnamefont {Y.}~\bibnamefont {Barlas}}, \bibinfo
  {author} {\bibfnamefont {T.}~\bibnamefont {Pereg-Barnea}},\ and\ \bibinfo
  {author} {\bibfnamefont {A.}~\bibnamefont {MacDonald}},\ }\href
  {https://doi.org/10.1016/j.ssc.2007.04.035} {\bibfield  {journal} {\bibinfo
  {journal} {Solid State Communications}\ }\textbf {\bibinfo {volume} {143}},\
  \bibinfo {pages} {58–62} (\bibinfo {year} {2007})}\BibitemShut {NoStop}%
\bibitem [{\citenamefont {Kotov}\ \emph {et~al.}(2012)\citenamefont {Kotov},
  \citenamefont {Uchoa}, \citenamefont {Pereira}, \citenamefont {Guinea},\ and\
  \citenamefont {Castro~Neto}}]{kotov12}%
  \BibitemOpen
  \bibfield  {author} {\bibinfo {author} {\bibfnamefont {V.~N.}\ \bibnamefont
  {Kotov}}, \bibinfo {author} {\bibfnamefont {B.}~\bibnamefont {Uchoa}},
  \bibinfo {author} {\bibfnamefont {V.~M.}\ \bibnamefont {Pereira}}, \bibinfo
  {author} {\bibfnamefont {F.}~\bibnamefont {Guinea}},\ and\ \bibinfo {author}
  {\bibfnamefont {A.~H.}\ \bibnamefont {Castro~Neto}},\ }\href
  {https://doi.org/10.1103/RevModPhys.84.1067} {\bibfield  {journal} {\bibinfo
  {journal} {Rev. Mod. Phys.}\ }\textbf {\bibinfo {volume} {84}},\ \bibinfo
  {pages} {1067} (\bibinfo {year} {2012})}\BibitemShut {NoStop}%
\bibitem [{\citenamefont {Elias}\ \emph {et~al.}(2011)\citenamefont {Elias},
  \citenamefont {Gorbachev}, \citenamefont {Mayorov}, \citenamefont {Morozov},
  \citenamefont {Zhukov}, \citenamefont {Blake}, \citenamefont {Ponomarenko},
  \citenamefont {Grigorieva}, \citenamefont {Novoselov}, \citenamefont
  {Guinea},\ and\ \citenamefont {Geim}}]{Elias2011}%
  \BibitemOpen
  \bibfield  {author} {\bibinfo {author} {\bibfnamefont {D.~C.}\ \bibnamefont
  {Elias}}, \bibinfo {author} {\bibfnamefont {R.~V.}\ \bibnamefont
  {Gorbachev}}, \bibinfo {author} {\bibfnamefont {A.~S.}\ \bibnamefont
  {Mayorov}}, \bibinfo {author} {\bibfnamefont {S.~V.}\ \bibnamefont
  {Morozov}}, \bibinfo {author} {\bibfnamefont {A.~A.}\ \bibnamefont {Zhukov}},
  \bibinfo {author} {\bibfnamefont {P.}~\bibnamefont {Blake}}, \bibinfo
  {author} {\bibfnamefont {L.~A.}\ \bibnamefont {Ponomarenko}}, \bibinfo
  {author} {\bibfnamefont {I.~V.}\ \bibnamefont {Grigorieva}}, \bibinfo
  {author} {\bibfnamefont {K.~S.}\ \bibnamefont {Novoselov}}, \bibinfo {author}
  {\bibfnamefont {F.}~\bibnamefont {Guinea}},\ and\ \bibinfo {author}
  {\bibfnamefont {A.~K.}\ \bibnamefont {Geim}},\ }\href
  {https://doi.org/10.1038/nphys2049} {\bibfield  {journal} {\bibinfo
  {journal} {Nature Physics}\ }\textbf {\bibinfo {volume} {7}},\ \bibinfo
  {pages} {701–704} (\bibinfo {year} {2011})}\BibitemShut {NoStop}%
\bibitem [{\citenamefont {Chae}\ \emph {et~al.}(2012)\citenamefont {Chae},
  \citenamefont {Jung}, \citenamefont {Young}, \citenamefont {Dean},
  \citenamefont {Wang}, \citenamefont {Gao}, \citenamefont {Watanabe},
  \citenamefont {Taniguchi}, \citenamefont {Hone}, \citenamefont {Shepard},
  \citenamefont {Kim}, \citenamefont {Zhitenev},\ and\ \citenamefont
  {Stroscio}}]{Chae12}%
  \BibitemOpen
  \bibfield  {author} {\bibinfo {author} {\bibfnamefont {J.}~\bibnamefont
  {Chae}}, \bibinfo {author} {\bibfnamefont {S.}~\bibnamefont {Jung}}, \bibinfo
  {author} {\bibfnamefont {A.~F.}\ \bibnamefont {Young}}, \bibinfo {author}
  {\bibfnamefont {C.~R.}\ \bibnamefont {Dean}}, \bibinfo {author}
  {\bibfnamefont {L.}~\bibnamefont {Wang}}, \bibinfo {author} {\bibfnamefont
  {Y.}~\bibnamefont {Gao}}, \bibinfo {author} {\bibfnamefont {K.}~\bibnamefont
  {Watanabe}}, \bibinfo {author} {\bibfnamefont {T.}~\bibnamefont {Taniguchi}},
  \bibinfo {author} {\bibfnamefont {J.}~\bibnamefont {Hone}}, \bibinfo {author}
  {\bibfnamefont {K.~L.}\ \bibnamefont {Shepard}}, \bibinfo {author}
  {\bibfnamefont {P.}~\bibnamefont {Kim}}, \bibinfo {author} {\bibfnamefont
  {N.~B.}\ \bibnamefont {Zhitenev}},\ and\ \bibinfo {author} {\bibfnamefont
  {J.~A.}\ \bibnamefont {Stroscio}},\ }\href
  {https://doi.org/10.1103/PhysRevLett.109.116802} {\bibfield  {journal}
  {\bibinfo  {journal} {Phys. Rev. Lett.}\ }\textbf {\bibinfo {volume} {109}},\
  \bibinfo {pages} {116802} (\bibinfo {year} {2012})}\BibitemShut {NoStop}%
\bibitem [{\citenamefont {Siegel}\ \emph {et~al.}(2011)\citenamefont {Siegel},
  \citenamefont {Park}, \citenamefont {Hwang}, \citenamefont {Deslippe},
  \citenamefont {Fedorov}, \citenamefont {Louie},\ and\ \citenamefont
  {Lanzara}}]{Siegel2011}%
  \BibitemOpen
  \bibfield  {author} {\bibinfo {author} {\bibfnamefont {D.~A.}\ \bibnamefont
  {Siegel}}, \bibinfo {author} {\bibfnamefont {C.-H.}\ \bibnamefont {Park}},
  \bibinfo {author} {\bibfnamefont {C.}~\bibnamefont {Hwang}}, \bibinfo
  {author} {\bibfnamefont {J.}~\bibnamefont {Deslippe}}, \bibinfo {author}
  {\bibfnamefont {A.~V.}\ \bibnamefont {Fedorov}}, \bibinfo {author}
  {\bibfnamefont {S.~G.}\ \bibnamefont {Louie}},\ and\ \bibinfo {author}
  {\bibfnamefont {A.}~\bibnamefont {Lanzara}},\ }\href
  {https://doi.org/10.1073/pnas.1100242108} {\bibfield  {journal} {\bibinfo
  {journal} {Proceedings of the National Academy of Sciences}\ }\textbf
  {\bibinfo {volume} {108}},\ \bibinfo {pages} {11365–11369} (\bibinfo {year}
  {2011})}\BibitemShut {NoStop}%
\bibitem [{\citenamefont {Ryu}\ \emph {et~al.}(2017)\citenamefont {Ryu},
  \citenamefont {Hwang}, \citenamefont {Wang}, \citenamefont {Disa},
  \citenamefont {Denlinger}, \citenamefont {Zhang}, \citenamefont {Mo},
  \citenamefont {Hwang},\ and\ \citenamefont {Lanzara}}]{Ryu2017}%
  \BibitemOpen
  \bibfield  {author} {\bibinfo {author} {\bibfnamefont {H.}~\bibnamefont
  {Ryu}}, \bibinfo {author} {\bibfnamefont {J.}~\bibnamefont {Hwang}}, \bibinfo
  {author} {\bibfnamefont {D.}~\bibnamefont {Wang}}, \bibinfo {author}
  {\bibfnamefont {A.~S.}\ \bibnamefont {Disa}}, \bibinfo {author}
  {\bibfnamefont {J.}~\bibnamefont {Denlinger}}, \bibinfo {author}
  {\bibfnamefont {Y.}~\bibnamefont {Zhang}}, \bibinfo {author} {\bibfnamefont
  {S.-K.}\ \bibnamefont {Mo}}, \bibinfo {author} {\bibfnamefont
  {C.}~\bibnamefont {Hwang}},\ and\ \bibinfo {author} {\bibfnamefont
  {A.}~\bibnamefont {Lanzara}},\ }\href
  {https://doi.org/10.1021/acs.nanolett.7b01650} {\bibfield  {journal}
  {\bibinfo  {journal} {Nano Letters}\ }\textbf {\bibinfo {volume} {17}},\
  \bibinfo {pages} {5914–5918} (\bibinfo {year} {2017})}\BibitemShut
  {NoStop}%
\bibitem [{\citenamefont {Yu}\ \emph {et~al.}(2013)\citenamefont {Yu},
  \citenamefont {Jalil}, \citenamefont {Belle}, \citenamefont {Mayorov},
  \citenamefont {Blake}, \citenamefont {Schedin}, \citenamefont {Morozov},
  \citenamefont {Ponomarenko}, \citenamefont {Chiappini}, \citenamefont
  {Wiedmann}, \citenamefont {Zeitler}, \citenamefont {Katsnelson},
  \citenamefont {Geim}, \citenamefont {Novoselov},\ and\ \citenamefont
  {Elias}}]{Yu2013}%
  \BibitemOpen
  \bibfield  {author} {\bibinfo {author} {\bibfnamefont {G.~L.}\ \bibnamefont
  {Yu}}, \bibinfo {author} {\bibfnamefont {R.}~\bibnamefont {Jalil}}, \bibinfo
  {author} {\bibfnamefont {B.}~\bibnamefont {Belle}}, \bibinfo {author}
  {\bibfnamefont {A.~S.}\ \bibnamefont {Mayorov}}, \bibinfo {author}
  {\bibfnamefont {P.}~\bibnamefont {Blake}}, \bibinfo {author} {\bibfnamefont
  {F.}~\bibnamefont {Schedin}}, \bibinfo {author} {\bibfnamefont {S.~V.}\
  \bibnamefont {Morozov}}, \bibinfo {author} {\bibfnamefont {L.~A.}\
  \bibnamefont {Ponomarenko}}, \bibinfo {author} {\bibfnamefont
  {F.}~\bibnamefont {Chiappini}}, \bibinfo {author} {\bibfnamefont
  {S.}~\bibnamefont {Wiedmann}}, \bibinfo {author} {\bibfnamefont
  {U.}~\bibnamefont {Zeitler}}, \bibinfo {author} {\bibfnamefont {M.~I.}\
  \bibnamefont {Katsnelson}}, \bibinfo {author} {\bibfnamefont {A.~K.}\
  \bibnamefont {Geim}}, \bibinfo {author} {\bibfnamefont {K.~S.}\ \bibnamefont
  {Novoselov}},\ and\ \bibinfo {author} {\bibfnamefont {D.~C.}\ \bibnamefont
  {Elias}},\ }\href {https://doi.org/10.1073/pnas.1300599110} {\bibfield
  {journal} {\bibinfo  {journal} {Proceedings of the National Academy of
  Sciences}\ }\textbf {\bibinfo {volume} {110}},\ \bibinfo {pages}
  {3282–3286} (\bibinfo {year} {2013})}\BibitemShut {NoStop}%
\bibitem [{\citenamefont {Knox}\ \emph {et~al.}(2011)\citenamefont {Knox},
  \citenamefont {Locatelli}, \citenamefont {Yilmaz}, \citenamefont {Cvetko},
  \citenamefont {Mente\ifmmode~\mbox{\c{s}}\else \c{s}\fi{}}, \citenamefont
  {Ni\~no}, \citenamefont {Kim}, \citenamefont {Morgante},\ and\ \citenamefont
  {Osgood}}]{knox11}%
  \BibitemOpen
  \bibfield  {author} {\bibinfo {author} {\bibfnamefont {K.~R.}\ \bibnamefont
  {Knox}}, \bibinfo {author} {\bibfnamefont {A.}~\bibnamefont {Locatelli}},
  \bibinfo {author} {\bibfnamefont {M.~B.}\ \bibnamefont {Yilmaz}}, \bibinfo
  {author} {\bibfnamefont {D.}~\bibnamefont {Cvetko}}, \bibinfo {author}
  {\bibfnamefont {T.~O.}\ \bibnamefont {Mente\ifmmode~\mbox{\c{s}}\else
  \c{s}\fi{}}}, \bibinfo {author} {\bibfnamefont {M.~A.}\ \bibnamefont
  {Ni\~no}}, \bibinfo {author} {\bibfnamefont {P.}~\bibnamefont {Kim}},
  \bibinfo {author} {\bibfnamefont {A.}~\bibnamefont {Morgante}},\ and\
  \bibinfo {author} {\bibfnamefont {R.~M.}\ \bibnamefont {Osgood}},\ }\href
  {https://doi.org/10.1103/PhysRevB.84.115401} {\bibfield  {journal} {\bibinfo
  {journal} {Phys. Rev. B}\ }\textbf {\bibinfo {volume} {84}},\ \bibinfo
  {pages} {115401} (\bibinfo {year} {2011})}\BibitemShut {NoStop}%
\bibitem [{\citenamefont {Hwang}\ \emph {et~al.}(2012)\citenamefont {Hwang},
  \citenamefont {Siegel}, \citenamefont {Mo}, \citenamefont {Regan},
  \citenamefont {Ismach}, \citenamefont {Zhang}, \citenamefont {Zettl},\ and\
  \citenamefont {Lanzara}}]{Hwang2012}%
  \BibitemOpen
  \bibfield  {author} {\bibinfo {author} {\bibfnamefont {C.}~\bibnamefont
  {Hwang}}, \bibinfo {author} {\bibfnamefont {D.~A.}\ \bibnamefont {Siegel}},
  \bibinfo {author} {\bibfnamefont {S.-K.}\ \bibnamefont {Mo}}, \bibinfo
  {author} {\bibfnamefont {W.}~\bibnamefont {Regan}}, \bibinfo {author}
  {\bibfnamefont {A.}~\bibnamefont {Ismach}}, \bibinfo {author} {\bibfnamefont
  {Y.}~\bibnamefont {Zhang}}, \bibinfo {author} {\bibfnamefont
  {A.}~\bibnamefont {Zettl}},\ and\ \bibinfo {author} {\bibfnamefont
  {A.}~\bibnamefont {Lanzara}},\ }\bibfield  {journal} {\bibinfo  {journal}
  {Scientific Reports}\ }\textbf {\bibinfo {volume} {2}},\ \href
  {https://doi.org/10.1038/srep00590} {10.1038/srep00590} (\bibinfo {year}
  {2012})\BibitemShut {NoStop}%
\bibitem [{\citenamefont {Lee}\ \emph {et~al.}(2026)\citenamefont {Lee},
  \citenamefont {Das}, \citenamefont {Herzog-Arbeitman}, \citenamefont {Papp},
  \citenamefont {Li}, \citenamefont {Daschner}, \citenamefont {Zhou},
  \citenamefont {Bhatt}, \citenamefont {Currle}, \citenamefont {Yu},
  \citenamefont {Jiang}, \citenamefont {Becherer}, \citenamefont {Mittermeier},
  \citenamefont {Altpeter}, \citenamefont {Obermayer}, \citenamefont {Lorenz},
  \citenamefont {Chavez}, \citenamefont {Le}, \citenamefont {Williams},
  \citenamefont {Watanabe}, \citenamefont {Taniguchi}, \citenamefont
  {Bernevig},\ and\ \citenamefont {Efetov}}]{Lee2026}%
  \BibitemOpen
  \bibfield  {author} {\bibinfo {author} {\bibfnamefont {M.}~\bibnamefont
  {Lee}}, \bibinfo {author} {\bibfnamefont {I.}~\bibnamefont {Das}}, \bibinfo
  {author} {\bibfnamefont {J.}~\bibnamefont {Herzog-Arbeitman}}, \bibinfo
  {author} {\bibfnamefont {J.}~\bibnamefont {Papp}}, \bibinfo {author}
  {\bibfnamefont {J.}~\bibnamefont {Li}}, \bibinfo {author} {\bibfnamefont
  {M.}~\bibnamefont {Daschner}}, \bibinfo {author} {\bibfnamefont
  {Z.}~\bibnamefont {Zhou}}, \bibinfo {author} {\bibfnamefont {M.}~\bibnamefont
  {Bhatt}}, \bibinfo {author} {\bibfnamefont {M.}~\bibnamefont {Currle}},
  \bibinfo {author} {\bibfnamefont {J.}~\bibnamefont {Yu}}, \bibinfo {author}
  {\bibfnamefont {Y.}~\bibnamefont {Jiang}}, \bibinfo {author} {\bibfnamefont
  {M.}~\bibnamefont {Becherer}}, \bibinfo {author} {\bibfnamefont
  {R.}~\bibnamefont {Mittermeier}}, \bibinfo {author} {\bibfnamefont
  {P.}~\bibnamefont {Altpeter}}, \bibinfo {author} {\bibfnamefont
  {C.}~\bibnamefont {Obermayer}}, \bibinfo {author} {\bibfnamefont
  {H.}~\bibnamefont {Lorenz}}, \bibinfo {author} {\bibfnamefont
  {G.}~\bibnamefont {Chavez}}, \bibinfo {author} {\bibfnamefont {B.~T.}\
  \bibnamefont {Le}}, \bibinfo {author} {\bibfnamefont {J.}~\bibnamefont
  {Williams}}, \bibinfo {author} {\bibfnamefont {K.}~\bibnamefont {Watanabe}},
  \bibinfo {author} {\bibfnamefont {T.}~\bibnamefont {Taniguchi}}, \bibinfo
  {author} {\bibfnamefont {B.~A.}\ \bibnamefont {Bernevig}},\ and\ \bibinfo
  {author} {\bibfnamefont {D.~K.}\ \bibnamefont {Efetov}},\ }\bibfield
  {journal} {\bibinfo  {journal} {Nano Letters}\ }\href
  {https://doi.org/10.1021/acs.nanolett.5c05015} {10.1021/acs.nanolett.5c05015}
  (\bibinfo {year} {2026})\BibitemShut {NoStop}%
\bibitem [{\citenamefont {Vafek}\ and\ \citenamefont {Kang}(2020)}]{kang20}%
  \BibitemOpen
  \bibfield  {author} {\bibinfo {author} {\bibfnamefont {O.}~\bibnamefont
  {Vafek}}\ and\ \bibinfo {author} {\bibfnamefont {J.}~\bibnamefont {Kang}},\
  }\href {https://doi.org/10.1103/PhysRevLett.125.257602} {\bibfield  {journal}
  {\bibinfo  {journal} {Phys. Rev. Lett.}\ }\textbf {\bibinfo {volume} {125}},\
  \bibinfo {pages} {257602} (\bibinfo {year} {2020})}\BibitemShut {NoStop}%
\bibitem [{\citenamefont {S\'anchez~S\'anchez}\ \emph
  {et~al.}(2025)\citenamefont {S\'anchez~S\'anchez}, \citenamefont
  {Gonz\'alez},\ and\ \citenamefont {Stauber}}]{sanchez25}%
  \BibitemOpen
  \bibfield  {author} {\bibinfo {author} {\bibfnamefont {M.}~\bibnamefont
  {S\'anchez~S\'anchez}}, \bibinfo {author} {\bibfnamefont {J.}~\bibnamefont
  {Gonz\'alez}},\ and\ \bibinfo {author} {\bibfnamefont {T.}~\bibnamefont
  {Stauber}},\ }\href {https://doi.org/10.1103/PhysRevB.111.205133} {\bibfield
  {journal} {\bibinfo  {journal} {Phys. Rev. B}\ }\textbf {\bibinfo {volume}
  {111}},\ \bibinfo {pages} {205133} (\bibinfo {year} {2025})}\BibitemShut
  {NoStop}%
\bibitem [{\citenamefont {S\'anchez}\ and\ \citenamefont
  {Stauber}(2026)}]{sanchez26}%
  \BibitemOpen
  \bibfield  {author} {\bibinfo {author} {\bibfnamefont {M.~S.}\ \bibnamefont
  {S\'anchez}}\ and\ \bibinfo {author} {\bibfnamefont {T.}~\bibnamefont
  {Stauber}},\ }\href {https://doi.org/10.1103/qlp6-hjf2} {\bibfield  {journal}
  {\bibinfo  {journal} {Phys. Rev. B}\ }\textbf {\bibinfo {volume} {113}},\
  \bibinfo {pages} {035155} (\bibinfo {year} {2026})}\BibitemShut {NoStop}%
\bibitem [{\citenamefont {Huang}\ \emph {et~al.}(2025)\citenamefont {Huang},
  \citenamefont {Chou},\ and\ \citenamefont {Das~Sarma}}]{huang25}%
  \BibitemOpen
  \bibfield  {author} {\bibinfo {author} {\bibfnamefont {Y.}~\bibnamefont
  {Huang}}, \bibinfo {author} {\bibfnamefont {Y.-Z.}\ \bibnamefont {Chou}},\
  and\ \bibinfo {author} {\bibfnamefont {S.}~\bibnamefont {Das~Sarma}},\ }\href
  {https://doi.org/10.1103/2px8-h7c2} {\bibfield  {journal} {\bibinfo
  {journal} {Phys. Rev. B}\ }\textbf {\bibinfo {volume} {112}},\ \bibinfo
  {pages} {245132} (\bibinfo {year} {2025})}\BibitemShut {NoStop}%
\bibitem [{\citenamefont {Lu}\ \emph {et~al.}(2025)\citenamefont {Lu},
  \citenamefont {Xie},\ and\ \citenamefont
  {Liu}}]{lu2025numericalperspectivemoiresuperlattices}%
  \BibitemOpen
  \bibfield  {author} {\bibinfo {author} {\bibfnamefont {X.}~\bibnamefont
  {Lu}}, \bibinfo {author} {\bibfnamefont {B.}~\bibnamefont {Xie}},\ and\
  \bibinfo {author} {\bibfnamefont {J.}~\bibnamefont {Liu}},\ }\href
  {https://arxiv.org/abs/2512.07115} {\bibinfo {title} {A numerical perspective
  on moir\'e superlattices: From single-particle properties to many-body
  physics}} (\bibinfo {year} {2025}),\ \Eprint
  {https://arxiv.org/abs/2512.07115} {arXiv:2512.07115 [cond-mat.str-el]}
  \BibitemShut {NoStop}%
\bibitem [{\citenamefont {Lu}\ \emph {et~al.}(2026)\citenamefont {Lu},
  \citenamefont {Yang}, \citenamefont {Guo},\ and\ \citenamefont
  {Liu}}]{lu2026generalmanybodyperturbationframework}%
  \BibitemOpen
  \bibfield  {author} {\bibinfo {author} {\bibfnamefont {X.}~\bibnamefont
  {Lu}}, \bibinfo {author} {\bibfnamefont {Y.}~\bibnamefont {Yang}}, \bibinfo
  {author} {\bibfnamefont {Z.}~\bibnamefont {Guo}},\ and\ \bibinfo {author}
  {\bibfnamefont {J.}~\bibnamefont {Liu}},\ }\href
  {https://arxiv.org/abs/2509.19764} {\bibinfo {title} {General many-body
  perturbation framework for moir\'e systems}} (\bibinfo {year} {2026}),\
  \Eprint {https://arxiv.org/abs/2509.19764} {arXiv:2509.19764
  [cond-mat.str-el]} \BibitemShut {NoStop}%
\bibitem [{\citenamefont {Guo}\ \emph {et~al.}(2024)\citenamefont {Guo},
  \citenamefont {Lu}, \citenamefont {Xie},\ and\ \citenamefont {Liu}}]{guo24}%
  \BibitemOpen
  \bibfield  {author} {\bibinfo {author} {\bibfnamefont {Z.}~\bibnamefont
  {Guo}}, \bibinfo {author} {\bibfnamefont {X.}~\bibnamefont {Lu}}, \bibinfo
  {author} {\bibfnamefont {B.}~\bibnamefont {Xie}},\ and\ \bibinfo {author}
  {\bibfnamefont {J.}~\bibnamefont {Liu}},\ }\href
  {https://doi.org/10.1103/PhysRevB.110.075109} {\bibfield  {journal} {\bibinfo
   {journal} {Phys. Rev. B}\ }\textbf {\bibinfo {volume} {110}},\ \bibinfo
  {pages} {075109} (\bibinfo {year} {2024})}\BibitemShut {NoStop}%
\bibitem [{\citenamefont {Stauber}\ \emph {et~al.}(2025)\citenamefont
  {Stauber}, \citenamefont {Sánchez~Sánchez}, \citenamefont {Vasilevskiy},
  \citenamefont {González}, \citenamefont {Mouriño~Gallego}, \citenamefont
  {Wackerl}, \citenamefont {Wenk},\ and\ \citenamefont
  {Schliemann}}]{forgezenodo}%
  \BibitemOpen
  \bibfield  {author} {\bibinfo {author} {\bibfnamefont {T.}~\bibnamefont
  {Stauber}}, \bibinfo {author} {\bibfnamefont {M.}~\bibnamefont
  {Sánchez~Sánchez}}, \bibinfo {author} {\bibfnamefont {I.}~\bibnamefont
  {Vasilevskiy}}, \bibinfo {author} {\bibfnamefont {J.}~\bibnamefont
  {González}}, \bibinfo {author} {\bibfnamefont {J.~C.}\ \bibnamefont
  {Mouriño~Gallego}}, \bibinfo {author} {\bibfnamefont {M.}~\bibnamefont
  {Wackerl}}, \bibinfo {author} {\bibfnamefont {P.}~\bibnamefont {Wenk}},\ and\
  \bibinfo {author} {\bibfnamefont {J.}~\bibnamefont {Schliemann}},\ }\href
  {https://doi.org/10.5281/ZENODO.17368106} {\bibinfo {title} {Forge: Fock
  optimization of real-space giant environment}},\ \bibinfo {howpublished}
  {\url{https://zenodo.org/doi/10.5281/zenodo.17368106}} (\bibinfo {year}
  {2025})\BibitemShut {NoStop}%
\bibitem [{\citenamefont {Carr}\ \emph {et~al.}(2018)\citenamefont {Carr},
  \citenamefont {Massatt}, \citenamefont {Torrisi}, \citenamefont {Cazeaux},
  \citenamefont {Luskin},\ and\ \citenamefont {Kaxiras}}]{carr18}%
  \BibitemOpen
  \bibfield  {author} {\bibinfo {author} {\bibfnamefont {S.}~\bibnamefont
  {Carr}}, \bibinfo {author} {\bibfnamefont {D.}~\bibnamefont {Massatt}},
  \bibinfo {author} {\bibfnamefont {S.~B.}\ \bibnamefont {Torrisi}}, \bibinfo
  {author} {\bibfnamefont {P.}~\bibnamefont {Cazeaux}}, \bibinfo {author}
  {\bibfnamefont {M.}~\bibnamefont {Luskin}},\ and\ \bibinfo {author}
  {\bibfnamefont {E.}~\bibnamefont {Kaxiras}},\ }\href
  {https://doi.org/10.1103/PhysRevB.98.224102} {\bibfield  {journal} {\bibinfo
  {journal} {Phys. Rev. B}\ }\textbf {\bibinfo {volume} {98}},\ \bibinfo
  {pages} {224102} (\bibinfo {year} {2018})}\BibitemShut {NoStop}%
\bibitem [{\citenamefont {Carr}\ \emph
  {et~al.}(2019{\natexlab{a}})\citenamefont {Carr}, \citenamefont {Fang},
  \citenamefont {Zhu},\ and\ \citenamefont {Kaxiras}}]{carr19}%
  \BibitemOpen
  \bibfield  {author} {\bibinfo {author} {\bibfnamefont {S.}~\bibnamefont
  {Carr}}, \bibinfo {author} {\bibfnamefont {S.}~\bibnamefont {Fang}}, \bibinfo
  {author} {\bibfnamefont {Z.}~\bibnamefont {Zhu}},\ and\ \bibinfo {author}
  {\bibfnamefont {E.}~\bibnamefont {Kaxiras}},\ }\href
  {https://doi.org/10.1103/PhysRevResearch.1.013001} {\bibfield  {journal}
  {\bibinfo  {journal} {Phys. Rev. Res.}\ }\textbf {\bibinfo {volume} {1}},\
  \bibinfo {pages} {013001} (\bibinfo {year} {2019}{\natexlab{a}})}\BibitemShut
  {NoStop}%
\bibitem [{\citenamefont {Nam}\ and\ \citenamefont
  {Koshino}(2017)}]{koshino20_1}%
  \BibitemOpen
  \bibfield  {author} {\bibinfo {author} {\bibfnamefont {N.~N.~T.}\
  \bibnamefont {Nam}}\ and\ \bibinfo {author} {\bibfnamefont {M.}~\bibnamefont
  {Koshino}},\ }\href {https://doi.org/10.1103/PhysRevB.96.075311} {\bibfield
  {journal} {\bibinfo  {journal} {Phys. Rev. B}\ }\textbf {\bibinfo {volume}
  {96}},\ \bibinfo {pages} {075311} (\bibinfo {year} {2017})}\BibitemShut
  {NoStop}%
\bibitem [{\citenamefont {Nam}\ and\ \citenamefont
  {Koshino}(2020)}]{koshino20_2}%
  \BibitemOpen
  \bibfield  {author} {\bibinfo {author} {\bibfnamefont {N.~N.~T.}\
  \bibnamefont {Nam}}\ and\ \bibinfo {author} {\bibfnamefont {M.}~\bibnamefont
  {Koshino}},\ }\href {https://doi.org/10.1103/PhysRevB.101.099901} {\bibfield
  {journal} {\bibinfo  {journal} {Phys. Rev. B}\ }\textbf {\bibinfo {volume}
  {101}},\ \bibinfo {pages} {099901} (\bibinfo {year} {2020})}\BibitemShut
  {NoStop}%
\bibitem [{\citenamefont {Kang}\ and\ \citenamefont {Vafek}(2023)}]{kang23}%
  \BibitemOpen
  \bibfield  {author} {\bibinfo {author} {\bibfnamefont {J.}~\bibnamefont
  {Kang}}\ and\ \bibinfo {author} {\bibfnamefont {O.}~\bibnamefont {Vafek}},\
  }\href {https://doi.org/10.1103/PhysRevB.107.075408} {\bibfield  {journal}
  {\bibinfo  {journal} {Phys. Rev. B}\ }\textbf {\bibinfo {volume} {107}},\
  \bibinfo {pages} {075408} (\bibinfo {year} {2023})}\BibitemShut {NoStop}%
\bibitem [{\citenamefont {Bi}\ \emph {et~al.}(2019)\citenamefont {Bi},
  \citenamefont {Yuan},\ and\ \citenamefont {Fu}}]{bi19}%
  \BibitemOpen
  \bibfield  {author} {\bibinfo {author} {\bibfnamefont {Z.}~\bibnamefont
  {Bi}}, \bibinfo {author} {\bibfnamefont {N.~F.~Q.}\ \bibnamefont {Yuan}},\
  and\ \bibinfo {author} {\bibfnamefont {L.}~\bibnamefont {Fu}},\ }\href
  {https://doi.org/10.1103/PhysRevB.100.035448} {\bibfield  {journal} {\bibinfo
   {journal} {Phys. Rev. B}\ }\textbf {\bibinfo {volume} {100}},\ \bibinfo
  {pages} {035448} (\bibinfo {year} {2019})}\BibitemShut {NoStop}%
\bibitem [{\citenamefont {Herzog-Arbeitman}\ \emph {et~al.}(2025)\citenamefont
  {Herzog-Arbeitman}, \citenamefont {Yu}, \citenamefont {C\ifmmode \u{a}\else
  \u{a}\fi{}lug\ifmmode~\u{a}\else \u{a}\fi{}ru}, \citenamefont {Hu},
  \citenamefont {Regnault}, \citenamefont {Vafek}, \citenamefont {Kang},\ and\
  \citenamefont {Bernevig}}]{arbeitman25}%
  \BibitemOpen
  \bibfield  {author} {\bibinfo {author} {\bibfnamefont {J.}~\bibnamefont
  {Herzog-Arbeitman}}, \bibinfo {author} {\bibfnamefont {J.}~\bibnamefont
  {Yu}}, \bibinfo {author} {\bibfnamefont {D.}~\bibnamefont {C\ifmmode
  \u{a}\else \u{a}\fi{}lug\ifmmode~\u{a}\else \u{a}\fi{}ru}}, \bibinfo {author}
  {\bibfnamefont {H.}~\bibnamefont {Hu}}, \bibinfo {author} {\bibfnamefont
  {N.}~\bibnamefont {Regnault}}, \bibinfo {author} {\bibfnamefont
  {O.}~\bibnamefont {Vafek}}, \bibinfo {author} {\bibfnamefont
  {J.}~\bibnamefont {Kang}},\ and\ \bibinfo {author} {\bibfnamefont {B.~A.}\
  \bibnamefont {Bernevig}},\ }\href {https://doi.org/10.1103/xv3m-vtlr}
  {\bibfield  {journal} {\bibinfo  {journal} {Phys. Rev. B}\ }\textbf {\bibinfo
  {volume} {112}},\ \bibinfo {pages} {125128} (\bibinfo {year}
  {2025})}\BibitemShut {NoStop}%
\bibitem [{\citenamefont {Bistritzer}\ and\ \citenamefont
  {MacDonald}(2011)}]{Bistritzer2011}%
  \BibitemOpen
  \bibfield  {author} {\bibinfo {author} {\bibfnamefont {R.}~\bibnamefont
  {Bistritzer}}\ and\ \bibinfo {author} {\bibfnamefont {A.~H.}\ \bibnamefont
  {MacDonald}},\ }\href {https://doi.org/10.1073/pnas.1108174108} {\bibfield
  {journal} {\bibinfo  {journal} {Proceedings of the National Academy of
  Sciences}\ }\textbf {\bibinfo {volume} {108}},\ \bibinfo {pages}
  {12233–12237} (\bibinfo {year} {2011})}\BibitemShut {NoStop}%
\bibitem [{\citenamefont {Fang}\ and\ \citenamefont {Kaxiras}(2016)}]{fang16}%
  \BibitemOpen
  \bibfield  {author} {\bibinfo {author} {\bibfnamefont {S.}~\bibnamefont
  {Fang}}\ and\ \bibinfo {author} {\bibfnamefont {E.}~\bibnamefont {Kaxiras}},\
  }\href {https://doi.org/10.1103/PhysRevB.93.235153} {\bibfield  {journal}
  {\bibinfo  {journal} {Phys. Rev. B}\ }\textbf {\bibinfo {volume} {93}},\
  \bibinfo {pages} {235153} (\bibinfo {year} {2016})}\BibitemShut {NoStop}%
\bibitem [{\citenamefont {Kang}\ and\ \citenamefont {Vafek}(2025)}]{kang25}%
  \BibitemOpen
  \bibfield  {author} {\bibinfo {author} {\bibfnamefont {J.}~\bibnamefont
  {Kang}}\ and\ \bibinfo {author} {\bibfnamefont {O.}~\bibnamefont {Vafek}},\
  }\href {https://doi.org/10.1103/s3s7-513d} {\bibfield  {journal} {\bibinfo
  {journal} {Phys. Rev. B}\ }\textbf {\bibinfo {volume} {112}},\ \bibinfo
  {pages} {125138} (\bibinfo {year} {2025})}\BibitemShut {NoStop}%
\bibitem [{\citenamefont {Kazmierczak}\ \emph {et~al.}(2021)\citenamefont
  {Kazmierczak}, \citenamefont {Van~Winkle}, \citenamefont {Ophus},
  \citenamefont {Bustillo}, \citenamefont {Carr}, \citenamefont {Brown},
  \citenamefont {Ciston}, \citenamefont {Taniguchi}, \citenamefont {Watanabe},\
  and\ \citenamefont {Bediako}}]{Kazmierczak2021}%
  \BibitemOpen
  \bibfield  {author} {\bibinfo {author} {\bibfnamefont {N.~P.}\ \bibnamefont
  {Kazmierczak}}, \bibinfo {author} {\bibfnamefont {M.}~\bibnamefont
  {Van~Winkle}}, \bibinfo {author} {\bibfnamefont {C.}~\bibnamefont {Ophus}},
  \bibinfo {author} {\bibfnamefont {K.~C.}\ \bibnamefont {Bustillo}}, \bibinfo
  {author} {\bibfnamefont {S.}~\bibnamefont {Carr}}, \bibinfo {author}
  {\bibfnamefont {H.~G.}\ \bibnamefont {Brown}}, \bibinfo {author}
  {\bibfnamefont {J.}~\bibnamefont {Ciston}}, \bibinfo {author} {\bibfnamefont
  {T.}~\bibnamefont {Taniguchi}}, \bibinfo {author} {\bibfnamefont
  {K.}~\bibnamefont {Watanabe}},\ and\ \bibinfo {author} {\bibfnamefont
  {D.~K.}\ \bibnamefont {Bediako}},\ }\href
  {https://doi.org/10.1038/s41563-021-00973-w} {\bibfield  {journal} {\bibinfo
  {journal} {Nature Materials}\ }\textbf {\bibinfo {volume} {20}},\ \bibinfo
  {pages} {956–963} (\bibinfo {year} {2021})}\BibitemShut {NoStop}%
\bibitem [{\citenamefont {Wehling}\ \emph {et~al.}(2011)\citenamefont
  {Wehling}, \citenamefont {\ifmmode \mbox{\c{S}}\else \c{S}\fi{}a\ifmmode
  \mbox{\c{s}}\else \c{s}\fi{}\ifmmode \imath \else \i
  \fi{}o\ifmmode~\breve{g}\else \u{g}\fi{}lu}, \citenamefont {Friedrich},
  \citenamefont {Lichtenstein}, \citenamefont {Katsnelson},\ and\ \citenamefont
  {Bl\"ugel}}]{wehling11}%
  \BibitemOpen
  \bibfield  {author} {\bibinfo {author} {\bibfnamefont {T.~O.}\ \bibnamefont
  {Wehling}}, \bibinfo {author} {\bibfnamefont {E.}~\bibnamefont {\ifmmode
  \mbox{\c{S}}\else \c{S}\fi{}a\ifmmode \mbox{\c{s}}\else \c{s}\fi{}\ifmmode
  \imath \else \i \fi{}o\ifmmode~\breve{g}\else \u{g}\fi{}lu}}, \bibinfo
  {author} {\bibfnamefont {C.}~\bibnamefont {Friedrich}}, \bibinfo {author}
  {\bibfnamefont {A.~I.}\ \bibnamefont {Lichtenstein}}, \bibinfo {author}
  {\bibfnamefont {M.~I.}\ \bibnamefont {Katsnelson}},\ and\ \bibinfo {author}
  {\bibfnamefont {S.}~\bibnamefont {Bl\"ugel}},\ }\href
  {https://doi.org/10.1103/PhysRevLett.106.236805} {\bibfield  {journal}
  {\bibinfo  {journal} {Phys. Rev. Lett.}\ }\textbf {\bibinfo {volume} {106}},\
  \bibinfo {pages} {236805} (\bibinfo {year} {2011})}\BibitemShut {NoStop}%
\bibitem [{\citenamefont {Sch\"uler}\ \emph {et~al.}(2013)\citenamefont
  {Sch\"uler}, \citenamefont {R\"osner}, \citenamefont {Wehling}, \citenamefont
  {Lichtenstein},\ and\ \citenamefont {Katsnelson}}]{schuler13}%
  \BibitemOpen
  \bibfield  {author} {\bibinfo {author} {\bibfnamefont {M.}~\bibnamefont
  {Sch\"uler}}, \bibinfo {author} {\bibfnamefont {M.}~\bibnamefont {R\"osner}},
  \bibinfo {author} {\bibfnamefont {T.~O.}\ \bibnamefont {Wehling}}, \bibinfo
  {author} {\bibfnamefont {A.~I.}\ \bibnamefont {Lichtenstein}},\ and\ \bibinfo
  {author} {\bibfnamefont {M.~I.}\ \bibnamefont {Katsnelson}},\ }\href
  {https://doi.org/10.1103/PhysRevLett.111.036601} {\bibfield  {journal}
  {\bibinfo  {journal} {Phys. Rev. Lett.}\ }\textbf {\bibinfo {volume} {111}},\
  \bibinfo {pages} {036601} (\bibinfo {year} {2013})}\BibitemShut {NoStop}%
\bibitem [{sup()}]{supplement}%
  \BibitemOpen
  \href@noop {} {}\bibinfo {note} {See Supplementary Materials, including Refs.
  ~\cite{carr18,dossantos12,zou18,fang16,kang23,guinea19,Giuliani_Vignale_2005,Kudin2002,bernevig21,sanchez25,fang2019angledependentitabinitio,koshino20,Ceferino2024,Coleman_2015,sodemann12,barnes14,wolframeta,koshino20_1,koshino20_2},
  for details on the Hartree-Fock tight-binding approach, the generalized BM
  model and the Fermi velocity and interlayer couplings renormalization, and
  for results for an alternative tight-binding model.}\BibitemShut {Stop}%
\bibitem [{\citenamefont {Fang}\ \emph {et~al.}(2019)\citenamefont {Fang},
  \citenamefont {Carr}, \citenamefont {Zhu}, \citenamefont {Massatt},\ and\
  \citenamefont {Kaxiras}}]{fang2019angledependentitabinitio}%
  \BibitemOpen
  \bibfield  {author} {\bibinfo {author} {\bibfnamefont {S.}~\bibnamefont
  {Fang}}, \bibinfo {author} {\bibfnamefont {S.}~\bibnamefont {Carr}}, \bibinfo
  {author} {\bibfnamefont {Z.}~\bibnamefont {Zhu}}, \bibinfo {author}
  {\bibfnamefont {D.}~\bibnamefont {Massatt}},\ and\ \bibinfo {author}
  {\bibfnamefont {E.}~\bibnamefont {Kaxiras}},\ }\href
  {https://arxiv.org/abs/1908.00058} {\bibinfo {title} {Angle-dependent {\it ab
  initio} low-energy hamiltonians for a relaxed twisted bilayer graphene
  heterostructure}} (\bibinfo {year} {2019}),\ \Eprint
  {https://arxiv.org/abs/1908.00058} {arXiv:1908.00058 [cond-mat.mes-hall]}
  \BibitemShut {NoStop}%
\bibitem [{\citenamefont {Kong}\ \emph {et~al.}(2025)\citenamefont {Kong},
  \citenamefont {Watson}, \citenamefont {Luskin},\ and\ \citenamefont
  {Stubbs}}]{Kong2025}%
  \BibitemOpen
  \bibfield  {author} {\bibinfo {author} {\bibfnamefont {T.}~\bibnamefont
  {Kong}}, \bibinfo {author} {\bibfnamefont {A.~B.}\ \bibnamefont {Watson}},
  \bibinfo {author} {\bibfnamefont {M.}~\bibnamefont {Luskin}},\ and\ \bibinfo
  {author} {\bibfnamefont {K.~D.}\ \bibnamefont {Stubbs}},\ }\href
  {https://doi.org/10.1088/2516-1075/adeb25} {\bibfield  {journal} {\bibinfo
  {journal} {Electronic Structure}\ }\textbf {\bibinfo {volume} {7}},\ \bibinfo
  {pages} {035001} (\bibinfo {year} {2025})}\BibitemShut {NoStop}%
\bibitem [{\citenamefont {Kwan}\ \emph {et~al.}(2021)\citenamefont {Kwan},
  \citenamefont {Wagner}, \citenamefont {Soejima}, \citenamefont {Zaletel},
  \citenamefont {Simon}, \citenamefont {Parameswaran},\ and\ \citenamefont
  {Bultinck}}]{kwan21}%
  \BibitemOpen
  \bibfield  {author} {\bibinfo {author} {\bibfnamefont {Y.~H.}\ \bibnamefont
  {Kwan}}, \bibinfo {author} {\bibfnamefont {G.}~\bibnamefont {Wagner}},
  \bibinfo {author} {\bibfnamefont {T.}~\bibnamefont {Soejima}}, \bibinfo
  {author} {\bibfnamefont {M.~P.}\ \bibnamefont {Zaletel}}, \bibinfo {author}
  {\bibfnamefont {S.~H.}\ \bibnamefont {Simon}}, \bibinfo {author}
  {\bibfnamefont {S.~A.}\ \bibnamefont {Parameswaran}},\ and\ \bibinfo {author}
  {\bibfnamefont {N.}~\bibnamefont {Bultinck}},\ }\href
  {https://doi.org/10.1103/PhysRevX.11.041063} {\bibfield  {journal} {\bibinfo
  {journal} {Phys. Rev. X}\ }\textbf {\bibinfo {volume} {11}},\ \bibinfo
  {pages} {041063} (\bibinfo {year} {2021})}\BibitemShut {NoStop}%
\bibitem [{\citenamefont {Crippa}\ \emph {et~al.}(2026)\citenamefont {Crippa},
  \citenamefont {Rai}, \citenamefont {Călugăru}, \citenamefont {Hu},
  \citenamefont {Herzog-Arbeitman}, \citenamefont {Bernevig}, \citenamefont
  {Valentí}, \citenamefont {Sangiovanni},\ and\ \citenamefont
  {Wehling}}]{crippa2026interplaymanybodycorrelationsstrain}%
  \BibitemOpen
  \bibfield  {author} {\bibinfo {author} {\bibfnamefont {L.}~\bibnamefont
  {Crippa}}, \bibinfo {author} {\bibfnamefont {G.}~\bibnamefont {Rai}},
  \bibinfo {author} {\bibfnamefont {D.}~\bibnamefont {Călugăru}}, \bibinfo
  {author} {\bibfnamefont {H.}~\bibnamefont {Hu}}, \bibinfo {author}
  {\bibfnamefont {J.}~\bibnamefont {Herzog-Arbeitman}}, \bibinfo {author}
  {\bibfnamefont {B.~A.}\ \bibnamefont {Bernevig}}, \bibinfo {author}
  {\bibfnamefont {R.}~\bibnamefont {Valentí}}, \bibinfo {author}
  {\bibfnamefont {G.}~\bibnamefont {Sangiovanni}},\ and\ \bibinfo {author}
  {\bibfnamefont {T.}~\bibnamefont {Wehling}},\ }\href
  {https://arxiv.org/abs/2509.19436} {\bibinfo {title} {Interplay between
  many-body correlations, strain and lattice relaxation in twisted bilayer
  graphene}} (\bibinfo {year} {2026}),\ \Eprint
  {https://arxiv.org/abs/2509.19436} {arXiv:2509.19436 [cond-mat.str-el]}
  \BibitemShut {NoStop}%
\bibitem [{\citenamefont {Nuckolls}\ \emph {et~al.}(2023)\citenamefont
  {Nuckolls}, \citenamefont {Lee}, \citenamefont {Oh}, \citenamefont {Wong},
  \citenamefont {Soejima}, \citenamefont {Hong}, \citenamefont {Călugăru},
  \citenamefont {Herzog-Arbeitman}, \citenamefont {Bernevig}, \citenamefont
  {Watanabe}, \citenamefont {Taniguchi}, \citenamefont {Regnault},
  \citenamefont {Zaletel},\ and\ \citenamefont {Yazdani}}]{Nuckolls2023}%
  \BibitemOpen
  \bibfield  {author} {\bibinfo {author} {\bibfnamefont {K.~P.}\ \bibnamefont
  {Nuckolls}}, \bibinfo {author} {\bibfnamefont {R.~L.}\ \bibnamefont {Lee}},
  \bibinfo {author} {\bibfnamefont {M.}~\bibnamefont {Oh}}, \bibinfo {author}
  {\bibfnamefont {D.}~\bibnamefont {Wong}}, \bibinfo {author} {\bibfnamefont
  {T.}~\bibnamefont {Soejima}}, \bibinfo {author} {\bibfnamefont {J.~P.}\
  \bibnamefont {Hong}}, \bibinfo {author} {\bibfnamefont {D.}~\bibnamefont
  {Călugăru}}, \bibinfo {author} {\bibfnamefont {J.}~\bibnamefont
  {Herzog-Arbeitman}}, \bibinfo {author} {\bibfnamefont {B.~A.}\ \bibnamefont
  {Bernevig}}, \bibinfo {author} {\bibfnamefont {K.}~\bibnamefont {Watanabe}},
  \bibinfo {author} {\bibfnamefont {T.}~\bibnamefont {Taniguchi}}, \bibinfo
  {author} {\bibfnamefont {N.}~\bibnamefont {Regnault}}, \bibinfo {author}
  {\bibfnamefont {M.~P.}\ \bibnamefont {Zaletel}},\ and\ \bibinfo {author}
  {\bibfnamefont {A.}~\bibnamefont {Yazdani}},\ }\href
  {https://doi.org/10.1038/s41586-023-06226-x} {\bibfield  {journal} {\bibinfo
  {journal} {Nature}\ }\textbf {\bibinfo {volume} {620}},\ \bibinfo {pages}
  {525–532} (\bibinfo {year} {2023})}\BibitemShut {NoStop}%
\bibitem [{\citenamefont {Yoo}\ \emph {et~al.}(2019)\citenamefont {Yoo},
  \citenamefont {Engelke}, \citenamefont {Carr}, \citenamefont {Fang},
  \citenamefont {Zhang}, \citenamefont {Cazeaux}, \citenamefont {Sung},
  \citenamefont {Hovden}, \citenamefont {Tsen}, \citenamefont {Taniguchi},
  \citenamefont {Watanabe}, \citenamefont {Yi}, \citenamefont {Kim},
  \citenamefont {Luskin}, \citenamefont {Tadmor}, \citenamefont {Kaxiras},\
  and\ \citenamefont {Kim}}]{Yoo2019}%
  \BibitemOpen
  \bibfield  {author} {\bibinfo {author} {\bibfnamefont {H.}~\bibnamefont
  {Yoo}}, \bibinfo {author} {\bibfnamefont {R.}~\bibnamefont {Engelke}},
  \bibinfo {author} {\bibfnamefont {S.}~\bibnamefont {Carr}}, \bibinfo {author}
  {\bibfnamefont {S.}~\bibnamefont {Fang}}, \bibinfo {author} {\bibfnamefont
  {K.}~\bibnamefont {Zhang}}, \bibinfo {author} {\bibfnamefont
  {P.}~\bibnamefont {Cazeaux}}, \bibinfo {author} {\bibfnamefont {S.~H.}\
  \bibnamefont {Sung}}, \bibinfo {author} {\bibfnamefont {R.}~\bibnamefont
  {Hovden}}, \bibinfo {author} {\bibfnamefont {A.~W.}\ \bibnamefont {Tsen}},
  \bibinfo {author} {\bibfnamefont {T.}~\bibnamefont {Taniguchi}}, \bibinfo
  {author} {\bibfnamefont {K.}~\bibnamefont {Watanabe}}, \bibinfo {author}
  {\bibfnamefont {G.-C.}\ \bibnamefont {Yi}}, \bibinfo {author} {\bibfnamefont
  {M.}~\bibnamefont {Kim}}, \bibinfo {author} {\bibfnamefont {M.}~\bibnamefont
  {Luskin}}, \bibinfo {author} {\bibfnamefont {E.~B.}\ \bibnamefont {Tadmor}},
  \bibinfo {author} {\bibfnamefont {E.}~\bibnamefont {Kaxiras}},\ and\ \bibinfo
  {author} {\bibfnamefont {P.}~\bibnamefont {Kim}},\ }\href
  {https://doi.org/10.1038/s41563-019-0346-z} {\bibfield  {journal} {\bibinfo
  {journal} {Nature Materials}\ }\textbf {\bibinfo {volume} {18}},\ \bibinfo
  {pages} {448–453} (\bibinfo {year} {2019})}\BibitemShut {NoStop}%
\bibitem [{\citenamefont {Pierce}\ \emph {et~al.}(2021)\citenamefont {Pierce},
  \citenamefont {Xie}, \citenamefont {Park}, \citenamefont {Khalaf},
  \citenamefont {Lee}, \citenamefont {Cao}, \citenamefont {Parker},
  \citenamefont {Forrester}, \citenamefont {Chen}, \citenamefont {Watanabe},
  \citenamefont {Taniguchi}, \citenamefont {Vishwanath}, \citenamefont
  {Jarillo-Herrero},\ and\ \citenamefont {Yacoby}}]{Pierce2021}%
  \BibitemOpen
  \bibfield  {author} {\bibinfo {author} {\bibfnamefont {A.~T.}\ \bibnamefont
  {Pierce}}, \bibinfo {author} {\bibfnamefont {Y.}~\bibnamefont {Xie}},
  \bibinfo {author} {\bibfnamefont {J.~M.}\ \bibnamefont {Park}}, \bibinfo
  {author} {\bibfnamefont {E.}~\bibnamefont {Khalaf}}, \bibinfo {author}
  {\bibfnamefont {S.~H.}\ \bibnamefont {Lee}}, \bibinfo {author} {\bibfnamefont
  {Y.}~\bibnamefont {Cao}}, \bibinfo {author} {\bibfnamefont {D.~E.}\
  \bibnamefont {Parker}}, \bibinfo {author} {\bibfnamefont {P.~R.}\
  \bibnamefont {Forrester}}, \bibinfo {author} {\bibfnamefont {S.}~\bibnamefont
  {Chen}}, \bibinfo {author} {\bibfnamefont {K.}~\bibnamefont {Watanabe}},
  \bibinfo {author} {\bibfnamefont {T.}~\bibnamefont {Taniguchi}}, \bibinfo
  {author} {\bibfnamefont {A.}~\bibnamefont {Vishwanath}}, \bibinfo {author}
  {\bibfnamefont {P.}~\bibnamefont {Jarillo-Herrero}},\ and\ \bibinfo {author}
  {\bibfnamefont {A.}~\bibnamefont {Yacoby}},\ }\href
  {https://doi.org/10.1038/s41567-021-01347-4} {\bibfield  {journal} {\bibinfo
  {journal} {Nature Physics}\ }\textbf {\bibinfo {volume} {17}},\ \bibinfo
  {pages} {1210–1215} (\bibinfo {year} {2021})}\BibitemShut {NoStop}%
\bibitem [{\citenamefont {Xie}\ \emph {et~al.}(2019)\citenamefont {Xie},
  \citenamefont {Lian}, \citenamefont {J\"{a}ck}, \citenamefont {Liu},
  \citenamefont {Chiu}, \citenamefont {Watanabe}, \citenamefont {Taniguchi},
  \citenamefont {Bernevig},\ and\ \citenamefont {Yazdani}}]{Xie2019}%
  \BibitemOpen
  \bibfield  {author} {\bibinfo {author} {\bibfnamefont {Y.}~\bibnamefont
  {Xie}}, \bibinfo {author} {\bibfnamefont {B.}~\bibnamefont {Lian}}, \bibinfo
  {author} {\bibfnamefont {B.}~\bibnamefont {J\"{a}ck}}, \bibinfo {author}
  {\bibfnamefont {X.}~\bibnamefont {Liu}}, \bibinfo {author} {\bibfnamefont
  {C.-L.}\ \bibnamefont {Chiu}}, \bibinfo {author} {\bibfnamefont
  {K.}~\bibnamefont {Watanabe}}, \bibinfo {author} {\bibfnamefont
  {T.}~\bibnamefont {Taniguchi}}, \bibinfo {author} {\bibfnamefont {B.~A.}\
  \bibnamefont {Bernevig}},\ and\ \bibinfo {author} {\bibfnamefont
  {A.}~\bibnamefont {Yazdani}},\ }\href
  {https://doi.org/10.1038/s41586-019-1422-x} {\bibfield  {journal} {\bibinfo
  {journal} {Nature}\ }\textbf {\bibinfo {volume} {572}},\ \bibinfo {pages}
  {101–105} (\bibinfo {year} {2019})}\BibitemShut {NoStop}%
\bibitem [{\citenamefont {Zhang}\ \emph {et~al.}(2025)\citenamefont {Zhang},
  \citenamefont {Wu}, \citenamefont {Călugăru}, \citenamefont {Hu},
  \citenamefont {Taniguchi}, \citenamefont {Wanatabe}, \citenamefont
  {Bernevig},\ and\ \citenamefont
  {Andrei}}]{zhang2025heavyfermionsmassrenormalization}%
  \BibitemOpen
  \bibfield  {author} {\bibinfo {author} {\bibfnamefont {Z.}~\bibnamefont
  {Zhang}}, \bibinfo {author} {\bibfnamefont {S.}~\bibnamefont {Wu}}, \bibinfo
  {author} {\bibfnamefont {D.}~\bibnamefont {Călugăru}}, \bibinfo {author}
  {\bibfnamefont {H.}~\bibnamefont {Hu}}, \bibinfo {author} {\bibfnamefont
  {T.}~\bibnamefont {Taniguchi}}, \bibinfo {author} {\bibfnamefont
  {K.}~\bibnamefont {Wanatabe}}, \bibinfo {author} {\bibfnamefont {A.~B.}\
  \bibnamefont {Bernevig}},\ and\ \bibinfo {author} {\bibfnamefont {E.~Y.}\
  \bibnamefont {Andrei}},\ }\href {https://arxiv.org/abs/2503.17875} {\bibinfo
  {title} {Heavy fermions, mass renormalization and local moments in
  magic-angle twisted bilayer graphene via planar tunneling spectroscopy}}
  (\bibinfo {year} {2025}),\ \Eprint {https://arxiv.org/abs/2503.17875}
  {arXiv:2503.17875 [cond-mat.mes-hall]} \BibitemShut {NoStop}%
\bibitem [{\citenamefont {Miao}\ \emph {et~al.}(2023)\citenamefont {Miao},
  \citenamefont {Li}, \citenamefont {Han}, \citenamefont {Pan},\ and\
  \citenamefont {Dai}}]{miao23}%
  \BibitemOpen
  \bibfield  {author} {\bibinfo {author} {\bibfnamefont {W.}~\bibnamefont
  {Miao}}, \bibinfo {author} {\bibfnamefont {C.}~\bibnamefont {Li}}, \bibinfo
  {author} {\bibfnamefont {X.}~\bibnamefont {Han}}, \bibinfo {author}
  {\bibfnamefont {D.}~\bibnamefont {Pan}},\ and\ \bibinfo {author}
  {\bibfnamefont {X.}~\bibnamefont {Dai}},\ }\href
  {https://doi.org/10.1103/PhysRevB.107.125112} {\bibfield  {journal} {\bibinfo
   {journal} {Phys. Rev. B}\ }\textbf {\bibinfo {volume} {107}},\ \bibinfo
  {pages} {125112} (\bibinfo {year} {2023})}\BibitemShut {NoStop}%
\bibitem [{\citenamefont {Guinea}\ and\ \citenamefont
  {Walet}(2019)}]{guinea19}%
  \BibitemOpen
  \bibfield  {author} {\bibinfo {author} {\bibfnamefont {F.}~\bibnamefont
  {Guinea}}\ and\ \bibinfo {author} {\bibfnamefont {N.~R.}\ \bibnamefont
  {Walet}},\ }\href {https://doi.org/10.1103/PhysRevB.99.205134} {\bibfield
  {journal} {\bibinfo  {journal} {Phys. Rev. B}\ }\textbf {\bibinfo {volume}
  {99}},\ \bibinfo {pages} {205134} (\bibinfo {year} {2019})}\BibitemShut
  {NoStop}%
\bibitem [{Note1()}]{Note1}%
  \BibitemOpen
  \bibinfo {note} {Note that $\protect \bm {q}_{1}$ is not a reciprocal lattice
  vector. This is because the origin of momentum in Eq. (\ref {bwaves}) is
  different for different layers.}\BibitemShut {Stop}%
\bibitem [{\citenamefont {Koshino}\ and\ \citenamefont
  {Nam}(2020)}]{koshino20}%
  \BibitemOpen
  \bibfield  {author} {\bibinfo {author} {\bibfnamefont {M.}~\bibnamefont
  {Koshino}}\ and\ \bibinfo {author} {\bibfnamefont {N.~N.~T.}\ \bibnamefont
  {Nam}},\ }\href {https://doi.org/10.1103/PhysRevB.101.195425} {\bibfield
  {journal} {\bibinfo  {journal} {Phys. Rev. B}\ }\textbf {\bibinfo {volume}
  {101}},\ \bibinfo {pages} {195425} (\bibinfo {year} {2020})}\BibitemShut
  {NoStop}%
\bibitem [{Note2()}]{Note2}%
  \BibitemOpen
  \bibinfo {note} {Keeping momenta $\protect \bm {q}$ with $q \leq \protect
  \sqrt {19}q_1$ is sufficient for convergence.}\BibitemShut {Stop}%
\bibitem [{\citenamefont {Escudero}(2024)}]{escudero24}%
  \BibitemOpen
  \bibfield  {author} {\bibinfo {author} {\bibfnamefont {F.}~\bibnamefont
  {Escudero}},\ }\href {https://doi.org/10.1103/PhysRevB.110.045442} {\bibfield
   {journal} {\bibinfo  {journal} {Phys. Rev. B}\ }\textbf {\bibinfo {volume}
  {110}},\ \bibinfo {pages} {045442} (\bibinfo {year} {2024})}\BibitemShut
  {NoStop}%
\bibitem [{Note3()}]{Note3}%
  \BibitemOpen
  \bibinfo {note} {Keeping momenta $\protect \bm {g}$ with $g \leq 3g_1$ is
  sufficient for convergence.}\BibitemShut {Stop}%
\bibitem [{\citenamefont {Ceferino}\ and\ \citenamefont
  {Guinea}(2024)}]{Ceferino2024}%
  \BibitemOpen
  \bibfield  {author} {\bibinfo {author} {\bibfnamefont {A.}~\bibnamefont
  {Ceferino}}\ and\ \bibinfo {author} {\bibfnamefont {F.}~\bibnamefont
  {Guinea}},\ }\href {https://doi.org/10.1088/2053-1583/ad3b0e} {\bibfield
  {journal} {\bibinfo  {journal} {2D Materials}\ }\textbf {\bibinfo {volume}
  {11}},\ \bibinfo {pages} {035015} (\bibinfo {year} {2024})}\BibitemShut
  {NoStop}%
\bibitem [{Note4()}]{Note4}%
  \BibitemOpen
  \bibinfo {note} {As a remark, even if $A_{0,3}^{\ell ,\protect \bm {g}}$ are
  negligible, the renormalizations $\delta A_{0,3}^{\ell ,\protect \bm {g}}$
  can be nonzero in general}\BibitemShut {NoStop}%
\bibitem [{Note5()}]{Note5}%
  \BibitemOpen
  \bibinfo {note} {Note that $v_F = 2.416\protect \text { eV}\cdot a$ is $10
  \%$ larger than the $\protect \textit {ab initio}$ value. This choice
  effectively includes Fermi velocity renormalization and places the magic
  angle of the BM model at $1.08^\circ $. We stress that, when comparing the
  bare and renormalized results for, say, the flat Fermi velocity at
  intermediate twist angles, the $\protect \textit {ab initio}$ $v_F$ must be
  used for consistency.}\BibitemShut {Stop}%
\bibitem [{\citenamefont {Esparza}\ and\ \citenamefont {Juri\ifmmode
  \check{c}\else \v{c}\fi{}i\ifmmode~\acute{c}\else
  \'{c}\fi{}}(2025)}]{esparza25}%
  \BibitemOpen
  \bibfield  {author} {\bibinfo {author} {\bibfnamefont {J.~P.}\ \bibnamefont
  {Esparza}}\ and\ \bibinfo {author} {\bibfnamefont {V.}~\bibnamefont
  {Juri\ifmmode \check{c}\else \v{c}\fi{}i\ifmmode~\acute{c}\else
  \'{c}\fi{}}},\ }\href {https://doi.org/10.1103/dl59-vl7v} {\bibfield
  {journal} {\bibinfo  {journal} {Phys. Rev. Lett.}\ }\textbf {\bibinfo
  {volume} {134}},\ \bibinfo {pages} {226602} (\bibinfo {year}
  {2025})}\BibitemShut {NoStop}%
\bibitem [{\citenamefont {Stepanov}\ \emph {et~al.}(2020)\citenamefont
  {Stepanov}, \citenamefont {Das}, \citenamefont {Lu}, \citenamefont
  {Fahimniya}, \citenamefont {Watanabe}, \citenamefont {Taniguchi},
  \citenamefont {Koppens}, \citenamefont {Lischner}, \citenamefont {Levitov},\
  and\ \citenamefont {Efetov}}]{Stepanov2020}%
  \BibitemOpen
  \bibfield  {author} {\bibinfo {author} {\bibfnamefont {P.}~\bibnamefont
  {Stepanov}}, \bibinfo {author} {\bibfnamefont {I.}~\bibnamefont {Das}},
  \bibinfo {author} {\bibfnamefont {X.}~\bibnamefont {Lu}}, \bibinfo {author}
  {\bibfnamefont {A.}~\bibnamefont {Fahimniya}}, \bibinfo {author}
  {\bibfnamefont {K.}~\bibnamefont {Watanabe}}, \bibinfo {author}
  {\bibfnamefont {T.}~\bibnamefont {Taniguchi}}, \bibinfo {author}
  {\bibfnamefont {F.~H.~L.}\ \bibnamefont {Koppens}}, \bibinfo {author}
  {\bibfnamefont {J.}~\bibnamefont {Lischner}}, \bibinfo {author}
  {\bibfnamefont {L.}~\bibnamefont {Levitov}},\ and\ \bibinfo {author}
  {\bibfnamefont {D.~K.}\ \bibnamefont {Efetov}},\ }\href
  {https://doi.org/10.1038/s41586-020-2459-6} {\bibfield  {journal} {\bibinfo
  {journal} {Nature}\ }\textbf {\bibinfo {volume} {583}},\ \bibinfo {pages}
  {375–378} (\bibinfo {year} {2020})}\BibitemShut {NoStop}%
\bibitem [{\citenamefont {Luican}\ \emph {et~al.}(2011)\citenamefont {Luican},
  \citenamefont {Li}, \citenamefont {Reina}, \citenamefont {Kong},
  \citenamefont {Nair}, \citenamefont {Novoselov}, \citenamefont {Geim},\ and\
  \citenamefont {Andrei}}]{luican11}%
  \BibitemOpen
  \bibfield  {author} {\bibinfo {author} {\bibfnamefont {A.}~\bibnamefont
  {Luican}}, \bibinfo {author} {\bibfnamefont {G.}~\bibnamefont {Li}}, \bibinfo
  {author} {\bibfnamefont {A.}~\bibnamefont {Reina}}, \bibinfo {author}
  {\bibfnamefont {J.}~\bibnamefont {Kong}}, \bibinfo {author} {\bibfnamefont
  {R.~R.}\ \bibnamefont {Nair}}, \bibinfo {author} {\bibfnamefont {K.~S.}\
  \bibnamefont {Novoselov}}, \bibinfo {author} {\bibfnamefont {A.~K.}\
  \bibnamefont {Geim}},\ and\ \bibinfo {author} {\bibfnamefont {E.~Y.}\
  \bibnamefont {Andrei}},\ }\href
  {https://doi.org/10.1103/PhysRevLett.106.126802} {\bibfield  {journal}
  {\bibinfo  {journal} {Phys. Rev. Lett.}\ }\textbf {\bibinfo {volume} {106}},\
  \bibinfo {pages} {126802} (\bibinfo {year} {2011})}\BibitemShut {NoStop}%
\bibitem [{\citenamefont {Cao}\ \emph {et~al.}(2016)\citenamefont {Cao},
  \citenamefont {Luo}, \citenamefont {Fatemi}, \citenamefont {Fang},
  \citenamefont {Sanchez-Yamagishi}, \citenamefont {Watanabe}, \citenamefont
  {Taniguchi}, \citenamefont {Kaxiras},\ and\ \citenamefont
  {Jarillo-Herrero}}]{cao16}%
  \BibitemOpen
  \bibfield  {author} {\bibinfo {author} {\bibfnamefont {Y.}~\bibnamefont
  {Cao}}, \bibinfo {author} {\bibfnamefont {J.~Y.}\ \bibnamefont {Luo}},
  \bibinfo {author} {\bibfnamefont {V.}~\bibnamefont {Fatemi}}, \bibinfo
  {author} {\bibfnamefont {S.}~\bibnamefont {Fang}}, \bibinfo {author}
  {\bibfnamefont {J.~D.}\ \bibnamefont {Sanchez-Yamagishi}}, \bibinfo {author}
  {\bibfnamefont {K.}~\bibnamefont {Watanabe}}, \bibinfo {author}
  {\bibfnamefont {T.}~\bibnamefont {Taniguchi}}, \bibinfo {author}
  {\bibfnamefont {E.}~\bibnamefont {Kaxiras}},\ and\ \bibinfo {author}
  {\bibfnamefont {P.}~\bibnamefont {Jarillo-Herrero}},\ }\href
  {https://doi.org/10.1103/PhysRevLett.117.116804} {\bibfield  {journal}
  {\bibinfo  {journal} {Phys. Rev. Lett.}\ }\textbf {\bibinfo {volume} {117}},\
  \bibinfo {pages} {116804} (\bibinfo {year} {2016})}\BibitemShut {NoStop}%
\bibitem [{\citenamefont {Pizarro}\ \emph {et~al.}(2019)\citenamefont
  {Pizarro}, \citenamefont {R\"osner}, \citenamefont {Thomale}, \citenamefont
  {Valent\'{\i}},\ and\ \citenamefont {Wehling}}]{pizarro19}%
  \BibitemOpen
  \bibfield  {author} {\bibinfo {author} {\bibfnamefont {J.~M.}\ \bibnamefont
  {Pizarro}}, \bibinfo {author} {\bibfnamefont {M.}~\bibnamefont {R\"osner}},
  \bibinfo {author} {\bibfnamefont {R.}~\bibnamefont {Thomale}}, \bibinfo
  {author} {\bibfnamefont {R.}~\bibnamefont {Valent\'{\i}}},\ and\ \bibinfo
  {author} {\bibfnamefont {T.~O.}\ \bibnamefont {Wehling}},\ }\href
  {https://doi.org/10.1103/PhysRevB.100.161102} {\bibfield  {journal} {\bibinfo
   {journal} {Phys. Rev. B}\ }\textbf {\bibinfo {volume} {100}},\ \bibinfo
  {pages} {161102} (\bibinfo {year} {2019})}\BibitemShut {NoStop}%
\bibitem [{\citenamefont {Vanhala}\ and\ \citenamefont
  {Pollet}(2020)}]{vanhala20}%
  \BibitemOpen
  \bibfield  {author} {\bibinfo {author} {\bibfnamefont {T.~I.}\ \bibnamefont
  {Vanhala}}\ and\ \bibinfo {author} {\bibfnamefont {L.}~\bibnamefont
  {Pollet}},\ }\href {https://doi.org/10.1103/PhysRevB.102.035154} {\bibfield
  {journal} {\bibinfo  {journal} {Phys. Rev. B}\ }\textbf {\bibinfo {volume}
  {102}},\ \bibinfo {pages} {035154} (\bibinfo {year} {2020})}\BibitemShut
  {NoStop}%
\bibitem [{\citenamefont {Goodwin}\ \emph {et~al.}(2019)\citenamefont
  {Goodwin}, \citenamefont {Corsetti}, \citenamefont {Mostofi},\ and\
  \citenamefont {Lischner}}]{goodwin19}%
  \BibitemOpen
  \bibfield  {author} {\bibinfo {author} {\bibfnamefont {Z.~A.~H.}\
  \bibnamefont {Goodwin}}, \bibinfo {author} {\bibfnamefont {F.}~\bibnamefont
  {Corsetti}}, \bibinfo {author} {\bibfnamefont {A.~A.}\ \bibnamefont
  {Mostofi}},\ and\ \bibinfo {author} {\bibfnamefont {J.}~\bibnamefont
  {Lischner}},\ }\href {https://doi.org/10.1103/PhysRevB.100.235424} {\bibfield
   {journal} {\bibinfo  {journal} {Phys. Rev. B}\ }\textbf {\bibinfo {volume}
  {100}},\ \bibinfo {pages} {235424} (\bibinfo {year} {2019})}\BibitemShut
  {NoStop}%
\bibitem [{\citenamefont {Zhang}\ \emph {et~al.}(2022)\citenamefont {Zhang},
  \citenamefont {Lu},\ and\ \citenamefont {Liu}}]{zhang22}%
  \BibitemOpen
  \bibfield  {author} {\bibinfo {author} {\bibfnamefont {S.}~\bibnamefont
  {Zhang}}, \bibinfo {author} {\bibfnamefont {X.}~\bibnamefont {Lu}},\ and\
  \bibinfo {author} {\bibfnamefont {J.}~\bibnamefont {Liu}},\ }\href
  {https://doi.org/10.1103/PhysRevLett.128.247402} {\bibfield  {journal}
  {\bibinfo  {journal} {Phys. Rev. Lett.}\ }\textbf {\bibinfo {volume} {128}},\
  \bibinfo {pages} {247402} (\bibinfo {year} {2022})}\BibitemShut {NoStop}%
\bibitem [{\citenamefont {Cao}\ \emph {et~al.}(2018{\natexlab{a}})\citenamefont
  {Cao}, \citenamefont {Fatemi}, \citenamefont {Fang}, \citenamefont
  {Watanabe}, \citenamefont {Taniguchi}, \citenamefont {Kaxiras},\ and\
  \citenamefont {Jarillo-Herrero}}]{Cao2018}%
  \BibitemOpen
  \bibfield  {author} {\bibinfo {author} {\bibfnamefont {Y.}~\bibnamefont
  {Cao}}, \bibinfo {author} {\bibfnamefont {V.}~\bibnamefont {Fatemi}},
  \bibinfo {author} {\bibfnamefont {S.}~\bibnamefont {Fang}}, \bibinfo {author}
  {\bibfnamefont {K.}~\bibnamefont {Watanabe}}, \bibinfo {author}
  {\bibfnamefont {T.}~\bibnamefont {Taniguchi}}, \bibinfo {author}
  {\bibfnamefont {E.}~\bibnamefont {Kaxiras}},\ and\ \bibinfo {author}
  {\bibfnamefont {P.}~\bibnamefont {Jarillo-Herrero}},\ }\href
  {https://doi.org/10.1038/nature26160} {\bibfield  {journal} {\bibinfo
  {journal} {Nature}\ }\textbf {\bibinfo {volume} {556}},\ \bibinfo {pages}
  {43–50} (\bibinfo {year} {2018}{\natexlab{a}})}\BibitemShut {NoStop}%
\bibitem [{\citenamefont {Cao}\ \emph {et~al.}(2021)\citenamefont {Cao},
  \citenamefont {Rodan-Legrain}, \citenamefont {Park}, \citenamefont {Yuan},
  \citenamefont {Watanabe}, \citenamefont {Taniguchi}, \citenamefont
  {Fernandes}, \citenamefont {Fu},\ and\ \citenamefont
  {Jarillo-Herrero}}]{Cao2021}%
  \BibitemOpen
  \bibfield  {author} {\bibinfo {author} {\bibfnamefont {Y.}~\bibnamefont
  {Cao}}, \bibinfo {author} {\bibfnamefont {D.}~\bibnamefont {Rodan-Legrain}},
  \bibinfo {author} {\bibfnamefont {J.~M.}\ \bibnamefont {Park}}, \bibinfo
  {author} {\bibfnamefont {N.~F.~Q.}\ \bibnamefont {Yuan}}, \bibinfo {author}
  {\bibfnamefont {K.}~\bibnamefont {Watanabe}}, \bibinfo {author}
  {\bibfnamefont {T.}~\bibnamefont {Taniguchi}}, \bibinfo {author}
  {\bibfnamefont {R.~M.}\ \bibnamefont {Fernandes}}, \bibinfo {author}
  {\bibfnamefont {L.}~\bibnamefont {Fu}},\ and\ \bibinfo {author}
  {\bibfnamefont {P.}~\bibnamefont {Jarillo-Herrero}},\ }\href
  {https://doi.org/10.1126/science.abc2836} {\bibfield  {journal} {\bibinfo
  {journal} {Science}\ }\textbf {\bibinfo {volume} {372}},\ \bibinfo {pages}
  {264–271} (\bibinfo {year} {2021})}\BibitemShut {NoStop}%
\bibitem [{\citenamefont {Lu}\ \emph {et~al.}(2019)\citenamefont {Lu},
  \citenamefont {Stepanov}, \citenamefont {Yang}, \citenamefont {Xie},
  \citenamefont {Aamir}, \citenamefont {Das}, \citenamefont {Urgell},
  \citenamefont {Watanabe}, \citenamefont {Taniguchi}, \citenamefont {Zhang},
  \citenamefont {Bachtold}, \citenamefont {MacDonald},\ and\ \citenamefont
  {Efetov}}]{Lu2019}%
  \BibitemOpen
  \bibfield  {author} {\bibinfo {author} {\bibfnamefont {X.}~\bibnamefont
  {Lu}}, \bibinfo {author} {\bibfnamefont {P.}~\bibnamefont {Stepanov}},
  \bibinfo {author} {\bibfnamefont {W.}~\bibnamefont {Yang}}, \bibinfo {author}
  {\bibfnamefont {M.}~\bibnamefont {Xie}}, \bibinfo {author} {\bibfnamefont
  {M.~A.}\ \bibnamefont {Aamir}}, \bibinfo {author} {\bibfnamefont
  {I.}~\bibnamefont {Das}}, \bibinfo {author} {\bibfnamefont {C.}~\bibnamefont
  {Urgell}}, \bibinfo {author} {\bibfnamefont {K.}~\bibnamefont {Watanabe}},
  \bibinfo {author} {\bibfnamefont {T.}~\bibnamefont {Taniguchi}}, \bibinfo
  {author} {\bibfnamefont {G.}~\bibnamefont {Zhang}}, \bibinfo {author}
  {\bibfnamefont {A.}~\bibnamefont {Bachtold}}, \bibinfo {author}
  {\bibfnamefont {A.~H.}\ \bibnamefont {MacDonald}},\ and\ \bibinfo {author}
  {\bibfnamefont {D.~K.}\ \bibnamefont {Efetov}},\ }\href
  {https://doi.org/10.1038/s41586-019-1695-0} {\bibfield  {journal} {\bibinfo
  {journal} {Nature}\ }\textbf {\bibinfo {volume} {574}},\ \bibinfo {pages}
  {653–657} (\bibinfo {year} {2019})}\BibitemShut {NoStop}%
\bibitem [{\citenamefont {Yankowitz}\ \emph {et~al.}(2019)\citenamefont
  {Yankowitz}, \citenamefont {Chen}, \citenamefont {Polshyn}, \citenamefont
  {Zhang}, \citenamefont {Watanabe}, \citenamefont {Taniguchi}, \citenamefont
  {Graf}, \citenamefont {Young},\ and\ \citenamefont {Dean}}]{Yankowitz2019}%
  \BibitemOpen
  \bibfield  {author} {\bibinfo {author} {\bibfnamefont {M.}~\bibnamefont
  {Yankowitz}}, \bibinfo {author} {\bibfnamefont {S.}~\bibnamefont {Chen}},
  \bibinfo {author} {\bibfnamefont {H.}~\bibnamefont {Polshyn}}, \bibinfo
  {author} {\bibfnamefont {Y.}~\bibnamefont {Zhang}}, \bibinfo {author}
  {\bibfnamefont {K.}~\bibnamefont {Watanabe}}, \bibinfo {author}
  {\bibfnamefont {T.}~\bibnamefont {Taniguchi}}, \bibinfo {author}
  {\bibfnamefont {D.}~\bibnamefont {Graf}}, \bibinfo {author} {\bibfnamefont
  {A.~F.}\ \bibnamefont {Young}},\ and\ \bibinfo {author} {\bibfnamefont
  {C.~R.}\ \bibnamefont {Dean}},\ }\href
  {https://doi.org/10.1126/science.aav1910} {\bibfield  {journal} {\bibinfo
  {journal} {Science}\ }\textbf {\bibinfo {volume} {363}},\ \bibinfo {pages}
  {1059–1064} (\bibinfo {year} {2019})}\BibitemShut {NoStop}%
\bibitem [{\citenamefont {Cao}\ \emph {et~al.}(2018{\natexlab{b}})\citenamefont
  {Cao}, \citenamefont {Fatemi}, \citenamefont {Demir}, \citenamefont {Fang},
  \citenamefont {Tomarken}, \citenamefont {Luo}, \citenamefont
  {Sanchez-Yamagishi}, \citenamefont {Watanabe}, \citenamefont {Taniguchi},
  \citenamefont {Kaxiras}, \citenamefont {Ashoori},\ and\ \citenamefont
  {Jarillo-Herrero}}]{Cao2018_2}%
  \BibitemOpen
  \bibfield  {author} {\bibinfo {author} {\bibfnamefont {Y.}~\bibnamefont
  {Cao}}, \bibinfo {author} {\bibfnamefont {V.}~\bibnamefont {Fatemi}},
  \bibinfo {author} {\bibfnamefont {A.}~\bibnamefont {Demir}}, \bibinfo
  {author} {\bibfnamefont {S.}~\bibnamefont {Fang}}, \bibinfo {author}
  {\bibfnamefont {S.~L.}\ \bibnamefont {Tomarken}}, \bibinfo {author}
  {\bibfnamefont {J.~Y.}\ \bibnamefont {Luo}}, \bibinfo {author} {\bibfnamefont
  {J.~D.}\ \bibnamefont {Sanchez-Yamagishi}}, \bibinfo {author} {\bibfnamefont
  {K.}~\bibnamefont {Watanabe}}, \bibinfo {author} {\bibfnamefont
  {T.}~\bibnamefont {Taniguchi}}, \bibinfo {author} {\bibfnamefont
  {E.}~\bibnamefont {Kaxiras}}, \bibinfo {author} {\bibfnamefont {R.~C.}\
  \bibnamefont {Ashoori}},\ and\ \bibinfo {author} {\bibfnamefont
  {P.}~\bibnamefont {Jarillo-Herrero}},\ }\href
  {https://doi.org/10.1038/nature26154} {\bibfield  {journal} {\bibinfo
  {journal} {Nature}\ }\textbf {\bibinfo {volume} {556}},\ \bibinfo {pages}
  {80–84} (\bibinfo {year} {2018}{\natexlab{b}})}\BibitemShut {NoStop}%
\bibitem [{\citenamefont {Wong}\ \emph {et~al.}(2020)\citenamefont {Wong},
  \citenamefont {Nuckolls}, \citenamefont {Oh}, \citenamefont {Lian},
  \citenamefont {Xie}, \citenamefont {Jeon}, \citenamefont {Watanabe},
  \citenamefont {Taniguchi}, \citenamefont {Bernevig},\ and\ \citenamefont
  {Yazdani}}]{Wong2020}%
  \BibitemOpen
  \bibfield  {author} {\bibinfo {author} {\bibfnamefont {D.}~\bibnamefont
  {Wong}}, \bibinfo {author} {\bibfnamefont {K.~P.}\ \bibnamefont {Nuckolls}},
  \bibinfo {author} {\bibfnamefont {M.}~\bibnamefont {Oh}}, \bibinfo {author}
  {\bibfnamefont {B.}~\bibnamefont {Lian}}, \bibinfo {author} {\bibfnamefont
  {Y.}~\bibnamefont {Xie}}, \bibinfo {author} {\bibfnamefont {S.}~\bibnamefont
  {Jeon}}, \bibinfo {author} {\bibfnamefont {K.}~\bibnamefont {Watanabe}},
  \bibinfo {author} {\bibfnamefont {T.}~\bibnamefont {Taniguchi}}, \bibinfo
  {author} {\bibfnamefont {B.~A.}\ \bibnamefont {Bernevig}},\ and\ \bibinfo
  {author} {\bibfnamefont {A.}~\bibnamefont {Yazdani}},\ }\href
  {https://doi.org/10.1038/s41586-020-2339-0} {\bibfield  {journal} {\bibinfo
  {journal} {Nature}\ }\textbf {\bibinfo {volume} {582}},\ \bibinfo {pages}
  {198–202} (\bibinfo {year} {2020})}\BibitemShut {NoStop}%
\bibitem [{\citenamefont {Zondiner}\ \emph {et~al.}(2020)\citenamefont
  {Zondiner}, \citenamefont {Rozen}, \citenamefont {Rodan-Legrain},
  \citenamefont {Cao}, \citenamefont {Queiroz}, \citenamefont {Taniguchi},
  \citenamefont {Watanabe}, \citenamefont {Oreg}, \citenamefont {von Oppen},
  \citenamefont {Stern}, \citenamefont {Berg}, \citenamefont
  {Jarillo-Herrero},\ and\ \citenamefont {Ilani}}]{Zondiner2020}%
  \BibitemOpen
  \bibfield  {author} {\bibinfo {author} {\bibfnamefont {U.}~\bibnamefont
  {Zondiner}}, \bibinfo {author} {\bibfnamefont {A.}~\bibnamefont {Rozen}},
  \bibinfo {author} {\bibfnamefont {D.}~\bibnamefont {Rodan-Legrain}}, \bibinfo
  {author} {\bibfnamefont {Y.}~\bibnamefont {Cao}}, \bibinfo {author}
  {\bibfnamefont {R.}~\bibnamefont {Queiroz}}, \bibinfo {author} {\bibfnamefont
  {T.}~\bibnamefont {Taniguchi}}, \bibinfo {author} {\bibfnamefont
  {K.}~\bibnamefont {Watanabe}}, \bibinfo {author} {\bibfnamefont
  {Y.}~\bibnamefont {Oreg}}, \bibinfo {author} {\bibfnamefont {F.}~\bibnamefont
  {von Oppen}}, \bibinfo {author} {\bibfnamefont {A.}~\bibnamefont {Stern}},
  \bibinfo {author} {\bibfnamefont {E.}~\bibnamefont {Berg}}, \bibinfo {author}
  {\bibfnamefont {P.}~\bibnamefont {Jarillo-Herrero}},\ and\ \bibinfo {author}
  {\bibfnamefont {S.}~\bibnamefont {Ilani}},\ }\href
  {https://doi.org/10.1038/s41586-020-2373-y} {\bibfield  {journal} {\bibinfo
  {journal} {Nature}\ }\textbf {\bibinfo {volume} {582}},\ \bibinfo {pages}
  {203–208} (\bibinfo {year} {2020})}\BibitemShut {NoStop}%
\bibitem [{\citenamefont {Saito}\ \emph {et~al.}(2021)\citenamefont {Saito},
  \citenamefont {Yang}, \citenamefont {Ge}, \citenamefont {Liu}, \citenamefont
  {Taniguchi}, \citenamefont {Watanabe}, \citenamefont {Li}, \citenamefont
  {Berg},\ and\ \citenamefont {Young}}]{Saito2021}%
  \BibitemOpen
  \bibfield  {author} {\bibinfo {author} {\bibfnamefont {Y.}~\bibnamefont
  {Saito}}, \bibinfo {author} {\bibfnamefont {F.}~\bibnamefont {Yang}},
  \bibinfo {author} {\bibfnamefont {J.}~\bibnamefont {Ge}}, \bibinfo {author}
  {\bibfnamefont {X.}~\bibnamefont {Liu}}, \bibinfo {author} {\bibfnamefont
  {T.}~\bibnamefont {Taniguchi}}, \bibinfo {author} {\bibfnamefont
  {K.}~\bibnamefont {Watanabe}}, \bibinfo {author} {\bibfnamefont {J.~I.~A.}\
  \bibnamefont {Li}}, \bibinfo {author} {\bibfnamefont {E.}~\bibnamefont
  {Berg}},\ and\ \bibinfo {author} {\bibfnamefont {A.~F.}\ \bibnamefont
  {Young}},\ }\href {https://doi.org/10.1038/s41586-021-03409-2} {\bibfield
  {journal} {\bibinfo  {journal} {Nature}\ }\textbf {\bibinfo {volume} {592}},\
  \bibinfo {pages} {220–224} (\bibinfo {year} {2021})}\BibitemShut {NoStop}%
\bibitem [{\citenamefont {Rozen}\ \emph {et~al.}(2021)\citenamefont {Rozen},
  \citenamefont {Park}, \citenamefont {Zondiner}, \citenamefont {Cao},
  \citenamefont {Rodan-Legrain}, \citenamefont {Taniguchi}, \citenamefont
  {Watanabe}, \citenamefont {Oreg}, \citenamefont {Stern}, \citenamefont
  {Berg}, \citenamefont {Jarillo-Herrero},\ and\ \citenamefont
  {Ilani}}]{Rozen2021}%
  \BibitemOpen
  \bibfield  {author} {\bibinfo {author} {\bibfnamefont {A.}~\bibnamefont
  {Rozen}}, \bibinfo {author} {\bibfnamefont {J.~M.}\ \bibnamefont {Park}},
  \bibinfo {author} {\bibfnamefont {U.}~\bibnamefont {Zondiner}}, \bibinfo
  {author} {\bibfnamefont {Y.}~\bibnamefont {Cao}}, \bibinfo {author}
  {\bibfnamefont {D.}~\bibnamefont {Rodan-Legrain}}, \bibinfo {author}
  {\bibfnamefont {T.}~\bibnamefont {Taniguchi}}, \bibinfo {author}
  {\bibfnamefont {K.}~\bibnamefont {Watanabe}}, \bibinfo {author}
  {\bibfnamefont {Y.}~\bibnamefont {Oreg}}, \bibinfo {author} {\bibfnamefont
  {A.}~\bibnamefont {Stern}}, \bibinfo {author} {\bibfnamefont
  {E.}~\bibnamefont {Berg}}, \bibinfo {author} {\bibfnamefont {P.}~\bibnamefont
  {Jarillo-Herrero}},\ and\ \bibinfo {author} {\bibfnamefont {S.}~\bibnamefont
  {Ilani}},\ }\href {https://doi.org/10.1038/s41586-021-03319-3} {\bibfield
  {journal} {\bibinfo  {journal} {Nature}\ }\textbf {\bibinfo {volume} {592}},\
  \bibinfo {pages} {214–219} (\bibinfo {year} {2021})}\BibitemShut {NoStop}%
\bibitem [{\citenamefont {Xiao}\ \emph {et~al.}(2025)\citenamefont {Xiao},
  \citenamefont {Inbar}, \citenamefont {Birkbeck}, \citenamefont {Gershon},
  \citenamefont {Zamir}, \citenamefont {Taniguchi}, \citenamefont {Watanabe},
  \citenamefont {Berg},\ and\ \citenamefont
  {Ilani}}]{xiao2025interactingenergybandsmagic}%
  \BibitemOpen
  \bibfield  {author} {\bibinfo {author} {\bibfnamefont {J.}~\bibnamefont
  {Xiao}}, \bibinfo {author} {\bibfnamefont {A.}~\bibnamefont {Inbar}},
  \bibinfo {author} {\bibfnamefont {J.}~\bibnamefont {Birkbeck}}, \bibinfo
  {author} {\bibfnamefont {N.}~\bibnamefont {Gershon}}, \bibinfo {author}
  {\bibfnamefont {Y.}~\bibnamefont {Zamir}}, \bibinfo {author} {\bibfnamefont
  {T.}~\bibnamefont {Taniguchi}}, \bibinfo {author} {\bibfnamefont
  {K.}~\bibnamefont {Watanabe}}, \bibinfo {author} {\bibfnamefont
  {E.}~\bibnamefont {Berg}},\ and\ \bibinfo {author} {\bibfnamefont
  {S.}~\bibnamefont {Ilani}},\ }\href {https://arxiv.org/abs/2506.20738}
  {\bibinfo {title} {The interacting energy bands of magic angle twisted
  bilayer graphene revealed by the quantum twisting microscope}} (\bibinfo
  {year} {2025}),\ \Eprint {https://arxiv.org/abs/2506.20738} {arXiv:2506.20738
  [cond-mat.mes-hall]} \BibitemShut {NoStop}%
\bibitem [{\citenamefont {Zhu}\ \emph {et~al.}(2026)\citenamefont {Zhu},
  \citenamefont {Bennett}, \citenamefont {Larson}, \citenamefont {Ezzi},
  \citenamefont {Manousakis},\ and\ \citenamefont
  {Kaxiras}}]{zhu2026twistedbilayergraphenefirstprinciples}%
  \BibitemOpen
  \bibfield  {author} {\bibinfo {author} {\bibfnamefont {A.}~\bibnamefont
  {Zhu}}, \bibinfo {author} {\bibfnamefont {D.}~\bibnamefont {Bennett}},
  \bibinfo {author} {\bibfnamefont {D.~T.}\ \bibnamefont {Larson}}, \bibinfo
  {author} {\bibfnamefont {M.~M.~A.}\ \bibnamefont {Ezzi}}, \bibinfo {author}
  {\bibfnamefont {E.}~\bibnamefont {Manousakis}},\ and\ \bibinfo {author}
  {\bibfnamefont {E.}~\bibnamefont {Kaxiras}},\ }\href
  {https://arxiv.org/abs/2601.16851} {\bibinfo {title} {Twisted bilayer
  graphene from first-principles: structural and electronic properties}}
  (\bibinfo {year} {2026}),\ \Eprint {https://arxiv.org/abs/2601.16851}
  {arXiv:2601.16851 [cond-mat.mes-hall]} \BibitemShut {NoStop}%
\bibitem [{\citenamefont {Carr}\ \emph
  {et~al.}(2019{\natexlab{b}})\citenamefont {Carr}, \citenamefont {Fang},
  \citenamefont {Po}, \citenamefont {Vishwanath},\ and\ \citenamefont
  {Kaxiras}}]{carr19_2}%
  \BibitemOpen
  \bibfield  {author} {\bibinfo {author} {\bibfnamefont {S.}~\bibnamefont
  {Carr}}, \bibinfo {author} {\bibfnamefont {S.}~\bibnamefont {Fang}}, \bibinfo
  {author} {\bibfnamefont {H.~C.}\ \bibnamefont {Po}}, \bibinfo {author}
  {\bibfnamefont {A.}~\bibnamefont {Vishwanath}},\ and\ \bibinfo {author}
  {\bibfnamefont {E.}~\bibnamefont {Kaxiras}},\ }\href
  {https://doi.org/10.1103/PhysRevResearch.1.033072} {\bibfield  {journal}
  {\bibinfo  {journal} {Phys. Rev. Res.}\ }\textbf {\bibinfo {volume} {1}},\
  \bibinfo {pages} {033072} (\bibinfo {year} {2019}{\natexlab{b}})}\BibitemShut
  {NoStop}%
\bibitem [{\citenamefont {Berezinskii}(1972)}]{Berezinskii72}%
  \BibitemOpen
  \bibfield  {author} {\bibinfo {author} {\bibfnamefont {V.~.~L.}\ \bibnamefont
  {Berezinskii}},\ }\href@noop {} {\bibfield  {journal} {\bibinfo  {journal}
  {Sov. Phys. JETP}\ }\textbf {\bibinfo {volume} {34}},\ \bibinfo {pages} {610}
  (\bibinfo {year} {1972})}\BibitemShut {NoStop}%
\bibitem [{\citenamefont {Kosterlitz}\ and\ \citenamefont
  {Thouless}(1973)}]{Kosterlitz1973}%
  \BibitemOpen
  \bibfield  {author} {\bibinfo {author} {\bibfnamefont {J.~M.}\ \bibnamefont
  {Kosterlitz}}\ and\ \bibinfo {author} {\bibfnamefont {D.~J.}\ \bibnamefont
  {Thouless}},\ }\href {https://doi.org/10.1088/0022-3719/6/7/010} {\bibfield
  {journal} {\bibinfo  {journal} {Journal of Physics C: Solid State Physics}\
  }\textbf {\bibinfo {volume} {6}},\ \bibinfo {pages} {1181–1203} (\bibinfo
  {year} {1973})}\BibitemShut {NoStop}%
\bibitem [{\citenamefont {Kosterlitz}(1974)}]{Kosterlitz1974}%
  \BibitemOpen
  \bibfield  {author} {\bibinfo {author} {\bibfnamefont {J.~M.}\ \bibnamefont
  {Kosterlitz}},\ }\href {https://doi.org/10.1088/0022-3719/7/6/005} {\bibfield
   {journal} {\bibinfo  {journal} {Journal of Physics C: Solid State Physics}\
  }\textbf {\bibinfo {volume} {7}},\ \bibinfo {pages} {1046–1060} (\bibinfo
  {year} {1974})}\BibitemShut {NoStop}%
\bibitem [{\citenamefont {Nelson}\ and\ \citenamefont
  {Kosterlitz}(1977)}]{Nelson77}%
  \BibitemOpen
  \bibfield  {author} {\bibinfo {author} {\bibfnamefont {D.~R.}\ \bibnamefont
  {Nelson}}\ and\ \bibinfo {author} {\bibfnamefont {J.~M.}\ \bibnamefont
  {Kosterlitz}},\ }\href {https://doi.org/10.1103/PhysRevLett.39.1201}
  {\bibfield  {journal} {\bibinfo  {journal} {Phys. Rev. Lett.}\ }\textbf
  {\bibinfo {volume} {39}},\ \bibinfo {pages} {1201} (\bibinfo {year}
  {1977})}\BibitemShut {NoStop}%
\bibitem [{\citenamefont {Peotta}\ and\ \citenamefont
  {T\"{o}rm\"{a}}(2015)}]{Peotta2015}%
  \BibitemOpen
  \bibfield  {author} {\bibinfo {author} {\bibfnamefont {S.}~\bibnamefont
  {Peotta}}\ and\ \bibinfo {author} {\bibfnamefont {P.}~\bibnamefont
  {T\"{o}rm\"{a}}},\ }\bibfield  {journal} {\bibinfo  {journal} {Nature
  Communications}\ }\textbf {\bibinfo {volume} {6}},\ \href
  {https://doi.org/10.1038/ncomms9944} {10.1038/ncomms9944} (\bibinfo {year}
  {2015})\BibitemShut {NoStop}%
\bibitem [{\citenamefont {Scalapino}\ \emph {et~al.}(1993)\citenamefont
  {Scalapino}, \citenamefont {White},\ and\ \citenamefont
  {Zhang}}]{Scalapino93}%
  \BibitemOpen
  \bibfield  {author} {\bibinfo {author} {\bibfnamefont {D.~J.}\ \bibnamefont
  {Scalapino}}, \bibinfo {author} {\bibfnamefont {S.~R.}\ \bibnamefont
  {White}},\ and\ \bibinfo {author} {\bibfnamefont {S.}~\bibnamefont {Zhang}},\
  }\href {https://doi.org/10.1103/PhysRevB.47.7995} {\bibfield  {journal}
  {\bibinfo  {journal} {Phys. Rev. B}\ }\textbf {\bibinfo {volume} {47}},\
  \bibinfo {pages} {7995} (\bibinfo {year} {1993})}\BibitemShut {NoStop}%
\bibitem [{\citenamefont {Hu}\ \emph {et~al.}(2019)\citenamefont {Hu},
  \citenamefont {Hyart}, \citenamefont {Pikulin},\ and\ \citenamefont
  {Rossi}}]{Hu19}%
  \BibitemOpen
  \bibfield  {author} {\bibinfo {author} {\bibfnamefont {X.}~\bibnamefont
  {Hu}}, \bibinfo {author} {\bibfnamefont {T.}~\bibnamefont {Hyart}}, \bibinfo
  {author} {\bibfnamefont {D.~I.}\ \bibnamefont {Pikulin}},\ and\ \bibinfo
  {author} {\bibfnamefont {E.}~\bibnamefont {Rossi}},\ }\href
  {https://doi.org/10.1103/PhysRevLett.123.237002} {\bibfield  {journal}
  {\bibinfo  {journal} {Phys. Rev. Lett.}\ }\textbf {\bibinfo {volume} {123}},\
  \bibinfo {pages} {237002} (\bibinfo {year} {2019})}\BibitemShut {NoStop}%
\bibitem [{\citenamefont {Julku}\ \emph {et~al.}(2020)\citenamefont {Julku},
  \citenamefont {Peltonen}, \citenamefont {Liang}, \citenamefont {Heikkil\"a},\
  and\ \citenamefont {T\"orm\"a}}]{Julku20}%
  \BibitemOpen
  \bibfield  {author} {\bibinfo {author} {\bibfnamefont {A.}~\bibnamefont
  {Julku}}, \bibinfo {author} {\bibfnamefont {T.~J.}\ \bibnamefont {Peltonen}},
  \bibinfo {author} {\bibfnamefont {L.}~\bibnamefont {Liang}}, \bibinfo
  {author} {\bibfnamefont {T.~T.}\ \bibnamefont {Heikkil\"a}},\ and\ \bibinfo
  {author} {\bibfnamefont {P.}~\bibnamefont {T\"orm\"a}},\ }\href
  {https://doi.org/10.1103/PhysRevB.101.060505} {\bibfield  {journal} {\bibinfo
   {journal} {Phys. Rev. B}\ }\textbf {\bibinfo {volume} {101}},\ \bibinfo
  {pages} {060505} (\bibinfo {year} {2020})}\BibitemShut {NoStop}%
\bibitem [{\citenamefont {Gonz\'alez}\ and\ \citenamefont
  {Stauber}(2019)}]{Gonzalez19}%
  \BibitemOpen
  \bibfield  {author} {\bibinfo {author} {\bibfnamefont {J.}~\bibnamefont
  {Gonz\'alez}}\ and\ \bibinfo {author} {\bibfnamefont {T.}~\bibnamefont
  {Stauber}},\ }\href {https://doi.org/10.1103/PhysRevLett.122.026801}
  {\bibfield  {journal} {\bibinfo  {journal} {Phys. Rev. Lett.}\ }\textbf
  {\bibinfo {volume} {122}},\ \bibinfo {pages} {026801} (\bibinfo {year}
  {2019})}\BibitemShut {NoStop}%
\bibitem [{\citenamefont {Sánchez~Sánchez}\ \emph {et~al.}(2026)\citenamefont
  {Sánchez~Sánchez}, \citenamefont {González},\ and\ \citenamefont
  {Stauber}}]{figureszenodo}%
  \BibitemOpen
  \bibfield  {author} {\bibinfo {author} {\bibfnamefont {M.}~\bibnamefont
  {Sánchez~Sánchez}}, \bibinfo {author} {\bibfnamefont {J.}~\bibnamefont
  {González}},\ and\ \bibinfo {author} {\bibfnamefont {T.}~\bibnamefont
  {Stauber}},\ }\href {https://doi.org/https://doi.org/10.5281/zenodo.19768074}
  {\bibinfo {title} {Figures for "fermi velocity, interlayer couplings and
  magic angle renormalization in twisted bilayer graphene"}},\ \bibinfo
  {howpublished} {\url{https://doi.org/10.5281/zenodo.19768074}} (\bibinfo
  {year} {2026})\BibitemShut {NoStop}%
\bibitem [{\citenamefont {Lopes~dos Santos}\ \emph {et~al.}(2012)\citenamefont
  {Lopes~dos Santos}, \citenamefont {Peres},\ and\ \citenamefont
  {Castro~Neto}}]{dossantos12}%
  \BibitemOpen
  \bibfield  {author} {\bibinfo {author} {\bibfnamefont {J.~M.~B.}\
  \bibnamefont {Lopes~dos Santos}}, \bibinfo {author} {\bibfnamefont
  {N.~M.~R.}\ \bibnamefont {Peres}},\ and\ \bibinfo {author} {\bibfnamefont
  {A.~H.}\ \bibnamefont {Castro~Neto}},\ }\href
  {https://doi.org/10.1103/PhysRevB.86.155449} {\bibfield  {journal} {\bibinfo
  {journal} {Phys. Rev. B}\ }\textbf {\bibinfo {volume} {86}},\ \bibinfo
  {pages} {155449} (\bibinfo {year} {2012})}\BibitemShut {NoStop}%
\bibitem [{\citenamefont {Zou}\ \emph {et~al.}(2018)\citenamefont {Zou},
  \citenamefont {Po}, \citenamefont {Vishwanath},\ and\ \citenamefont
  {Senthil}}]{zou18}%
  \BibitemOpen
  \bibfield  {author} {\bibinfo {author} {\bibfnamefont {L.}~\bibnamefont
  {Zou}}, \bibinfo {author} {\bibfnamefont {H.~C.}\ \bibnamefont {Po}},
  \bibinfo {author} {\bibfnamefont {A.}~\bibnamefont {Vishwanath}},\ and\
  \bibinfo {author} {\bibfnamefont {T.}~\bibnamefont {Senthil}},\ }\href
  {https://doi.org/10.1103/PhysRevB.98.085435} {\bibfield  {journal} {\bibinfo
  {journal} {Phys. Rev. B}\ }\textbf {\bibinfo {volume} {98}},\ \bibinfo
  {pages} {085435} (\bibinfo {year} {2018})}\BibitemShut {NoStop}%
\bibitem [{\citenamefont {Giuliani}\ and\ \citenamefont
  {Vignale}(2005)}]{Giuliani_Vignale_2005}%
  \BibitemOpen
  \bibfield  {author} {\bibinfo {author} {\bibfnamefont {G.}~\bibnamefont
  {Giuliani}}\ and\ \bibinfo {author} {\bibfnamefont {G.}~\bibnamefont
  {Vignale}},\ }\href@noop {} {\emph {\bibinfo {title} {Quantum Theory of the
  Electron Liquid}}}\ (\bibinfo  {publisher} {Cambridge University Press},\
  \bibinfo {year} {2005})\BibitemShut {NoStop}%
\bibitem [{\citenamefont {Kudin}\ \emph {et~al.}(2002)\citenamefont {Kudin},
  \citenamefont {Scuseria},\ and\ \citenamefont {Cancès}}]{Kudin2002}%
  \BibitemOpen
  \bibfield  {author} {\bibinfo {author} {\bibfnamefont {K.~N.}\ \bibnamefont
  {Kudin}}, \bibinfo {author} {\bibfnamefont {G.~E.}\ \bibnamefont
  {Scuseria}},\ and\ \bibinfo {author} {\bibfnamefont {E.}~\bibnamefont
  {Cancès}},\ }\href {https://doi.org/10.1063/1.1470195} {\bibfield  {journal}
  {\bibinfo  {journal} {The Journal of Chemical Physics}\ }\textbf {\bibinfo
  {volume} {116}},\ \bibinfo {pages} {8255–8261} (\bibinfo {year}
  {2002})}\BibitemShut {NoStop}%
\bibitem [{\citenamefont {Bernevig}\ \emph {et~al.}(2021)\citenamefont
  {Bernevig}, \citenamefont {Song}, \citenamefont {Regnault},\ and\
  \citenamefont {Lian}}]{bernevig21}%
  \BibitemOpen
  \bibfield  {author} {\bibinfo {author} {\bibfnamefont {B.~A.}\ \bibnamefont
  {Bernevig}}, \bibinfo {author} {\bibfnamefont {Z.-D.}\ \bibnamefont {Song}},
  \bibinfo {author} {\bibfnamefont {N.}~\bibnamefont {Regnault}},\ and\
  \bibinfo {author} {\bibfnamefont {B.}~\bibnamefont {Lian}},\ }\href
  {https://doi.org/10.1103/PhysRevB.103.205413} {\bibfield  {journal} {\bibinfo
   {journal} {Phys. Rev. B}\ }\textbf {\bibinfo {volume} {103}},\ \bibinfo
  {pages} {205413} (\bibinfo {year} {2021})}\BibitemShut {NoStop}%
\bibitem [{\citenamefont {Coleman}(2015)}]{Coleman_2015}%
  \BibitemOpen
  \bibfield  {author} {\bibinfo {author} {\bibfnamefont {P.}~\bibnamefont
  {Coleman}},\ }\href@noop {} {\emph {\bibinfo {title} {Introduction to
  Many-Body Physics}}}\ (\bibinfo  {publisher} {Cambridge University Press},\
  \bibinfo {year} {2015})\BibitemShut {NoStop}%
\bibitem [{\citenamefont {Weisstein}(2026)}]{wolframeta}%
  \BibitemOpen
  \bibfield  {author} {\bibinfo {author} {\bibfnamefont {E.~W.}\ \bibnamefont
  {Weisstein}},\ }\href@noop {} {\bibinfo {title} {Dirichlet eta function}},\
  \bibinfo {howpublished}
  {\url{https://mathworld.wolfram.com/DirichletEtaFunction.html}} (\bibinfo
  {year} {2026}),\ \bibinfo {note} {accessed: 2026-24-04}\BibitemShut {NoStop}%
\end{thebibliography}%


\clearpage

\setcounter{equation}{0}
\renewcommand\theequation{S\arabic{equation}}

\setcounter{figure}{0}
\renewcommand\thefigure{S\arabic{figure}}    

\setcounter{table}{0}
\renewcommand\thetable{S\arabic{table}}

\makeatletter

\renewcommand{\theHequation}{S\arabic{equation}}
\renewcommand{\theHfigure}{S\arabic{figure}}
\renewcommand{\theHtable}{S\arabic{table}}
\makeatletter

\title{Supplementary Materials for "Fermi velocity, interlayer couplings, and magic angle renormalization in twisted bilayer graphene"}

\maketitle

\onecolumngrid
\tableofcontents

\section{Hartree-Fock tight-binding approach}

In a graphene layer the lattice vectors are $\boldsymbol{a}_1 = a(1/2,\sqrt{3}/2)$ and $\boldsymbol{a}_2 = a(-1/2,\sqrt{3}/2)$, with $a = 2.46\text{ \AA}$ the lattice constant. The unit cell contains two atoms denoted $A$ and $B$. The $A$ sublattice is displaced from the lattice positions by $-(\boldsymbol{a}_1 - \boldsymbol{a}_2)/3$, and the $B$ sublattice by $(\boldsymbol{a}_1 + \boldsymbol{a_2})/3$. The reciprocal lattice vectors are $\boldsymbol{G}_1 = \frac{4\pi}{\sqrt{3}a} (\sqrt{3}/2,1/2)$ and $\boldsymbol{G}_2 = \frac{4\pi}{\sqrt{3}a} (-\sqrt{3}/2,1/2)$, and the $K$ points are $K = (\boldsymbol{G_1} - \boldsymbol{G_2})/3$ and $K' = -K$. 

Starting from two aligned graphene layers stacked at the equilibrium interlayer distance, i.e. at vertical positions $z=\pm d/2 = \pm 3.35/2$ $\text{ \AA}$, the top layer gets rotated by the angle $\theta/2$ and the bottom layer by $-\theta/2$, with the center of rotation being the center of one of the carbon hexagons. The rigid lattice then undergoes structural relaxation which we obtain using the elastic theory from Ref. \cite{carr18}. We chose $\cos(\theta) = 1 - 1/(6n_\theta^2 +6n_\theta +2)$ with integer $n_\theta$ such that the structure is commensurate \cite{dossantos12}. The unit vectors of the resulting superlattice are $\boldsymbol{R}_1 = L_M(\sqrt{3}/2,1/2)$ and $\boldsymbol{R}_2 = L_M(-\sqrt{3}/2,1/2)$ with $L_M = a\sqrt{3n_\theta^2+3 n_\theta+ 1}$ the moiré lattice constant. The reciprocal lattice has the lattice vectors $\boldsymbol{g}_1 = \frac{4\pi}{\sqrt{3}L_M}(1/2,\sqrt{3}/2)$ and $\boldsymbol{g}_2 = \frac{4\pi}{\sqrt{3}L_M}(-1/2,\sqrt{3}/2)$.

The point group of this structure is $C_{6v}$, generated by six-fold rotations about the $z$ axis, $C_{6z}$, and two-fold rotations about the $x$, $C_{2x}$. It includes the three-fold rotations $C_{3z} = C_{6z}^2$, and the two-fold rotations, $C_{2z}=C_{6z}^3$. $C_{6z}$ is directly inherited from the parent monolayers, and $C_{2x}$ can be viewed as coming from the mirror symmetry of the monolayers: in the twisted structure the layers are not aligned so the mirror is not simultaneously a symmetry of both layers, but the mirror plus an interchange of layers (i.e. the two-fold rotation about the $x$ axis at $z=0$) is a symmetry \cite{zou18}.

Consider the interacting tight-binding Hamiltonian $\mathcal{H}$ in the TBG lattice,
\begin{align}
\begin{split}
    \mathcal{H} = & H_{\text{TB}} + {H}_\text{int} , \\ 
    H_{\text{TB}} = & \sum_{\boldsymbol{r}\boldsymbol{r'}s} t(\boldsymbol{r},\boldsymbol{r'}) c^\dagger_{\boldsymbol{r}s}c_{\boldsymbol{r'}s},   \\  
    {H}_\text{int} =&\frac{1} {2}\sum_{\boldsymbol{r}\neq \boldsymbol{r'},ss'} V(\boldsymbol{r}- \boldsymbol{r'}) \delta n_{\rr s}  \delta n_{\rr' s'} + \frac{1}{2}\sum_{\boldsymbol{r}s} U  \delta n_{\rr s} \delta n_{\rr \Bar{s}},    \end{split}
    \label{hamiltoniansm}
\end{align}
where $ H_{\text{TB}}$ is the non-interacting tight-binding hamiltonian and ${H}_\text{int}$ is the interacting part, $c^\dagger_{\rr s}$ ($c_{\rr s}$) is the creation (annihilation) operator of an electron with spin $s$ at site $\rr$, $\delta n_{\rr s} = c^\dagger_{\rr s} c_{\rr s} - \frac{1}{2}$ is the spin-$s$ density operator at site $\rr$ relative to half filling and $\Bar{s}$ denotes the spin projection opposite to $s$.

$t(\rr,\rr')$ is the hopping function from Refs. \cite{fang16,kang23}.
This hopping function accounts for 'multi-angular momentum channels' and makes $t(\rr,\rp)$ dependent on the local crystalline environment, $t(\rr,\rp) \neq t(\rr - \rp)$ --- this is the reason for $\gamma_3 \neq \gamma_4$ in the SWM parametrization of graphite \cite{guinea19}. Also $t(\rr, \rp)$ is real thus the system enjoys a spinless time-reversal symmetry $\mathcal{T}$

$V(\rr-\rr')$ is the Coulomb potential that we take to be the potential felt by electrons on a sample between two metallic gates with gate-to-gate distance $\xi=10$ nm,
\begin{align}
    V(\rr) = \frac{e^2}{\epsilon}\sum_{n=-\infty}^\infty \frac{(-1)^n}{||\rr + n\xi \boldsymbol{\hat{z}}||},    
\end{align}
with $\epsilon=10$ the effective dielectric constant, and $U=4$ eV the on-site Hubbard energy.

In this work we assume a spin-symmetric state, so we drop the spin index in what follows. The standard mean-field, or Hartree-Fock, decoupling outputs the mean-field hamiltonian \cite{Giuliani_Vignale_2005}
\begin{align}
    \mathcal{H} \approx {H}_{\text{MF}} =& \sum_{\boldsymbol{r}\boldsymbol{r'}} t(\boldsymbol{r} ,\boldsymbol{r'}) c^\dagger_{\boldsymbol{r}}c_{\boldsymbol{r'}} + 2\sum_{\boldsymbol{r}\neq \boldsymbol{r'}} V(\boldsymbol{r}- \boldsymbol{r'}) \bigg( \langle c^\dagger_{\boldsymbol{r'}} c_{\boldsymbol{r'}} \rangle_0  - \frac{1}{2} \bigg) c^\dagger_{\boldsymbol{r}} c_{\boldsymbol{r}} \nonumber \\ 
    &-\sum_{\boldsymbol{r}\neq \boldsymbol{r'}} V(\boldsymbol{r}- \boldsymbol{r'})  \langle c^\dagger_{\boldsymbol{r'}} c_{\boldsymbol{r}} \rangle_0 c^\dagger_{\boldsymbol{r}} c_{\boldsymbol{r'}} + \sum_{\boldsymbol{r}} U \bigg( \langle c^\dagger_{\boldsymbol{r}} c_{\boldsymbol{r}} \rangle_0  - \frac{1}{2} \bigg) c^\dagger_{\boldsymbol{r}} c_{\boldsymbol{r}} + \text{constant}.
    \label{H_MF}
\end{align}
$ \langle c^\dagger_{\boldsymbol{r'}} c_{\boldsymbol{r}} \rangle_0$ denotes the expectation value in a state of reference. Self-consistent Hartree-Fock translates to finding a solution where the reference state corresponds to the ground state of ${H}_{\text{MF}}$: $\langle c^\dagger_{\boldsymbol{r'}} c_{\boldsymbol{r}} \rangle_0= \langle c^\dagger_{\boldsymbol{r'}} c_{\boldsymbol{r}} \rangle$. As ${H}_{\text{MF}}$ is a single-particle hamiltonian, from now on we switch to bra-ket notation.

We impose translational symmetry and assume periodic boundary conditions. It is then natural to work in the momentum basis,
\begin{align}
| \boldsymbol{k}\boldsymbol{\delta} \rangle = \frac{1}{\sqrt{N_c}}\sum_{\boldsymbol{R}} e^{i\k \cdot (\boldsymbol{\delta} + \boldsymbol{R})} | \boldsymbol{\delta}+ \boldsymbol{R} \rangle,    
\end{align}
where $N_c$ is the number of unit cells in the system and $\k$ is the momentum inside the rhombic Brillouin zone, $ \k = (n_1/\sqrt{N_c}) \boldsymbol{g}_1 + (n_2/\sqrt{N_c}) \boldsymbol{g}_2$ for $n_1,n_2=0,...,\sqrt{N_c}-1$. The index $\boldsymbol{\delta}$ denotes an atomic position inside the Wigner-Seitz cell, and $\boldsymbol{R}$ denotes a lattice vector. Distinct momenta are decoupled in the mean-field Hamiltonian.

In the self-consistent algorithm we take the ground state of $H_\text{TB}$ as the initial seed for $\langle c^\dagger_{\boldsymbol{r'}} c_{\boldsymbol{r}} \rangle_0$, with a $6\times 6$ discretization of the Brillouin zone ($N_c=36$). The ODA algorithm \cite{Kudin2002} is used to find the self-consistent solution. Crucially, we impose the crystallographic symmetries, time-reversal symmetry and valley charge conservation at each iteration in order to avoid reaching symmetry-broken solutions, see discussion below.

We introduce the mean-field basis $| \boldsymbol{k}n \rangle $, that diagonalizes $H_{\text{MF}}$ in the self-consistent solution,
\begin{align}
    {H}_{\text{MF}}& = \sum_{n \k } \varepsilon_n(\k) | n \boldsymbol{k} \rangle \langle n \boldsymbol{k}| + \text{constant}, \\  
    | n \boldsymbol{k} \rangle &= \sum_{\boldsymbol{\delta}} u_{n \k}(\boldsymbol{\delta})  | \boldsymbol{k}\boldsymbol{\delta} \rangle =  \frac{1}{\sqrt{N_c}}\sum_{\boldsymbol{r}} u_{n \k} (\boldsymbol{\delta}) e^{i\k \cdot \rr} | \rr \rangle,
\end{align}
where $n$ is the band index, $\varepsilon_n(\k)$ are the (renormalized) band energies, $u_{n\k} (\boldsymbol{\delta}) = \langle \boldsymbol{k} \boldsymbol{\delta} | \boldsymbol{k} n \rangle $ the Bloch periodic functions satisfying $\sum_{\boldsymbol{\delta}} |u_{n \k}(\boldsymbol{\delta})|^2 = 1$, and $\frac{1}{\sqrt{N_c}} u_{n \k} (\boldsymbol{\delta}) e^{i\k \cdot \rr} = \langle \boldsymbol{r} | \boldsymbol{k} n \rangle$ the wave functions in real space, with $\rr =  \boldsymbol{\delta} +  \boldsymbol{R}$.

\subsection{Enforcing symmetries in the Hartree-Fock tight-binding approach}

The main object of the Hartree-Fock tight-binding approach is the correlation, or Fock, matrix
\begin{align}
 \langle c^\dagger_{\boldsymbol{r}} c_{\rp } \rangle =  \frac{1}{N_c} \sum_{\varepsilon_n(\k) \leq \varepsilon_F} \lambda_{n\k} u_{n \k}(\db)^* u_{n \k}(\dbp)e^{-i \boldsymbol{k} \cdot(\db - \dbp + \R - \Rp)},
\end{align}
where $\varepsilon_F$ is the Fermi level, the degeneracy factor $\lambda_\alpha=1$ if $\varepsilon_{n}(\k) < \varepsilon_{F}$ and $\lambda_\alpha = (N_e-N_{\varepsilon_F})/N_{\varepsilon_F}$ if $\varepsilon_n(\k) = \varepsilon_{F}$, with $N_e$ the number of electrons in the system and $N_{\varepsilon_F}$ the number of electrons with energy $\varepsilon_F$. 

By translational symmetry, $\langle c^\dagger_{\boldsymbol{r} + \R} c_{\rp + \R}\rangle$ $= \langle c^\dagger_{\boldsymbol{r}} c_{\rp}\rangle$, hence the correlation matrix only depends on $\db$, $\dbp$ and $\R - \Rp$. Accordingly, we define the function $F$ as  
\begin{align}
    F(\db,\dbp,\R-\Rp) \equiv \langle c^\dagger_{\boldsymbol{r}} c_{\rp}\rangle,
\end{align}
or 
\begin{align}
 F(\db,\dbp, \R) = \frac{1}{N_c} \sum_{\varepsilon_n(\k) \leq \varepsilon_F} \lambda_{n\k} u_{n \k}(\db)^* u_{n \k}(\dbp)e^{-i \boldsymbol{k} \cdot(\db - \dbp + \R)}.
  \label{Fmatrix}
\end{align}

The correlation matrx is convolved with the Coulomb potential, which is in general long-ranged, in the Fock term of $H_{\text{MF}}$. With a finite gate distance the potential is screened for large separations; moreover $F$ also decays with $|\R|$ in general. Hence in the numerics it suffices to compute and store $F$ for some finite number of unit vectors $|\R|\leq R_c$, with $R_C$ a fixed cutoff or, equivalently, for some number lattice shells surrounding the origin. With our choice of gate distance, $\xi=10$ nm, for small twist angles it suffices to keep just one shell of lattice vectors, i.e. $R_c = |\R_1|$. For larger $\xi$, one would need larger $R_c$. We note that properly characterizing the decay of $F$ can help reduce the computational and memory requirements for large $\xi$. Notice also that the hermiticity of $\langle c^\dagger_{\boldsymbol{r}} c_{\rp}\rangle $ implies $F(\db,\dbp,\boldsymbol{0})=F(\dbp,\db, \boldsymbol{0})^*$ and $F(\db,\dbp,\R)=F(\dbp,\db,-\R)^*$, then the $F$ matrix has to be stored for only half of the lattice vectors. 

We impose the translational and spin-rotation symmetries (by construction) as well as the crystallographic symmetries, time-reversal and valley charge conservation in the Fock matrix at each step of the self-consistency loop. We now describe how to enforce the symmetries.

Spinless time-reversal symmetry acts as complex conjugation. The correlation matrix transforms as
\begin{align}
    \mathcal{T}:  F(\db, \dbp, \R) \to F^\mathcal{T}(\db,\dbp,\R) = F(\db,\dbp,\R)^*.
    \label{FT}
\end{align}
A $\mathcal{T}$-symmetric Fock matrix can be constructed via
\begin{align}
    \mathcal{T} \text{-symmetric } F:  F(\db, \dbp, \R) \to \frac{1}{2}\big( F(\db,\dbp,\R) +  F^{\mathcal{T}}(\db,\dbp,\R) \big)
\end{align}

$C_{2z}$ symmetry sends $\db \to C_{2z}(\db)$, $\dbp \to C_{2z}(\dbp)$ and $\R \to -\R$, so the correlation matrix transforms as
\begin{align}
    C_{2z}:  F(\db, \dbp, \R) \to F^{C_{2z}}(\db,\dbp,\R) = F(C_{2z}(\db),C_{2z}(\dbp),-\R)^* = F(C_{2z}(\dbp),C_{2z}(\db),\R)^*,
    \label{FC2} 
\end{align}
and $C_{2z}$ symmetry is enforced via
\begin{align}
    C_{2z} \text{-symmetric } F :  F(\db, \dbp, \R) \to \frac{1}{2}\big( F(\db,\dbp,\R) +  F^{C_{2z}}(\db,\dbp,\R) \big).
\end{align}

The action of the product $C_{2z}\mathcal{T}$ follows immediately,
\begin{align}
    C_{2z}\mathcal{T}:  F(\db, \dbp, \R) \to F^{C_{2z}\mathcal{T}} (\db,\dbp,\R) = F(C_{2z}(\dbp),C_{2z}(\db),\R),    \label{FC2T}   \\
    C_{2z}\mathcal{T} \text{-symmetric } F :  F(\db, \dbp, \R) \to \frac{1}{2}\big( F(\db,\dbp,\R) +  F^{C_{2z}\mathcal{T}}(\db,\dbp,\R) \big).
\end{align}

\begin{figure}[H]
    \raggedright
    \hspace{0.5cm} (a) \hspace{3.8cm} (b) \hspace{9.8cm} (c)
    \\
    \centering
    \includegraphics[valign=bottom,width=0.24\linewidth]{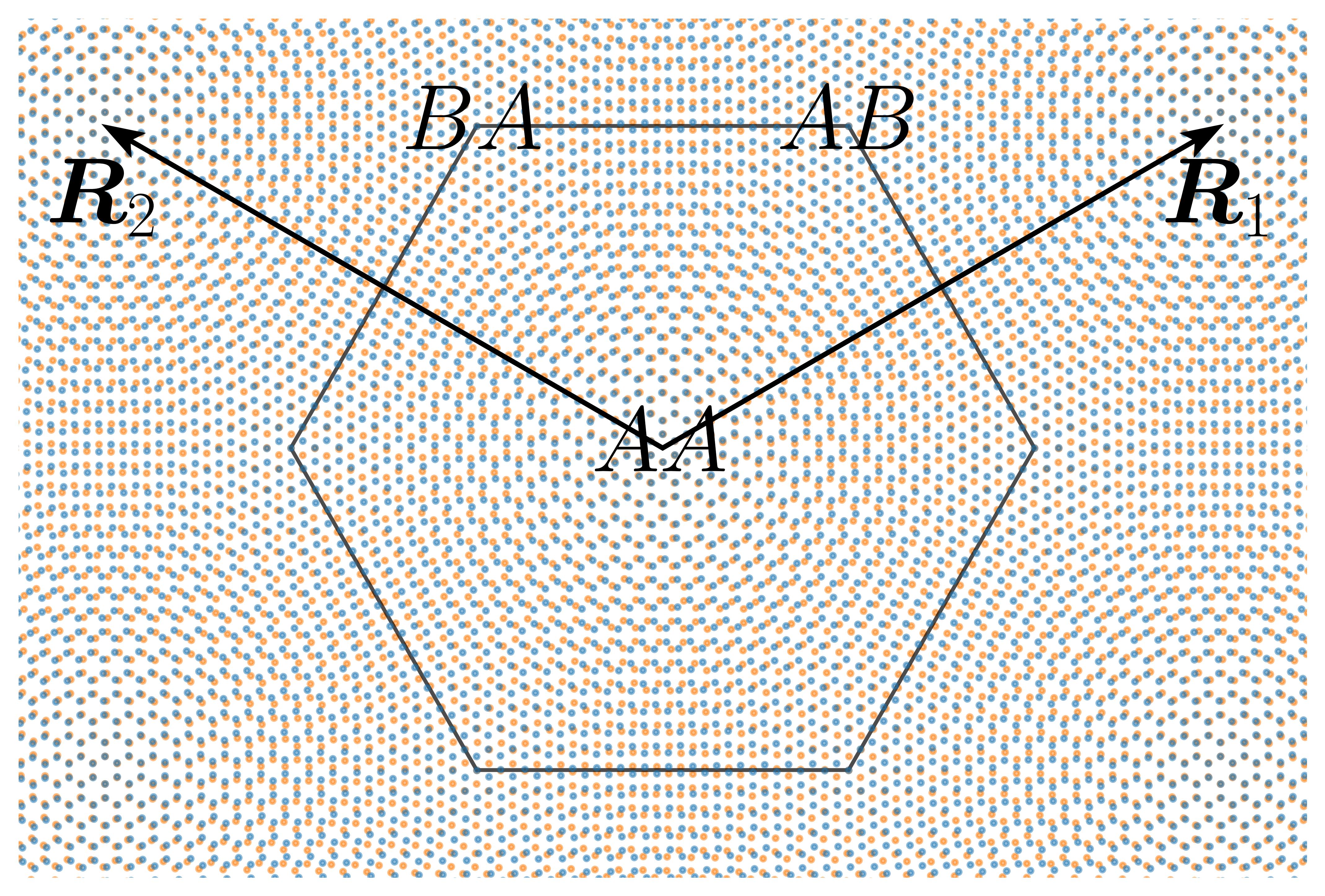}
    \includegraphics[valign=bottom,width=0.58\linewidth]{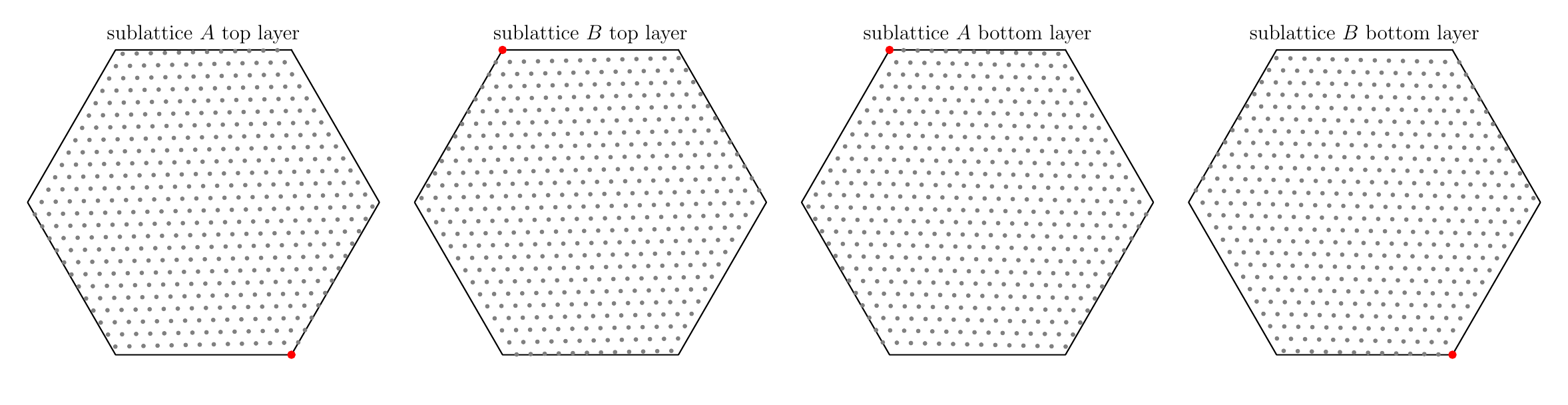}
    \includegraphics[valign=bottom,width=0.1\linewidth]{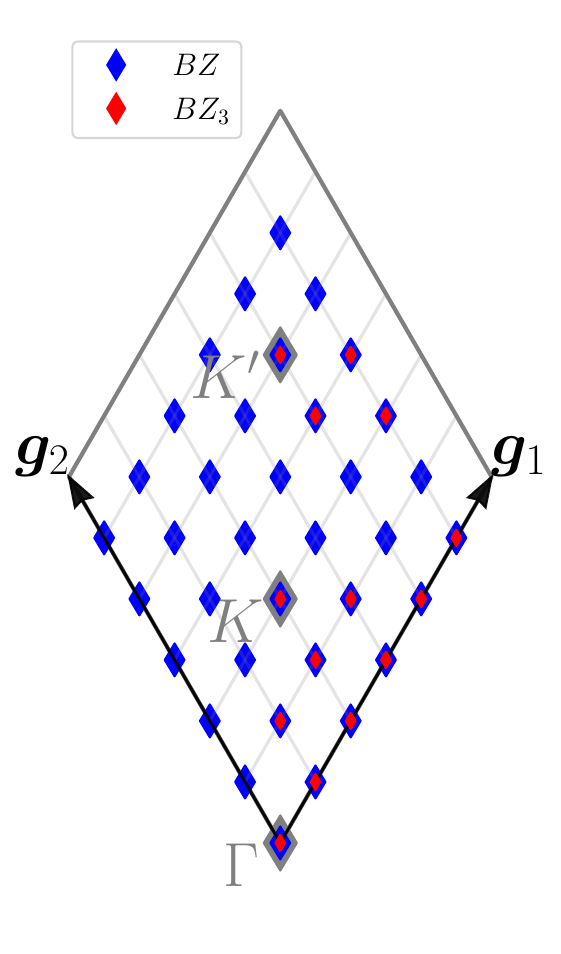}
    \caption{(a) Top view of TBG with twist angle $2.13^\circ$. The Wigner-Seitz cell, the lattice vectors and the regions where the local stacking is $AA$, $AB$ and $BA$ are labeled. (b) Atoms from the different sublattices and layers with twist angle $2.65^\circ$ on the Wigner-Seitz cell. In red, atoms on the corner of the unit cell. A three-fold rotation on the red atoms is equivalent to a lattice translation. (c) $6\times 6$ Brillouin zone ($BZ$) and reduced Brillouin zone ($BZ_3$) including the $C_{3z}$-invariant points $K$, $K'$ and $\Gamma$ and one representative out of the $C_{3z}$-related non-invariant points.}
    \label{tbsetup}
\end{figure}

The action of $C_{3z}$ on $F$ is more complicated. This is because the atoms at the corner of the unit cell (one atom per sublattice per layer) do not have a partner under $C_{3z}$ on the first unit cell, rather $C_{3z}(\db) = \db + \R_{\db}$ for some unit vector $\R_{\db}$ that depends on $\db$, see Fig. \ref{tbsetup}(b). Then, the transformed $F^{C_{3z}}$ cannot be expressed by a simple formula similar to Eqs. (\ref{FT}), (\ref{FC2}), (\ref{FC2T}) due to this geometrical obstruction. Moreover, the $C_{3z}$ symmetry cannot be exactly enforced in real space if the cutoff $R_c$ is finite. Consider for example $\db$ being one of the atoms on the corners of the unit cell, $\R$ one of the lattice vectors of the outermost shell stored in $F$ and $\dbp$ a generic atom not on one of the corners of the unit cell. In order to implement the $C_{3z}$ symmetry one would need to access $F(\db, C_{3z}(\dbp), C_{3z}(\R) + \R_{\db})$. But there always exist particular choices of $\R$ and $\db$ that yield $|C_{3z}(\R) + \R_{\db})| > R_c$, hence $F(\db, C_{3z}(\dbp), C_{3z}(\R) + \R_{\db})$ is not stored and one cannot enforce the $C_{3z}$ symmetry exactly.

In order to enforce the $C_{3z}$ symmetry, we turn to the mean-field basis. Under the rotation the wave function transforms as 
\begin{align}
    \big\langle \rr \big| C_{3z}^{\pm 1} | n \k \big\rangle &=  \frac{1}{\sqrt{N_c}}u_{n \k} \big( \tilde{C}_{3z}^{\mp 1}(\db) \big) e^{i C_{3z}^{\pm 1} (\boldsymbol{k}) \cdot(\db + \R)}
\end{align}
with $\tilde{C}_{3z}^{\pm 1}$ denoting the rotation by $\pm 2\pi/3$ modulo a lattice translation to the first unit cell; in other words $\tilde{C}_{3z}^{\pm 1}(\db) = C_{3z}^{\pm 1}(\db)$ for atoms inside the boundary of the unit cell and $\tilde{C}_{3z}^{\pm 1}(\db) = \db$ for the atoms on the corners of the unit cell. Then, defining
\begin{align}
     C_{3z}^{\pm1}: \ F(\db, \dbp, \R) \to F^{C_{3z}^{\pm 1}}(\db,\dbp,\R) = \frac{1}{N_c} \sum_{\varepsilon_n(\k) < \varepsilon_F}  \lambda_{n \k} u_{n \k}(\tilde{C}_{3z}^{\mp 1}(\db))^* u_{n \k}(\tilde{C}_{3z}^{\mp 1}(\dbp))e^{-i C_{3z}^{\pm 1}(\boldsymbol{k}) \cdot(\db - \dbp + \R)},
     \label{C3action}
\end{align}
the symmetry can be implemented with the substitution
\begin{align}
    C_{3z} \text{-symmetric } F : \ F(\db, \dbp, \R) \to \frac{1}{3}\big( F(\db,\dbp,\R) +  F^{C_{3z}}(\db,\dbp,\R) + F^{C_{3z}^{-1}}(\db,\dbp,\R) \big) 
    \label{FC3}
\end{align}

In practice, when the $C_{3z}$ symmetry will be imposed we take advantage of a computationally cheaper alternative. We consider momenta on a reduced Brillouin zone $BZ_3$ that contains the $C_{3z}$-invariant points $\Gamma_M, K_M$ and $K'_M$ and one representative momentum out of the three $C_{3z}$-related ones on the rest of the BZ, see Fig. \ref{tbsetup}(c). $F$ is constructed as
\begin{align}
    F(\db,\dbp,\R) = \frac{1}{N_c} \sum_{\substack{\varepsilon_n(\k) \leq \varepsilon_F \\ \k \in BZ_3}} \lambda_{n\k} u_{n \k}(\db)^* u_{n \k}(\dbp)e^{-i \boldsymbol{k} \cdot(\db - \dbp + \R)},
\end{align}
with the degeneracy factors $\lambda_{n \k} = 3$ if $\varepsilon_{n} (\k) < \varepsilon_F$ and $\k \neq \Gamma_M, K_M, K'_M$; $\lambda_{n \k} = 3(N-N_{\varepsilon_F})/N_{\varepsilon_F}$ if $\varepsilon_{n}( \k) = \varepsilon_F$ and $\k \neq \Gamma_M, K_M, K'_M$; $\lambda_{n \k} = 1$ if $\varepsilon_{n}( \k) < \varepsilon_F$ and $\k = \Gamma_M, K_M \text{ or } K'_M$; and $\lambda_{n \k} = (N-N_{\varepsilon_F})/N_{\varepsilon_F}$ if $\varepsilon_{n}( \k) = \varepsilon_F$ and $\k = \Gamma_M, K_M \text{ or } K'_M$. The factors of $3$ for the non-invariant points ensure proper counting, and $\varepsilon_F$ is determined assuming that the energies $\varepsilon_{n}(\k)$ satisfy $C_{3z}$ symmetry on the full Brillouin zone.  Finally, a symmetry preserving Fock matrix is obtained via the using Eqs. (\ref{C3action}) and (\ref{FC3}).

\subsubsection{Valley symmetry}

The valley is an emergent quantum number at low energies. It arises from the two valleys $K$ and $K'$ of the original graphene layers. If we assign a charge $+1$ and $-1$ to the valley $K$ and $K'$ respectively, then the valley charge is approximately conserved in the interacting system \cite{bernevig21}. Defining the valley charge is not immediate on the lattice, and we must resort to an effective definition. In this respect, the microscopic valley operator $\tilde{\tau}_z$ \cite{sanchez25} is defined on the lattice and has the property that states on valley $K(K')$ have expectation values of $\tilde{\tau}_z$ very close to $+1(-1)$ (typically of the order of $\pm 0.99$). This provides an operational definition of the valley charge.

Consider the definition of $F$ in Eq. (\ref{Fmatrix}). The set of occupied states can be decomposed into states with large negative energies, i.e. low band index, and low-energy states close to the Fermi level. Indeed the valley charge, being defined for states at low energies, is not meaningful for states deep in the Fermi sea for which the electrons do not belong to either valley. With this is mind, at each momentum we consider the set of bands with band index smaller than a particular cutoff value, $|n| \leq n_c$ (we count the band index from the charge neutrality point, such that $-N_{auc}/2 \leq n \leq N_{auc}/2$ with $N_{auc}$ the number of atoms in the unit cell). At this stage it is crucial to choose $n_c$ such that the set of the first $2n_c$ bands is isolated; this is, $\varepsilon_{n_c}(\k) < \varepsilon_{n_c+1}(\k)$ and $\varepsilon_{-n_c}( \k) > \varepsilon_{-n_c-1}( \k )$. Generally for TBG we can choose $n_c=2$ (the flat bands) or $n_c=10$ (as discussed in Apendix F of Ref. \cite{sanchez25}) for the whole BZ. 

The Fock matrix then decomposes into high and low energy parts,
\begin{align}
    &F(\db ,\dbp,\R) = F_{\text{high}}(\db ,\dbp,\R) + F_{\text{low}}(\db ,\dbp,\R)  \\
    &F_{\text{high}}(\db ,\dbp,\R)  = \frac{1}{N_c} \sum_{\substack{\varepsilon_n(\k) \leq \varepsilon_F \\ |n|>n_c}} \lambda_{n\k} u_{n \k}(\db)^* u_{n \k}(\dbp)e^{-i \boldsymbol{k} \cdot(\db - \dbp + \R)}  \\
    &F_{\text{low}}(\db ,\dbp,\R) = \frac{1}{N_c} \sum_{\substack{\varepsilon_n(\k) \leq \varepsilon_F \\ |n|\leq n_c}} \lambda_{n\k} u_{n \k}(\db)^* u_{n \k}(\dbp)e^{-i \boldsymbol{k} \cdot(\db - \dbp + \R)} 
\end{align}

We compute the projected valley operator on the central $2n_c$ bands as
\begin{align}
    [\tilde{\tau}_z(\k)]_{nm} = \langle \k n | \tilde{\tau}_z | \k m \rangle,
\end{align}
with $n,m=-n_c, ..., n_c$. The matrix $\tilde{\tau}(\k)$ is hermitian, with eigenvalues close to $\pm 1$. In order to define a symmetry operation, we transform $\tilde{\tau}_z(\k)$ to a unitary matrix ${\tau}_z(\k)$,
\begin{align}
    {\tau}_z(\k) \equiv {\tilde{\tau}}_z(\k)\big( \tilde{\tau}_z(\k)^2 \big)^{-1/2},
\end{align}
where $\big( \tilde{\tau}_z(\k)^2)^{-1/2}$ is the matrix with the same eigenvectors as $\tilde{\tau}_z(\k)$ and with eigenvalues $\{|\lambda|^{-1} \}$ if $\tilde{\tau}_z(\k)$ has eigenvalues $\{\lambda\}$. We denote the eigenstates transformed by the valley operator with a subscript and the corresponding periodic functions with a superscript, 
\begin{align}
    &|\k n \rangle_{\tau_z} =  [\tau_z(\k)]_{nm} |\k m \rangle, \\
    &\langle \rr | n \k \rangle_{\tau_z} =  \frac{1}{\sqrt{N_c}}u^{\tau_z}_{n \k}(\db) e^{i \boldsymbol{k} \cdot(\db + \R)} =  \frac{1}{\sqrt{N_c}} [\tau_z(\k)]_{nm} u_{m \k}(\db) e^{i \boldsymbol{k} \cdot(\db + \R)}.
\end{align}

The valley transformation of the low energy Fock matrix follows immediately, 
\begin{align}
    \tau_z:  \ F_{\text{low}}(\db, \dbp, \R) \to F_{\text{low}}^{\tau_z}(\db,\dbp,\R) = \frac{1}{N_c} \sum_{\substack{ \varepsilon_{n \k} \leq \varepsilon_F \\ |n|\leq n_c}} \lambda_{n \k} u^{\tau_z}_{n \k}(\db)^* u^{\tau_z}_{n \k}(\dbp)e^{-i \boldsymbol{k} \cdot(\db - \dbp + \R)},
\end{align}
and a Fock matrix that is invariant under the valley charge is obtained in the familiar way,
\begin{align}
    \tau_z \text{-symmetric } F:  \ F_{\text{low}}(\db, \dbp, \R) \to \frac{1}{2}\big( F_{\text{low}}(\db,\dbp,\R) +  F^{\tau_{z}}_{\text{low}}(\db,\dbp,\R) \big). 
    \label{FTZ}
\end{align}
The total Fock matrix is then the sum of $F_{\text{high}}$, which is unchanged, and the updated $F_{\text{low}}$.

Using this method we are able to avoid converging to states with spontaneous intervalley coherence that otherwise preserve the remaining symmetries. We have checked that the solutions with enforced valley symmetry are also solutions (fixed points of the self-consistency loop) if the symmetry is not enforced, such that the method does not produce unwanted instabilities. 

Enforcing all symmetries together is straightforward. Notice that we can obtain the valley and $C_{3z}$-transformed wave functions, $u_{n \k}^{\tau_z}(\tilde{C}_{3z}^{\pm 1}(\db))$, and simultaneously enforce $C_{3z}$ and valley conservation by combining Eqs. (\ref{FTZ}) and (\ref{FC3}). Once we have the valley and $C_{3z}$-symmetric $F$, $\mathcal{T}$ and $C_{2z}$ (and/or $C_{2z}\mathcal{T}$) can be applied sequentially from Eqs. (\ref{FT}) and (\ref{FC2}) (and/or (\ref{FC2T})). Finally, let us note that we do not account for the $C_{2x}$ symmetry in this discussion. In our simulations we find that the self-consistent solutions always preserve $C_{2x}$ if the remaining symmetries are enforced. 

\section{Generalized BM model}

Let us rewrite the plane waves introduced in the main text,
\begin{align}
    |\k,\ell,\sigma, \eta \rangle = \frac{2}{\sqrt{N_{at}}}\sum_{\rr \in \sigma \ell} e^{i (\eta \K_\ell + \k)\cdot \rr}| \rr \rangle,
\end{align}
with $\k =(k_x,k_y)$, $\ell =$ top($t,+$), bottom($b,-$) denotes the layer, $\sigma=A,B$ the graphene sublattice and $\eta=K(+1), K'(-1)$ the valley. $+(-) \K_\ell$ is the $K(K')$ point of layer $\ell$ (accounting for the rotation of the layers) and "$\rr \in \sigma \ell$" is short for the sum over atomic sites on sublattice $\sigma$ and layer $\ell$. $N_{at}$ is the number of atoms in the system, appearing in the normalization factor $(N_{at}/4)^{-1/2}$ ($N_{at}/4$ is the number of atoms in the system from each sublattice and layer). The plane waves form an orthonormal basis for the rigid TBG lattice. When lattice relaxation is included we must keep the unrelaxed atomic positions $\rr$ in $e^{i (\eta \K_\ell + \k)\cdot \rr}$ in order to preserve orthonormality. The states around $\k=0$ span the low-energy subspace of TBG, so it is natural to write the tight-binding and mean-field hamiltonians, $H_{\text{TB}}$ and $H_{\text{MF}}$, in this basis around $\k=\boldsymbol{0}$. In the following we discuss general symmetry properties, and use $H$ to refer to either one. We will explicitly mention when we address specific features of $H_{\text{TB}}$ or $H_{\text{MF}}$.   

The symmetries of TBG have the following action on the plane waves:
\begin{align}
    \hspace{0cm} C_{2z}&:\ \  \sigma_x \tau_x,  \hspace{-4cm} && \k \to C_{2z}(\k), \nonumber \\
    \hspace{0cm} C_{3z}&:\ \ e^{-2\pi i /3 \ \sigma_z \tau_z}, \hspace{-2cm} && \k \to C_{3z}(\k), \nonumber \\
    \hspace{0cm} C_{2x}&:\ \ \sigma_x \mu_x, \hspace{-4cm} && \k \to C_{2x}(\k), \nonumber \\
    \hspace{0cm} \mathcal{T}&: \ \ \tau_x, \hspace{-4cm} && \k \to -\k.
\end{align}
In these expressions, $\sigma_i,\tau_i,\mu_i$ denote the Pauli ($i=x,y,z$) matrices acting in the sublattice, valley and layer index, respectively (sublattice $A$, valley $K$ and the top layer are assigned and eigenvalue of $1$ under their respective Pauli $z$ matrix). The absence of a Pauli matrix is understood as the identity matrix. The symmetry operations acting on momentum implement their corresponding geometric transformations. $\mathcal{T}$ is antilinear, hence it involves an additional complex conjugation of the wave-function coefficients.

Moreover, the two valleys are decoupled, i.e. there is an additional emergent  $U(1)$ symmetry of conservation of the valley charge $\eta$. The symmetries within each valley sector are generated by $C_{3z}$, $C_{2x}$ and the product $C_{2z}\mathcal{T}$. From now on we focus on the $K$ valley and omit the $\eta$ label of the plane waves. The hamiltonian for valley $K$ can be obtained by $\mathcal{T}$ or $C_{2z}$.

\begin{figure}[H]
    \raggedright
    \hspace{.25cm} (a) \hspace{2.25cm} (b) \hspace{6.7cm} (c) \\
    \centering
    \includegraphics[valign=c, width=0.15\linewidth]{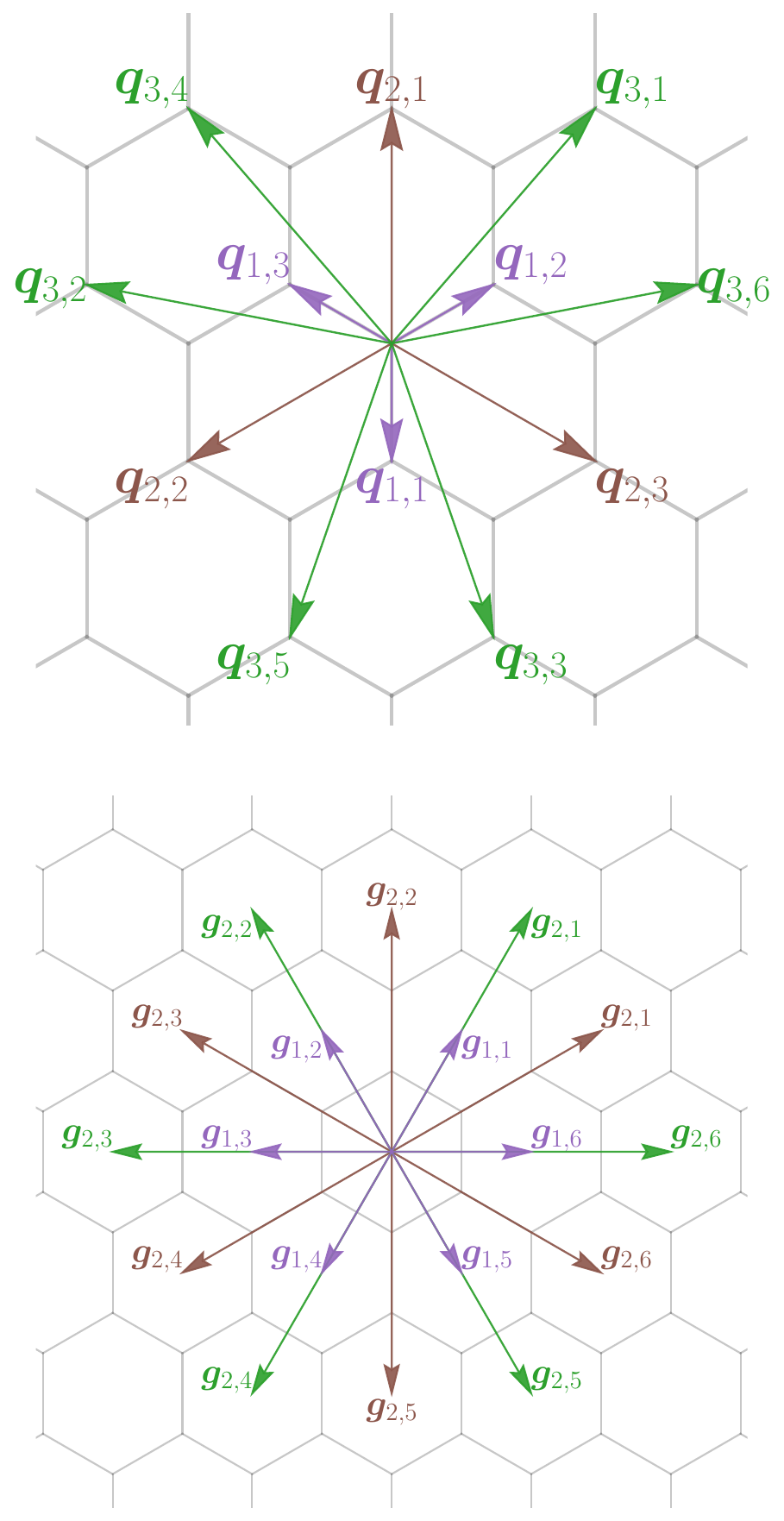}
    \includegraphics[valign=c, width=0.4\linewidth]{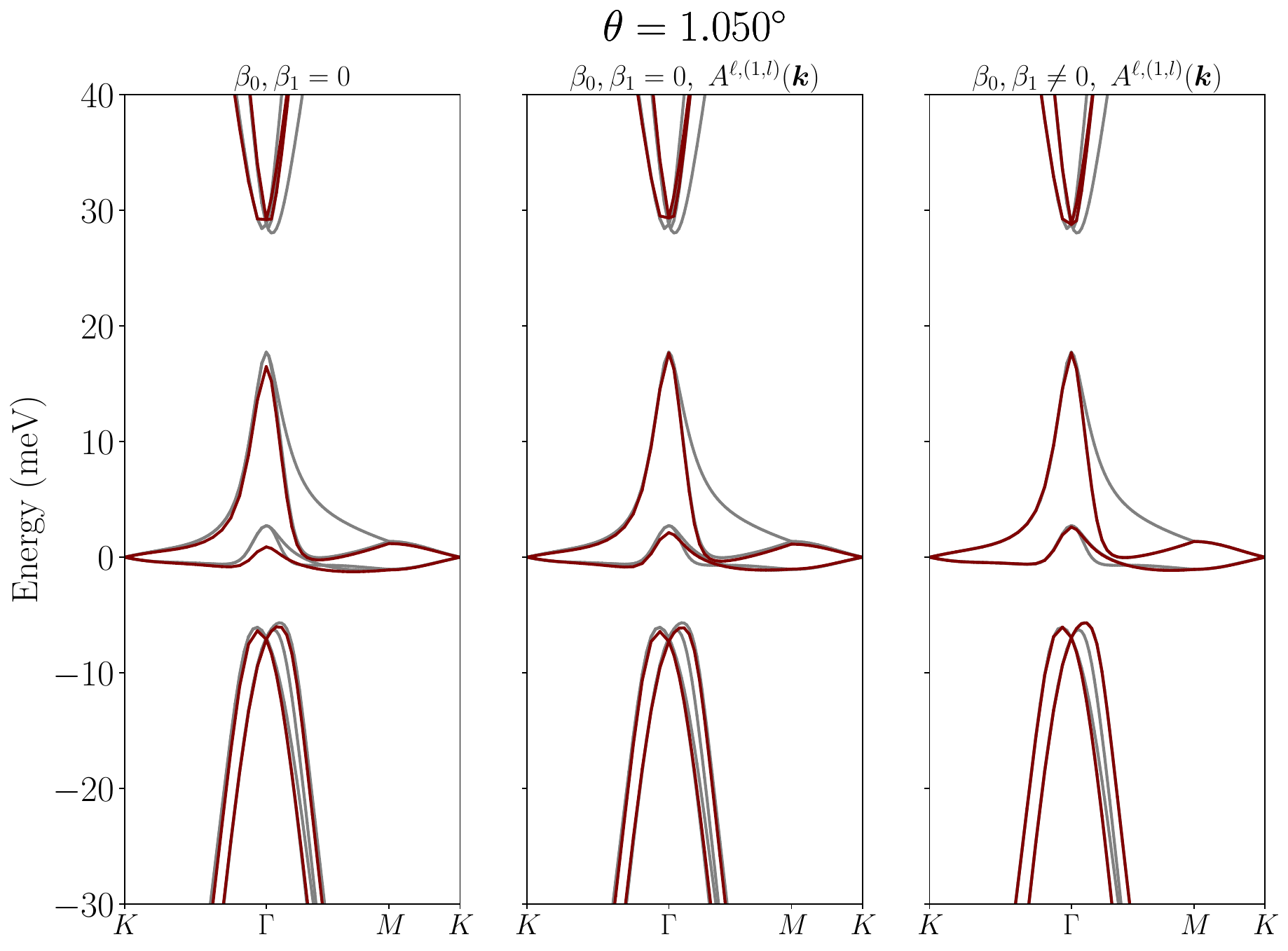}
    \includegraphics[valign=c, width=0.4\linewidth]{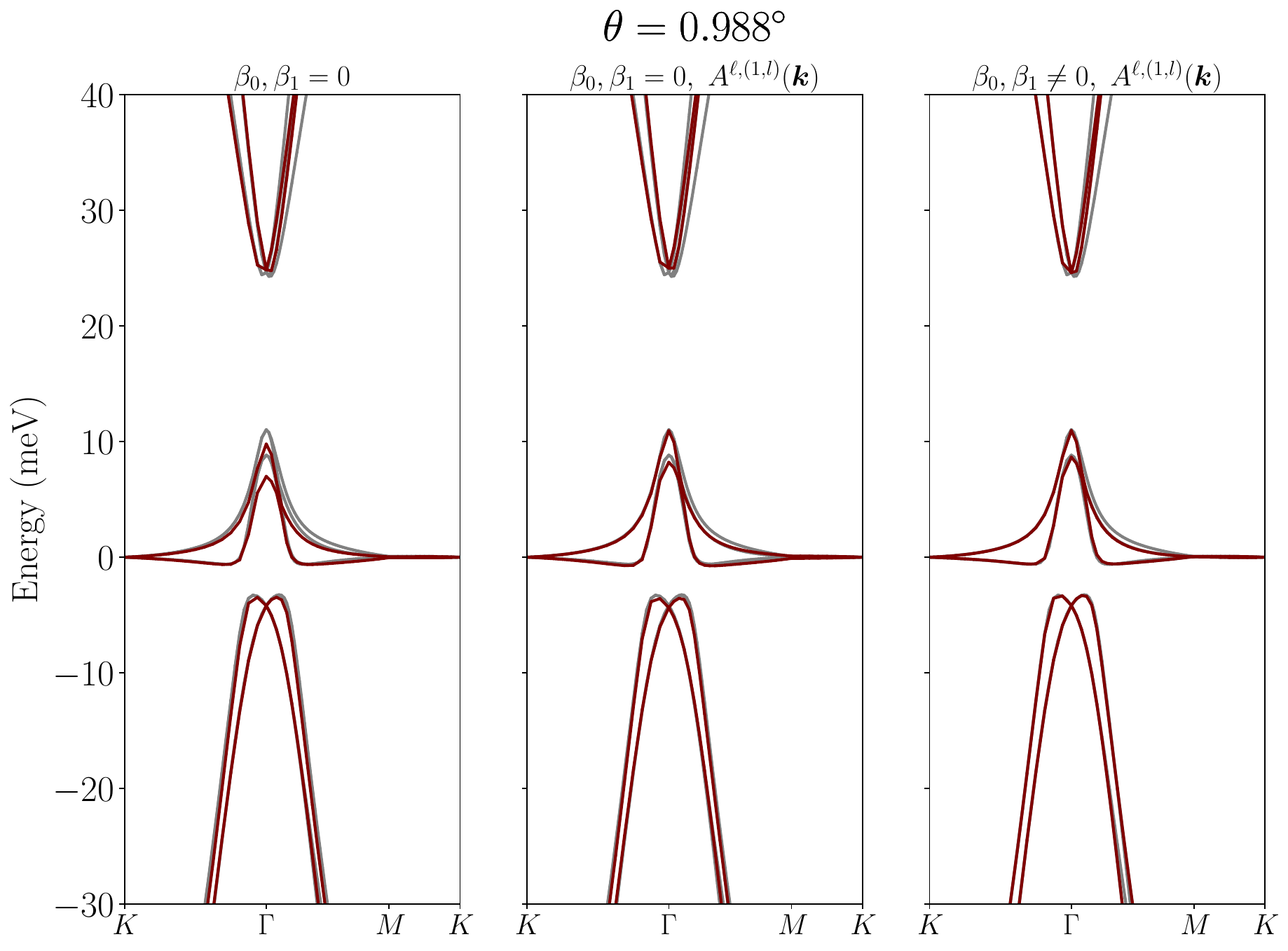}
    \caption{(a) We plot the first three momentum shells of momenta appearing in the interlayer hamiltonian, $\q_{\mu,l}$ (top), and in the intralayer hamiltonian, $\g_{\mu,l}$ (bottom). (b) TBG bands at $1.050^\circ$ and (c) at $0.988^\circ$. We plot in gray the tight-binding bands, and in red the valley $K$ bands for various approximations. Left: quadratic couplings $\beta_0,\beta_1=0$. Middle: quadratic couplings $\beta_0,\beta_1=0$ and full $\k$ dependence of the first shell of intralayer couplings. Right: quadratic couplings $\beta_0,\beta_1 \neq 0$ and full $\k$ dependence of the first shell of intralayer couplings. In all panels the interlayer couplings (local and non-local) are truncated at the sixth momentum shell and the local intralayer couplings at the fifth shell.}
    \label{generalizedBM}
\end{figure}

By moiré translational symmetry the nonzero matrix elements of the hamiltonian are of the form $\langle \k + \g, \ell,\sigma | H | \k, \ell,\sigma' \rangle$, $\langle \k + \q, t,\sigma | H | \k, b,\sigma' \rangle$ and $\langle \k - \q, b,\sigma | H | \k, t,\sigma' \rangle$, where $\q = \K_b - \K_t + \g$ and $\g$ is any moiré reciprocal lattice vector ($\K_b - \K_t = -(\g_1+\g_2)/3$). The momentum exchanges $\g$, $\q$ can be labeled $\g_{\mu,l}$, $\q_{\mu,l}$ where the index $\mu$ labels momentum shells with equal $|\g_{\mu,l}|$, $|\q_{\mu,l}|$ for equal $\mu$ (and increasing $|\g_{\mu,l}|$, $|\q_{\mu,l}|$ for increasing $\mu$) and $l$ labels the momenta within a shell, see Fig. \ref{generalizedBM}. $\g=\boldsymbol{0}$ is not included in $\{\g_{\mu,l}\}$ and will be treated separately. From now on, we will omit the sublattice index in the plane waves; the matrix elements are to be read as $2\times 2$ blocks in sublattice space.

By $C_{2z}\mathcal{T}$ symmetry we have $\langle \k', \ell' | H | \k, \ell \rangle = \sigma_x \langle \k', \ell' | H | \k, \ell \rangle^* \sigma_x$. This forces the interlayer and intralayer  blocks to be of the form 
\begin{align}
    \langle \k ,\ell | H | \k, \ell \rangle =&  h_0^\ell(\k) + h_1^\ell(\k)\sigma_x  + h^\ell_2(\k)\sigma_y, \nonumber \\
    \langle \k + \q_{\mu,l},t | H | \k,b \rangle =& w_0^{(\mu,l)}(\k) + w_1^{(\mu,l)}(\k)\sigma_x \nonumber + w_2^{(\mu,l)}(\k)\sigma_y + iw_3^{(\mu,l)}(\k)\sigma_z, \nonumber \\
    \langle \k + \g_{\mu,l},\ell | H | \k, \ell \rangle =&  A_0^{\ell,(\mu,l)}(\k) + A_1^{\ell,(\mu,l)}(\k)\sigma_x  + A_2^{\ell,(\mu.l)}(\k)\sigma_y + i A_3^{\ell,(\mu,l)}(\k)\sigma_z,
\end{align}
where $h_i^\ell(\k)$, $w_i^{(\mu,l)}(\k)$ and $A_i^{\ell,(\mu,l)}(\k)$ are real.

The 'graphene blocks' expanded up to terms quadratic in $\k$ read
\begin{align}
    \langle \k,\ell| H|\k,\ell \rangle = v_F \bar{\boldsymbol{\sigma}}_{\ell\theta/2} \cdot \k + \mu + \beta_0k^2 + \beta_1\big( (k_x^2-k_y^2)\sigma_x + 2k_xk_y\sigma_y\big) 
    \label{gblocksm}
\end{align}
with $k=|\k|$, $\bar{\boldsymbol{\sigma}} = (\sigma_x, -\sigma_y)$,  $\bar{\boldsymbol{\sigma}}_{\ell\theta/2} = e^{i \ell \theta \sigma_z /4} \bar{\boldsymbol{\sigma}} e^{-i \ell \theta  \sigma_z /4}$. The Dirac cones with Fermi velocity $v_F$, rotated by the angle $\ell \theta/2$, are the main components. In $H_{\text{TB}}$ we have $v_F=2.179  \text{ eV} \cdot a$, $\beta_0 = -0.1305  \text{ eV} \cdot a^2$ and $\beta_1 = -0.5673  \text{ eV} \cdot a^2$ (in units where $\hbar=1$) \cite{kang23}. In $H_{\text{MF}}$, $v_F=v_F(k)$ acquires momentum dependence. The chemical potential $\mu$ shifts rigidly the spectrum and is unimportant, and the quadratic couplings $\beta_0$, $\beta_1$ have a negligible effect, see Fig. \ref{generalizedBM}. For these reasons we keep only the Dirac cones in the discussion of the main text. Similarly, taking the renormalized $v_F$ in the bands of Fig. 4 of the main text is equivalent to taking the renormalized graphene blocks, Eq. (\ref{gblocksm}), for the remaining terms have a negligible effect (the interacting corrections to the various couplings are smaller than the couplings themselves, so we can safely say that the renormalized quadratic couplings have a negligible effect in the renormalized band structure). It can be readily checked that the symmetry relations from $C_{3z}$ and $C_{2x}$ are satisfied. We note in passing that to extract the renormalized $v_F(k)$ in Fig. 3(a) of the main text we fit the graphene block of the mean-field hamiltonian to the form of Eq. (\ref{gblocksm}) for $ 2.5|\g_1|>k > 1.5|\g_1|$ with constant $v_F$, $\beta_1$, $\beta_0$, $\mu$, and obtain $v_F(k)$ by subtracting the term proportional to $\beta_1$. The very small angular spread of the resulting $v_F(k)$ validates this method.

With respect to the interlayer blocks, by $C_{3z}$ we have
\begin{align}
     \langle \k + \q_{\mu,l},t | H | \k,b \rangle =  e^{-2\pi i\sigma_z \tau_z/3} \langle  C_{3z}(\k) + C_{3z}(\q_{\mu,l}),t | H | C_{3z}(\k)  ,b \rangle  e^{-2\pi i\sigma_z \tau_z/3},
\end{align}
and by $C_{2x}$ we have
\begin{align}
     \langle \k + \q_{\mu,l},t | H | \k,b \rangle &= \sigma_x \langle C_{2x}(\k) + C_{2x}(\q_{\mu,l}),b | H | C_{2x}(\k),t \rangle \sigma_x \nonumber \\
     &=   \sigma_x \langle  C_{2x}(\k),t | H | C_{2x}(\k) + C_{2x}(\q_{\mu,l}),b \rangle^\dagger \sigma_x \nonumber \\
     &=   \sigma_x \langle  C_{2x}(\k) + C_{2x}(\q_{\mu,l}) - C_{2x}(\q_{\mu,l}),t | H | C_{2x}(\k) + C_{2x}(\q_{\mu,l}),b \rangle^\dagger \sigma_x,
\end{align}
so matrix elements at $\k$ with momentum exchange $\q_{\mu,l}$ are related to matrix elements at $C_{3z}(\k)$ with momentum exchange $C_{3z}(\q_{\mu,l})$ by $C_{3z}$, and to  matrix elements at $C_{2x}(\k) + C_{2x}(\q_{\mu,l})$ with momentum exchange $-C_{2x}(\q_{\mu,l})$ by $C_{2x}$. Any approximation treating $w_i^{\mu,l}(\k)$ as constants independent of $\k$ must preserve these relations. From the identities $C_{3z}(-\q_{\mu,l}/2) =  -C_{3z}(\q_{\mu,l})/2$ and $C_{2x}(-\q_{\mu,l}/2) + C_{2x}(\q_{\mu,l}) = - (-C_{2x}(\q_{\mu,l}))/2$, we conclude that evaluating $w_i^{\mu,l}(\k)$ at $\k = -\q_{\mu,l}/2$ results in a valid approximation of constant matrix elements. In sum, defining 
\begin{align}
    w_i^{(\mu,l)} \equiv w_i^{(\mu,l)}(-\q_{\mu,l}/2)
\end{align}
allows the local approximation (we call it local because constant matrix elements are equivalent to a coupling between the layers that is local in space)
\begin{align}
    \langle \k + \q_{\mu,l},t | H | \k,b \rangle \approx w_0^{(\mu,l)} + w_1^{(\mu,l)}\sigma_x \nonumber + w_2^{(\mu,l)}\sigma_y + iw_3^{(\mu,l)}\sigma_z
    \label{Tfieldsm}
\end{align}
while respecting the symmetries of the problem. The remaining terms $w_i^{(\mu,l)}(\k) - w_i^{(\mu,l)}$, called 'non-local interlayer couplings', are well approximated by linear functions of $\k$ in the tight-binding model \cite{carr19,fang2019angledependentitabinitio,koshino20,kang23}.

Focusing on the first momentum shell, $C_{2x}$ forces $w_2^{(1,1)}=0$ and by $C_{3z}$ the terms for $l=2,3$ are written in terms of $w_i^{(1,1)}$,
\begin{align}
    &w_0^{(1,2)} = w_1^{(1,1)},   &&w_1^{(1,2)} = w_1^{(1,1)} \cos(2\pi/3),  
    &&&w_2^{(1,2)} = - w_1^{(1,1)} \sin(2\pi/3),  &&&&w_3^{(1,2)} = w_3^{(1,1)}, \nonumber \\
    &w_0^{(1,3)} = w_1^{(1,1)},  &&w_1^{(1,3)} = w_1^{(1,1)} \cos(-2\pi/3),  &&&w_2^{(1,2)} = - w_1^{(1,1)} \sin(-2\pi/3),  &&&&w_3^{(1,2)} = w_3^{(1,1)}.
\end{align}
$w_0^{(1,1)}$ and $w_1^{(1,1)}$ are readily identified as $w_0$ and $w_1$ of the BM model. The component $w_3^{(1,1)}\equiv w_3$ breaks the approximate particle-hole symmetry \cite{kang23} and does not appear in the BM model, but it is allowed by symmetry. In practice $w_3 \ll w_0,w_1$, see Fig. 2 of the main text for their bare (tight-binding) values.

The magnitude of the interlayer blocks decreases with increasing $\mu$. Convergence with the tight-binding band structure is achieved keeping the local couplings up to $\mu=6$ ($g_{3,l}=3g_1$) and the non-local couplings up to $\mu=3$. We choose however to keep all nonzero couplings up to $\mu=6$ for safety.

With respect to the intralayer blocks, a similar manipulation reveals that matrix elements at $\k$ with momentum exchange $\g_{\mu,l}$ are related to matrix elements at $C_{3z}(\k)$ with momentum exchange $C_{3z}(\g_{\mu,l})$ by $C_{3z}$, and to matrix elements at $C_{2x}(\k)$ with momentum exchange $C_{2x}(\g_{\mu,l})$ by $C_{2x}$. It can be seen that evaluating $A^{\ell,(\mu,l)}_i(\k)$ at $\k = \lambda \g_{\mu,l}$ for any $\lambda$ is a valid approximation of constant matrix elements. Moreover, by hermiticity 
\begin{align}
    \langle \k + \g_{\mu,l},\ell | H | \k, \ell \rangle = \langle \k, \ell | H | \k +  \g_{\mu,l}, \ell \rangle^\dagger =  \langle \k + \g_{\mu,l} - \g_{\mu,l}, \ell | H | \k +  \g_{\mu,l}, \ell \rangle^\dagger,
\end{align}
so matrix elements at $\k$ with momentum exchange $\g_{\mu,l}$ are related to matrix elements at $\k + \g_{\mu,l}$ with momentum exchange $-\g_{\mu,l}$. Requiring $\lambda \g_{\mu,l} + \g_{\mu,l}= \lambda(-\g_{\mu,l})$ sets $\lambda=-1/2$.  In sum, defining 
\begin{align}
    A_i^{\ell,(\mu,l)} \equiv A_i^{\ell, (\mu,l)}(-\g_{\mu,l}/2)
\end{align}
allows the local approximation
\begin{align}
    \langle \k + \g_{\mu,l}, \ell | H | \k, \ell \rangle \approx A_0^{\ell,(\mu,l)}  + A_1^{\ell,(\mu,l)} \sigma_x + A_2^{\ell,(\mu,l)} \sigma_y + i A_3^{\ell,(\mu,l)}  \sigma_z
    \label{Afieldsm}
\end{align}
while respecting the symmetries and hermiticity. In the tight-binding model $A_{0,3}^{\ell,(\mu,l)}$ are negligible and $A_{1,2}^{\ell,(\mu,l)}$ are conventionally written as $A_1^{\ell,(\mu,l)} + i A_2^{\ell,(\mu,l)} = \ell \gamma v_F e^{-i\ell \theta/2}( \mathcal{A}_x(\g_{\mu,l}) + i  \mathcal{A}_y(\g_{\mu,l})\big)$, where $\gamma v_F$ is a parameter related to the derivative of the hopping function and $\boldsymbol{\mathcal{A}}(\rr) = \sum_{\mu,l} \boldsymbol{\mathcal{A}}(\g_{\mu,l})e^{i \g_{\mu,l}\cdot \rr}$ is the lattice relaxation-induced pseudo-gauge field \cite{kang23,koshino20,Ceferino2024}. As $\gamma$ and $\boldsymbol{\mathcal{A}}$ appear together as a product, for simplicity we compress both objects together into $A$. It must be noted however that their distinction is physically meaningful.

Similar to $w_i^{(\mu,l)}(\k) - w_i^{(\mu,l)}$, the non-local terms $A_i^{\ell,(\mu,l)}(\k) - A_i^{\ell,(\mu,l)}$ are linear functions of $\k$ in the tight-binding hamiltonian \cite{kang23}. Essentially exact agreement with the tight-binding bands is achieved by keeping $A^{\ell,(\mu,l)}_i$ up to $\mu=5$ ($q_{5,l} = \sqrt{19}q_{1,1}$) and the first shell of non-local couplings, see Fig. \ref{generalizedBM}.

\section{Fermi velocity and interlayer couplings renormalization}\label{appa}

\begin{figure}[h]
\centering
\includegraphics[width=.75\linewidth]{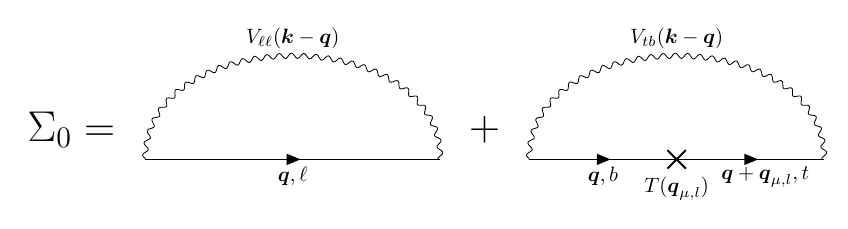}








\caption{Feynman diagrams for the self energy computed perturbatively to first order in the interlayer potential. There is a third diagram analogous to the second one of scatterings from the top ($t$) to the bottom ($b$) layer.}
\label{feynmandiag}
\end{figure}

In Hartree-Fock theory, the self-energy in the $K$ valley, in the plane-wave basis, reads (excluding the Hartree contribution which is inconsequential at charge neutrality in TBG) \cite{Coleman_2015} 
\begin{align}
\Sigma(\k + \boldsymbol{k'},\k)_{\ell \ell'} = i \int \frac{d\q}{(2\pi)^2} \int \frac{d\omega}{2\pi} G(\omega, \q + \boldsymbol{k'}, \q)_{\ell \ell'} V_{\ell \ell'}(\k-\q),    
\end{align}
with $V_{\ell\ell'}(\k)$ the Coulomb propagator in momentum space,
\begin{align}
    V_{\ell\ell'}(\k) &= \frac{2\pi e^2}{\epsilon} \Bigg\{ \frac{e^{-dk} + (1-e^{-dk})\delta_{\ell  \ell'}}{k}  -\frac{1 - \tanh\big(k\xi/2 \big)}{k}  \Bigg\} \ &&\text{double-gated potential},  \label{coulombpotd}  \\
    V_{\ell\ell'}(\k) &= \frac{2\pi e^2}{\epsilon} \Bigg\{ \frac{e^{-dk} + (1-e^{-dk})\delta_{\ell  \ell'}}{k}  -\frac{e^{-k\xi }}{k}  \Bigg\} \ &&\text{single-gated potential}. \label{coulombpots}
\end{align}
As a reminder $\epsilon$ is the effective dielectric constant, $d=3.35  \text{ \AA}$ is the inter-layer distance and $\xi$ is half distance from the sample to the gate(s). The second term (nonzero when $\xi < \infty$) accounts for the screening of the image charges \cite{bernevig21}, and we have assumed $\xi \gg d$. $G(\omega, \q + \boldsymbol{k'}, \q)_{\ell \ell'}$ is the electronic Green's function. In the Fock channel the two valleys are approximately decoupled because scatterings interchanging the valleys require very large momentum transfers and hence are suppressed \cite{bernevig21}; similarly the spins projections are decoupled in the Fock channel. The self-energy is computed for each valley and spin separately.

The quasiparticles in the system are described by the mean-field Hamiltonian $H_{\text{MF}} = H_{0} + \Sigma$, with $H_0$ the non-interacting hamiltonian, in our case $H_0=H_{\text{TB}}$. In self-consistent Hartree-Fock theory, $G$ is self-consistently determined by Dyson's equation $G^{-1} = G_0^{-1} - \Sigma[G]$, with $G_0$ the non interacting Green's function \cite{Coleman_2015}. Here we compute $\Sigma$ to first order; in other words we make the one-shot approximation $G^{-1} = G_0^{-1} - \Sigma[G_0]$. Furthermore, we approximate $G_0$ by the Green's function of the decoupled Dirac cones from each layer plus the first order perturbation from the interlayer potential $T(\q_{\mu,l}) = w_0^{(\mu,l)} +  w_1^{(\mu,l)}\sigma_x + w_2^{(\mu,l)}\sigma_y$, 
\begin{align}
G_0(\omega, \q+\boldsymbol{k'}, \q)_{\ell \ell'} \approx \delta_{\boldsymbol{k'},\boldsymbol{0}} G_{g}(\omega,\q) \delta_{\ell \ell'} +  \delta_{\boldsymbol{k'},\q_{\mu,l}} G_g(\omega,\q+\q_{\mu,l}) T(\q_{\mu,l})_{\ell \ell'} G_g(\omega,\q),
\end{align}
where
\begin{align}
    G_g(\omega,\boldsymbol{q}) = \sum_{s=\pm} \frac{ \frac{1}{2} (1 + s\bar{\boldsymbol{\sigma}} \cdot\q/q )}{\omega - sv_Fq + is\eta}    
\end{align}
is the Green's function of the uncoupled Dirac cones of each layer, with $\eta=0^+$ and $q=|\q|$. The $w_{3}^{(\mu,l)}$ terms, non-local couplings, pseudo-gauge fields, etc. are parametrically smaller and appear in subsequent perturbative orders of the expansion of $G_0$. Actually, $w_{0,1,2}^{(\mu>1,l)} \ll w_{0,1,2}^{(1,l)}$, corrections beyond the first shell are subleading; we will consider generic $(\mu,l)$ and later specialize to $\mu=1$. We also disregard the rotation of the Pauli matrices by the angle $\ell\theta/2$. The self energy $\Sigma[G_0] \approx \Sigma[G_g + G_g T G_g]\equiv \Sigma_0$ is diagrammatically depicted in Fig. \ref{feynmandiag}.

\subsection{Fermi velocity renormalization}
To zeroth order in the interlayer coupling, the self energy is diagonal in the layers, inducing the well-known Fermi velocity renormalization. For now, let us fix $\xi=\infty$.
\begin{align}
    \Sigma_{0,\xi=\infty}(\k,\k)_{\ell  \ell} &=  i \int \frac{d\q}{(2\pi)^2} \int \frac{d\omega}{2\pi} \sum_{s=\pm }\frac{\frac{1}{2}\big(1+s\bar{\boldsymbol{\sigma}}\cdot\q/q \big)}{\omega - sv_Fq + is\eta} \frac{2\pi e^2}{|\k - \q|} \\
    &= \frac{-e^2}{16 \pi^2 \epsilon} \int d\q \frac{1- \bar{\boldsymbol{\sigma}}\cdot \q/q}{|\k-\q|} 
\end{align}
Following Refs. \cite{sodemann12}, \cite{barnes14}, we set $\k = k \boldsymbol{x}$ ($k>0$) and readily solve the integral using the change of variables $\Big\{q_x = \frac{k}{2}\big( \cosh(\mu)\cos(\nu) +1 \big),\ q_y = \frac{k}{2}\big( \sinh(\mu)\sin(\nu) \big) \Big\}$. The result is
\begin{align}
   \Sigma_{0,\xi=\infty}(\k,\k)_{\ell  \ell}= \frac{-e^2 \Lambda}{8\pi \epsilon} + \frac{e^2 k }{e\epsilon} \sigma_x \log\bigg( \frac{4\Lambda}{\sqrt{e} k}\bigg)+ O(k/\Lambda).
\end{align}
In this formula, $\Lambda$ is a momentum cutoff, which might be determined by comparing to experiments or to microscopic calculations. $\Sigma_0$ can be encoded as a renormalization of the Fermi velocity, $\delta v_F(k)$. Recovering the generic dependence on $\k$ and disregarding the constant piece, we have
\begin{align}
    \Sigma_{0,\xi=\infty}(\k,\k)_{\ell  \ell} = \delta v_F(k) \bar{\boldsymbol{\sigma}} \cdot \k = \frac{\alpha v_F}{4} \log\bigg( \frac{4\Lambda}{\sqrt{e} k}\bigg)\bar{\boldsymbol{\sigma}} \cdot \k,
    \label{vfrxiinf}
\end{align}
where we defined the graphene fine structure constant
\begin{align}
    \alpha = \frac{e^2}{\epsilon v_F}.
\end{align}
Coming back to the general case $\xi < \infty$, the integral to compute is
\begin{align}
\Sigma_0(k \boldsymbol{x},k \boldsymbol{x})_{\ell \ell} &= \frac{\alpha v_F}{4\pi}\int d\q \Bigg(\frac{\tanh \big({|\k-\q|\xi}/{2} \big) -1}{|\k-\q|}\Bigg)\frac{\bar{\boldsymbol{\sigma}} \cdot  \q}{q} \ &&\text{double gate}\\
\Sigma_0(k \boldsymbol{x},k \boldsymbol{x})_{\ell \ell} &= \frac{\alpha v_F}{4\pi}\int d\q  \Bigg(\frac{- \exp\big({-|\k-\q|\xi} \big)}{|\k-\q|}\Bigg)\frac{\bar{\boldsymbol{\sigma}} \cdot  \q}{q} \ &&\text{single gate}.
\end{align}
Instead of attempting to solve it exactly, we will solve it in the $k=0$ and $k\xi \gg 1$ limits.

Starting for $k\xi \gg 1$, we decompose $\Sigma_0 = \Sigma_{0,\xi=\infty} + \delta \Sigma_0$. We make the change of coordinates $\q \to \k+\q$ and expand the spinor $\frac{ \bar{\boldsymbol{\sigma}} \cdot (\k+\q) }{|\k+\q|} = \frac{\bar{\boldsymbol{\sigma}}\cdot \k}{k} +  \frac{ \bar{\boldsymbol{\sigma}}\cdot \q }{k} - \frac{(\q\cdot \k)(\bar{\boldsymbol{\sigma}} \cdot \k)}{k^2} + O\big(\frac{q^2}{k^2} \big)$. The second and third terms vanish upon angular integration, and we get
\begin{align}
\delta \Sigma_0(k \boldsymbol{x},k \boldsymbol{x})_{\ell \ell} &= \frac{\alpha v_F}{4\pi}  \frac{\k \cdot \bar{\boldsymbol{\sigma}}}{k} \int d\q \Bigg(\frac{\tanh(q\xi/2)-1}{q}\Bigg) + O\big((k\xi)^{-2}\big) \ &&\text{double gate},\\ 
\delta\Sigma_0(k \boldsymbol{x},k \boldsymbol{x})_{\ell \ell} &= \frac{\alpha v_F}{4\pi}  \frac{\k \cdot \bar{\boldsymbol{\sigma}}}{k} \int d\q \Bigg(\frac{-\exp(q \xi)}{q}\Bigg) + O\big((k\xi)^{-2}\big) \  &&\text{single gate}.    
\end{align}

Then:
\begin{align}
    \delta\Sigma_0(k \boldsymbol{x},k \boldsymbol{x})_{\ell \ell} &= \frac{\alpha v_F}{4} \bigg(\frac{-4 \log(2)}{k\xi} \bigg) \bar{\boldsymbol{\sigma}} \cdot \k  &&\text{double gate}, \\
    \delta \Sigma_0(k \boldsymbol{x},k \boldsymbol{x})_{\ell \ell} &= \frac{\alpha v_F}{4} \bigg(\frac{-2}{k\xi}\bigg) \bar{\boldsymbol{\sigma}} \cdot \k &&\text{single gate},
\end{align}
where we used $\int_0^{\Lambda \xi/2} dk \ \tanh(k)-1 = - \log(2)$, $\int_0^{\Lambda \xi} dk \ -\exp(-k)  =  1$ for $\Lambda\xi \to \infty$.

For $k\xi \ll 1$, expand the Coulomb kernels for small $k$, with $\k = k \boldsymbol{x}$
\begin{align}
&\frac{\tanh \big({|\k-\q}|\xi/2\big)}{|\k-\q|} = \frac{\tanh ({q}\xi/2)}{q} - \bigg(\frac{\xi - \xi \tanh^2 (q\xi/2)}{2q} - \frac{\tanh(q\xi/2)}{q^2} \bigg) k \cos(\theta) + O\big((k\xi)^2\big), \\
&\frac{1 - \exp \big({-|\k-\q}|\xi\big)}{|\k-\q|} = \frac{1 - \exp({-q\xi})}{q} - \bigg(\frac{(q\xi+1)\exp(-q\xi)-1}{q^2} \bigg) k \cos(\theta) + O\big((k\xi)^2\big).     
\end{align}
The first term vanishes by angular integration, and we have, after recovering generic $\k$,
\begin{align}
    &\Sigma_0(k \boldsymbol{x},k \boldsymbol{x})_{\ell \ell}= \frac{\alpha v_F}{4} \Big( \log(\Lambda\xi/2) - 0.181 \Big) \bar{\boldsymbol{\sigma}} \cdot \k \ &&\text{double gate,} \\ 
    &\Sigma_0(k \boldsymbol{x},k \boldsymbol{x})_{\ell \ell} = \frac{\alpha v_F}{4} \Big( \log(\Lambda\xi/2) + 0.270 \Big) \bar{\boldsymbol{\sigma}} \cdot \k \ &&\text{single gate}. 
\end{align}
For the double gate we extracted log and constant parts using the integrals $\int_1^{\Lambda \xi/2} dk'/k' = \log(\Lambda\xi/2)$ and $\int_0^{\Lambda\xi/2} dk' (1-\tanh^2(k')) = 1$, $\int_0^1 dk' \tanh(k')/k' = 0.910$, $\int_1^{\Lambda\xi_/2} dk' (\tanh(k')-1)/k' = -0.0909$ for $\Lambda\xi \to \infty$. For the singled-gated case, $\log(\Lambda\xi/2) + 0.270 = \big( \lim_{x\to \infty} - \lim_{x\to 0} \big)\big(\text{Ei}(x) - \log(x) - e^{-x} \big)$.

We interpolate the solutions at $k\xi \ll 1$ and $k\xi \gg 1$ by introducing the $O(1)$ function $p(k\xi)$ into Eq. (\ref{vfrxiinf}),
\begin{empheq}[box=\fbox]{align}
    \delta v_F(k) =& \frac{ \alpha v_F}{4} \log \bigg( \frac{4\Lambda/\sqrt{e}}{k + p(k\xi)\xi^{-1}} \bigg), \label{vfrxi}\\
    p(k\xi) =& p_1 + p_2 e^{-p_3k\xi}
\end{empheq}
The parameters $p_1$, $p_2$, $p_3$ are fixed by requiring that $\delta v_F(k)$ has the same $(k\xi)^{-1}$ asymptotics for large $k\xi$ and the same value at $k=0$ as the exact solution, as well as zero slope at $k=0$. Their numerical values are
\begin{align}
    &p_1 = 4\log(2),  &&p_2 = 8e^{0.181 - 1/2} - 4\log(2),   &&& p_3 = \big( 8 e^{0.181-0.5} - 4\log(2) \big)^{-1}, &&&&\text{double gate}, \\
    &p_1 = 2,  &&p_2 = 8e^{-0.270 - 1/2} - 2, &&&p_3 = \big( 8 e^{-0.270-0.5} - 2 \big)^{-1}, &&&&\text{single gate}.
\end{align}
In Fig. \ref{renormalizedvfw0w1}(a) we plot the Fermi velocity from self-consistent Hartree-Fock on graphene. The formula (\ref{vfrxi}) matches the numerical data with the fitted momentum cutoff
\begin{align}
    \Lambda = 1.595 \text{ \AA}^{-1} = 3.92 a^{-1}.
\end{align}

\subsection{Interlayer couplings renormalization}

To first order in the interlayer couplings, the self energy is off-diagonal in the layers; let us fix $\ell=t, \ell'=b$. The Green's function reads 
\begin{align}
G_0(\omega, \q+\q_{\mu,l}, \q)_{tb} &\approx G_g(\omega,\q+\q_{\mu,l}) T(\q_{\mu,l})_{tb} G_g(\omega,\q)    
\nonumber \\ &= \sum_{s=\pm} \frac{\frac{1}{2} \big(1 + s\bar {\boldsymbol{\sigma}} \cdot (\q+\q_{\mu,l})/|\q+\q_{\mu,l}|\big)}{\omega - s v_F|\q+\q_{\mu,l}| + is\eta} T(\q_{\mu,l}) \sum_{s'=\pm} \frac{\frac{1}{2} \big(1 + s'\bar{\boldsymbol{\sigma}} \cdot \q/q \big)}{\omega - s' v_F q + is'\eta}.
\end{align}
Performing the frequency integral,
\begin{align}
    \int \frac{d\omega}{2\pi}  \frac{1}{\omega - s v_F|\q+\q_{\mu,l}| + is \eta}  \frac{1}{\omega - s' v_Fq + is'\eta}
    = \frac{-i}{v_F(|\q+\q_{\mu,l}|+q)}\bigg(\delta_{s,-1}\delta_{s',+1} +\delta_{s,+1}\delta_{s',-1}\bigg),
\end{align}
the self-energy  reads
\begin{align}
\Sigma_0(\k+\q_{\mu,l},\k)_{tb} =& \int \frac{d\q}{(2\pi)^2} \frac{ 1}{v_F (|\q+\q_{\mu,l}|+q)} \nonumber \\ & \times \Bigg\{ \frac{1}{4} \bigg(1 + \frac{\bar \sigma \cdot (\q+\q_{\mu,l})}{|\q+\q_{\mu,l}|} \bigg) T(\q_{\mu,l}) \bigg(1 - \frac{\bar \sigma \cdot \q}{q}\bigg) + \frac{1}{4}  \bigg(1 - \frac{\bar \sigma \cdot (\q+\q_{\mu,l})}{|\q+\q_{\mu,l}|} \bigg) T(\q_{\mu,l}) \bigg (1 + \frac{\bar \sigma \cdot \q}{q} \bigg) \Bigg\}V_{tb}(\k-\q) \\
 =& \frac{1}{2} \int \frac{d^2q}{(2\pi)^2} \frac{ V_{tb}(\k-\q)}{v_F (|\q+\q_{\mu,l}|+q)} \Bigg\{  T(\q_{\mu,l}) - \bigg( \frac{\bar \sigma \cdot (\q+\q_{\mu,l})}{|\q+\q_{\mu,l}|}\bigg) T(\q_{\mu,l}) \bigg( \frac{\bar \sigma \cdot \q}{q} \bigg) \Bigg\}V_{tb}(\k-\q)
 \end{align}
We now make the change of variables $\q \to \q - \q_{\mu,l}/2$ and substitute $T(\q_{\mu,l}) = w_0^{(\mu,l)} + w_1^{(\mu,l)} \sigma_x + w_2^{(\mu,l)} \sigma_y$. We have
 \begin{align}
\Sigma_0(\k+\q_{\mu.l},\k)_{tb} \equiv \Sigma_{01}(\k+\q_{\mu.l},\k)_{tb} + \Sigma_{02}(\k+\q_{\mu.l},\k)_{tb},
\end{align}
with
\begin{align}
\Sigma_{01}(\k + \q_{\mu,l},\k)_{tb }= \frac{1}{2} \int \frac{d^2q}{(2\pi)^2} \frac{w_0^{(\mu,l)}}{v_F (|\q+\q_{\mu,l}/2|+|\q-\q_{\mu,l}/2|)} \Bigg( 1 - \frac{q^2 - q_{\mu,l}^2/4}{|\q + \q_{\mu,l}/2| |\q - \q_{\mu,l}/2|}\Bigg) V_{tb}(\k + \q_{\mu,l}/2 -\q)
\end{align}
and
\begin{align}
\Sigma_{02}(\k+\q_{\mu.l},\k)_{tb} \equiv \Sigma_{02a}(\k+\q_{\mu.l},\k)_{tb} + \Sigma_{02b}(\k+\q_{\mu.l},\k)_{tb},
\end{align}
\begin{align}
 \Sigma_{02a}(\k + \q_{\mu,l},\k)_{tb }=  \frac{1}{2} \int & \frac{d^2q}{(2\pi)^2}  \frac{1}{v_F (|\q+\q_{\mu,l}/2|+ |\q-\q_{\mu,l}/2|)} \Big(w_1^{(\mu,l)}\sigma_x + w_2^{(\mu,l)}\sigma_y \Big)V_{tb}(\k + \q_{\mu,l}/2 -\q)     
\end{align} 
\begin{align}
\Sigma_{02b}(\k + \q_{\mu,l},\k)_{tb }= & -\frac{1}{2} \int  \frac{d^2 q}{(2\pi)^2}  \frac{1}{v_F (|\q+\q_{\mu,l}/2|+ |\q-\q_{\mu,l}/2|)}\frac{1}{|\q + \q_{\mu,l}/2||\q-\q_{\mu,l}/2|} \nonumber \\
\times & \Big\{ w_1^{(\mu,l)} \big( q_x^2 - q_y^2 - {q_{\mu,lx}^2}/{4} + {q_{\mu,ly}^2}/{4} \big) \sigma_x - w_1^{(\mu,l)} \big(2q_x q_y - 2q_{\mu,lx} q_{\mu,ly}/{4} \big) \sigma_y  \nonumber \\  
&- w_2^{(\mu,l)} \big( q_x^2 - q_y^2 - {q_{\mu,lx}^2}/{4} + {q_{\mu,ly}^2}/{4} \big) \sigma_y +  w_2^{(\mu,l)} \big( 2q_y q_y - 2q_{\mu,lx} q_{\mu,ly}/{4} \big) \sigma_x  \Big\} V_{tb}(\k + \q_{\mu,l}/2 -\q).
\end{align}
Making the change of variables $\q \to R_{-\varphi}(\q)$ where $R_{-\varphi}$ denotes the rotation by the angle $-\varphi = - \text{angle}\big(w_1^{\mu,l} + iw_{2}^{\mu,l} \big)/2$
\begin{align}
\Sigma_{02b}(\k + \q_{\mu,l},\k)_{tb }= -\frac{1}{2} \int & \frac{d^2 q}{(2\pi)^2}  \frac{1}{v_F (|\q+\tilde{\q}_{\mu,l}/2|+ |\q-\tilde{\q}_{\mu,l}/2|)}\frac{\sqrt{\big( w_1^{(\mu,l)} \big)^2 + \big( w_2^{(\mu,l)}\big)^2} }{|\q + \tilde{\q}_{\mu,l}/2||\q-\tilde{\q}_{\mu,l}/2|} \nonumber \\
\times & \Big\{ \big( q_x^2 - q_y^2 - {\tilde{q}_{\mu,l,x}^2}/{4} + {\tilde{q}_{\mu,l,y}^2}/{4} \big) \sigma_x - 2\big( q_y q_y - \tilde{q}_{\mu,l,x} \tilde{q}_{\mu,l,y}/{4} \big) \sigma_y \Big\} V_{tb}(\tilde{\k} + \tilde{\q}_{\mu,l}/2 -\q),
\end{align} 
with $\tilde{\q}_{\mu,l} = R_\varphi(\q_{\mu,l})$, $\tilde{\k} = R_\varphi(\k)$. This expression only depends on $\sqrt{\big( w_1^{(\mu,l)} \big)^2 + \big( w_2^{(\mu,l)}\big)^2}$; the relative magnitude of $w_1^{(\mu,l)}$ and $w_2^{(\mu,l)}$ is encoded in the coordinate rotation. Now, consider the function
\begin{align}
J(\k,\g)= \int  d^2 q \frac{1}{|\q+ \g/2|+ |\q-\g/2|}\frac{\bar{q}^2 - (\bar{g}/2)^2 }{|\q + \g/2||\q-\g/2|}  V_{tb}(\k + \g/2 -\q),
\end{align}
with $\bar{q} = q_x - iq_y$,  $\bar{g} = g_x - ig_y$. Notice that the real and imaginary parts of $J(\tilde{\k}, \tilde{\q}_{\mu,l})$ are equal to the $\sigma_x$ and $\sigma_y$ components of $\Sigma_{02b}(\k + \q_{\mu,l},\k)_{tb }$, up to numerical factors. $J$ obeys the relation
\begin{align}
    J(R_{\vartheta}(\k),R_\vartheta(\g)) = e^{-2i\vartheta} J(\k,\g),
\end{align}
which tells us that one only needs to compute  $\Sigma_{02b}(\k + \q_{\mu,l},\k)_{tb }$ (equivalently $J(\k,\g)$) for $\q_{\mu,l}$ ($\g$) along a given direction, say along $-\hat{\boldsymbol{y}}$, and the result for generic $\q_{\mu,l}$ ($\g$). It is easy to see that by similar manipulations we can draw the same conclusion for $\Sigma_{01}(\k + \q_{\mu,l},\k)_{tb }$ and $\Sigma_{02a}(\k + \q_{\mu,l},\k)_{tb }$. At this point, let us notice that the intralayer couplings in Eq. (\ref{Afieldsm}) have the same structure as the interlayer couplings, Eq. (\ref{Tfieldsm}). Hence the  first order corrections to $A^{\ell,(\mu,l)}$ are obtained replacing $\q_{\mu,l}$, $w_1^{\ell,(\mu,l)}$, $w_2^{\ell,(\mu,l)}$ by $\g_{\mu,l}$, $A_1^{\ell,(\mu,l)}$, $A_2^{\ell,(\mu,l)}$ above.

We are interested in the first shell of interlayer couplings with $\q_{1,1} \equiv \q_1 = -q_1 \hat{\boldsymbol{y}}$, $w_{0}^{(1,1)} \equiv w_{0}$, $w_{1}^{(1,1)} \equiv w_{1}$ and  $w_{2}^{(1,1)} \equiv w_{2}= 0$. The couplings with $\q_{1,2}$ and $\q_{1,3}$ are related by $C_{3z}$ symmetry, or equivalently by the aforementioned coordinate rotations. Introducing the notation of the generalized BM model,
\begin{align}
    \Sigma_0(\k + \q_1, \k)_{tb} = \delta w_0(\k) + \delta w_1(\k)\sigma_x + \delta w_1(\k)\sigma_y,
\end{align}
with
\begin{empheq}[box=\fbox]{align}
\delta w_0(\k) =& \frac{1}{2} \int \frac{d\q}{(2\pi)^2} \frac{w_0}{v_F\big( |\q+\q_1/2| + |\q-\q_1/2| \big)} \bigg( 1 - \frac{q^2 -q_1^2/4}{|\q+\q_1/2||\q-\q_1/2|}\bigg)V_{tb}(\k + \q_1/2 - \q), \label{dw0intfull} \\
\delta w_1(\k) =& \frac{1}{2} \int \frac{d\q}{(2\pi)^2} \frac{w_1}{v_F\big( |\q+\q_1/2| + |\q-\q_1/2|\big)} \bigg( 1 - \frac{q_x^2 - q_y^2 - (q_{1x}^2/4 - q_{1y}^2/4)}{|\q+\q_1/2||\q-\q_1/2|}\bigg)V_{tb}(\k + \q_1/2 -\q), \label{dw1intfull} \\
\delta w_2(\k) =& \frac{1}{2} \int \frac{d\q}{(2\pi)^2} \frac{w_1}{v_F\big( |\q+\q_1/2| + |\q-\q_1/2|\big)} \bigg(\frac{2(q_x q_y - q_{1x} q_{1y}/4)}{|\q-\q_1/2||\q+\q_1/2|}\bigg) V_{tb}(\k + \q_1/2 -\q). \label{dw2intfull}
\end{empheq}

\subsubsection{Local couplings}

The local couplings are defined by $\delta w_i \equiv \delta w_i(-\q_1/2)$. Notice that $\delta w_2(\k)$ is odd under a reflection across the $k_y=-q_{1y}/2$ line, as required by $C_{2x}$ symmetry, hence $\boxed{\delta w_2 = w_2 = 0}$. 

Substituting $V_{tb}(-\q)$ for Eqs. (\ref{coulombpotd}), (\ref{coulombpots}), we see that $\delta w_{0,1}$ can be written as the sum of a function of $q_1d$ plus a function of $q_1 \xi$.  Let us focus now on $\xi=\infty$ and study the corresponding dependence on $q_1d$. In that case,
\begin{align}
\delta w_0(\xi=\infty) =& \frac{\alpha w_0}{4\pi} \int {d\q} \frac{1}{ |\q - \hat{\boldsymbol{y}}/2| + |\q + \hat{\boldsymbol{y}}/2|} \bigg( 1 - \frac{q^2 - 1/4}{|\q-\hat{\boldsymbol{y}}/2||\q+\hat{\boldsymbol{y}}/2|}\bigg) \frac{e^{-q_1dq}}{q}, \label{dw0int} \\
\delta w_1(\xi=\infty) =& \frac{\alpha w_1}{4\pi} \int {d\q} \frac{1}{ |\q - \hat{\boldsymbol{y}}/2| + |\q + \hat{\boldsymbol{y}}/2|} \bigg( 1 - \frac{q_x^2 - q_y^2 + 1/4}{|\q-\hat{\boldsymbol{y}}_1/2||\q+\hat{\boldsymbol{y}}_1/2|}\bigg) \frac{e^{-q_1d q}}{q}, \label{dw1int}
\end{align}
where we made the change of variables $\q \to q_1\q$.

For $\delta w_0$, in the $q_1d \ll 1$ limit, appropriate near the magic angle ($q_1d \approx 0.1$ for a twist angle of $1.0^\circ$), let us expand $e^{-q_1d q}/q \approx 1/q - q_1d$ under the integral sign. Then, with the change of variables to elliptic coordinates $\Big\{ u= |\q + \hat{\boldsymbol{y}}/2| + |\q - \hat{\boldsymbol{y}}/2, v= |\q + \hat{\boldsymbol{y}}/2| - |\q - \hat{\boldsymbol{y}}/2| \Big\}$,
\begin{align}
    &\int {d\q} \frac{1}{ |\q - \hat{\boldsymbol{y}}/2| + |\q + \hat{\boldsymbol{y}}/2|} \bigg( 1 - \frac{q^2 - 1/4}{|\q-\hat{\boldsymbol{y}}/2||\q+\hat{\boldsymbol{y}}/2|}\bigg) \frac{1}{q} = \int_{-1}^{1} dv \int_{1}^\infty du \frac{2\sqrt{1 - v^2}}{u \sqrt{\big(u^2 - 1 \big) \big( u^2+v^2- 1 \big)}} = 2\pi, \\
    &\int {d\q} \frac{1}{ |\q - \hat{\boldsymbol{y}}/2| + |\q + \hat{\boldsymbol{y}}/2|} \bigg( 1 - \frac{q^2 - 1/4}{|\q-\hat{\boldsymbol{y}}/2||\q+\hat{\boldsymbol{y}}/2|}\bigg)  = \int_{-1}^{1} dv \sqrt{1 -v^2}  \int_{1}^\infty du \frac{1}{u \sqrt{u^2 - 1}} = \frac{\pi^2}{4},
\end{align}
and $\delta w_0$ can be approximated by a linear function of $q_1d$ with the constant and linear coefficients given above. 

\begin{figure}[H]
    \centering
    \raisebox{25ex}{(a)}
    \includegraphics[width=0.3\linewidth]{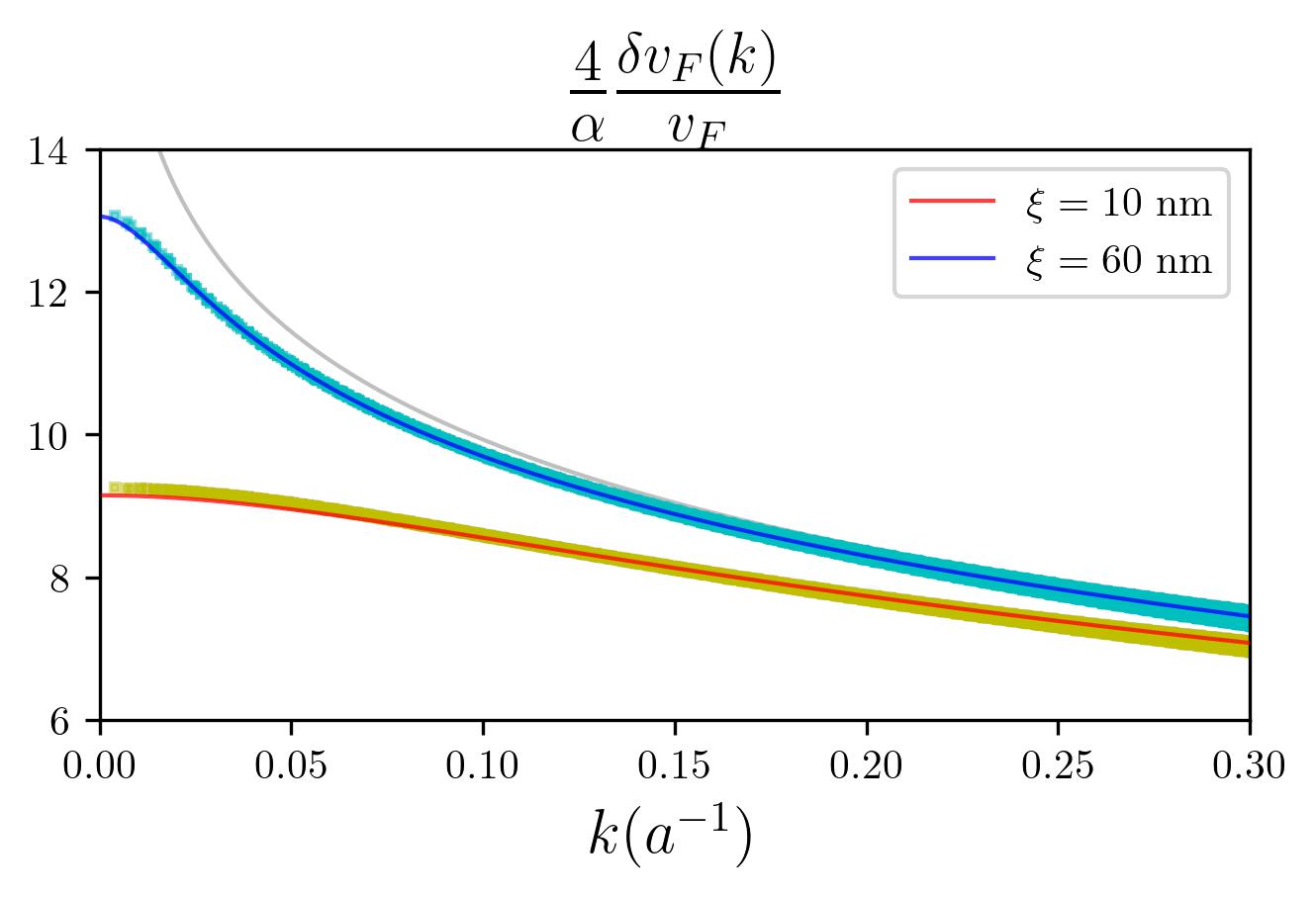}
    \raisebox{25ex}{(b)}
    \includegraphics[width=0.6\linewidth]{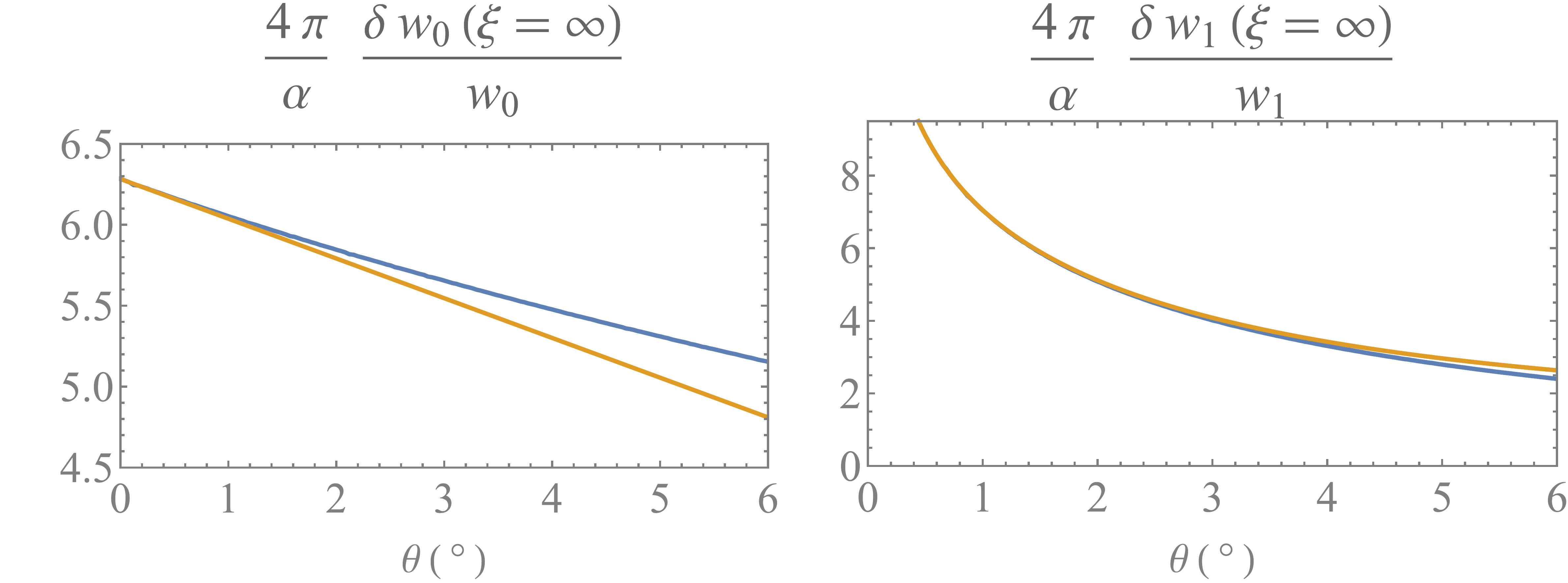}
    \caption{(a) Fermi velocity renormalization in monolayer graphene, with $\xi=10$ and $60$ nm in the double-gate setup. We plot the numerically extracted Fermi velocity (more precisely, $4\delta v_F(k)/(\alpha v_F)$) as squares and the analytical formula Eq. (\ref{vfrxi}) with the best fit value of $\Lambda = 3.92 a^{-1}$. We also plot in gray the $\xi=\infty$ curve. We have set $\epsilon=5$ in both calculations. (b) Renormalization of $w_0$ and $w_1$ without gate screening ($\xi=\infty$) as a function of the twist angle. We plot in blue the numerical values from Eqs. (\ref{dw0int}), (\ref{dw1int}) and in orange the analytical approximations from Eqs. (\ref{dw0}), (\ref{dw1}). For reference, the angle of $6^\circ$ corresponds to $q_1d \approx 0.6$.}
\label{renormalizedvfw0w1}
\end{figure}

For $\delta w_1$, expand the parenthesis in Eq. (\ref{dw1int}) and consider the first integral. We split it into three components, 
\begin{align}
    &\int {d\q} \frac{1}{ |\q - \hat{\boldsymbol{y}}/2| + |\q + \hat{\boldsymbol{y}}/2|} \frac{e^{-q_1d q}}{q} = \nonumber \\
    &\int_{q<1} {d\q} \frac{1}{ |\q - \hat{\boldsymbol{y}}/2| + |\q + \hat{\boldsymbol{y}}/2|} \frac{e^{-q_1d q}}{q}+ \int_{q>1} {d\q} \Bigg(\frac{1}{ |\q - \hat{\boldsymbol{y}}/2| + |\q + \hat{\boldsymbol{y}}/2|} - \frac{1}{2q}\Bigg) \frac{e^{-q_1d q}}{q}  + \int_{q>1} {d\q} \frac{e^{-q_1d q}}{2q^2}. 
\end{align}
The first and second integrals converge and might be computed again in the $q_1d \ll 1$ limit as linear functions of $q_1d$. By numerical evaluation we find $4.8343 - 2.0996 q_1d$ and $-0.0947 + 0.1917 q_1d$ for the first and second integral, respectively. The third integral is logarithmically divergent and evaluates exactly to $-\pi \ \text{Ei}(-q_1d) \approx \pi \big(-\gamma -\log(q_1d) + q_1d \big)$ for $q_1d \ll 1$, with $\text{Ei}$ the exponential integral and $\gamma$ the Euler-Mascheroni constant. Finally, we evaluate numerically the constant and linear coefficients of the second integral in Eq. (\ref{dw1int}) after expanding the parenthesis,
\begin{align}
 \int d\q\frac{1}{ |\q - \hat{\boldsymbol{y}}/2| + |\q + \hat{\boldsymbol{y}}/2|} \frac{q_x^2 - q_y^2 + 1/4}{|\q-\hat{\boldsymbol{y}}_1/2||\q+\hat{\boldsymbol{y}}_1/2|} \frac{e^{-q_1d q}}{q} \approx -3.3640 + 1.2337 q_1d.
\end{align}
All four terms together give
\begin{align}
    \frac{4\pi}{\alpha} \frac{\delta w_1 (\xi=\infty)}{w_1} \approx \pi \log \bigg(\frac{0.8699}{q_1d}\bigg) + 2.4674 q_1d.
\end{align}
In fact, $0.8699 \approx 8 e^{3/2 - \gamma - \pi}$ and $2.4674 \approx \pi^2/4$. An analysis of the integral Eq. (\ref{dw1int}) by the LLM Claude by Anthropic verifies that this is indeed the closed form of the coefficients. The final result for the interlayer couplings renormalizations, as reported in the main text is,
\begin{empheq}[box=\fbox]{align}
    &\frac{4\pi}{\alpha} \frac{\delta w_0(\xi=\infty)}{w_0} \approx 2\pi -\frac{\pi^2}{4} q_1d, \label{dw0} \\
    &\frac{4\pi}{\alpha} \frac{\delta w_1 (\xi=\infty)}{w_1} \approx \pi \log \bigg(\frac{8 e^{3/2 - \gamma - \pi}}{q_1d}\bigg) + \frac{\pi^2}{4} q_1d. \label{dw1}
\end{empheq}
In Fig. \ref{renormalizedvfw0w1}(b) we plot the numerical values of $\delta w_{0}$ and $\delta w_1$ alongside the analytical approximations for twist angles $< 6.0^\circ$. 

\begin{figure}[H]
    \centering
    \includegraphics[width=0.99\linewidth]{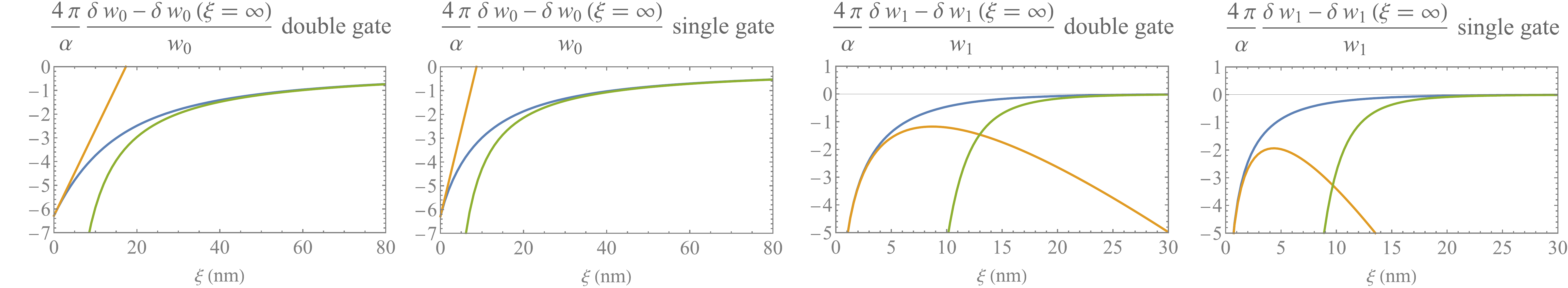}
    \caption{Change in the renormalized $w_0$ and $w_1$ due to the presence of the metallic gates as a function of the gate distance, for the twist angle of $1.0^\circ$. Notice that the change is negative. We plot in blue the numerical values from Eqs. (\ref{dw0intfull}), (\ref{dw1intfull}), in orange the analytical approximations for $q_1\xi \ll 1$ and in green the analytical approximations for $q_1\xi \gg 1$, from Eqs. (\ref{dwxi1}-\ref{dwxi2}). For reference, at the chosen twist angle, $\xi= 10$ nm corresponds to $q_1\xi = 2.97$.}
    \label{renormalizedw01gate}
\end{figure}

We proceed similarly to obtain approximations for the corrections due to the screening of the gates. For $q_1\xi \ll 1$ we obtain the leading logarithmic, constant and linear terms in $q_1 \xi$. In fact, in the single gate setup the change in the Coulomb kernel is the same as the kernel without the gate, after replacing $d$ by $\xi$ and adding a minus sign, see (\ref{coulombpots}). For the double gate we have used the identities $\int_x^\infty dt(1 - \tanh{(t)})/t = 2\sum_{n=1}^\infty (-1)^n \text{Ei}(-2nx) \approx - \log(2x) - \gamma + \log(\pi/2)  + x$ for $x \ll 1$ (this expression is obtained after expanding $\text{Ei}(-2nx)$ for small $x$ for each $n$ and performing the eta-regularized sums of the log, constant and linear coefficients \cite{wolframeta}) to obtain the $q_1 \xi \ll 1$ components.  For $q_1 \xi \gg 1$, expanding integrand (rather, the part of the integrand that is not the Coulomb kernel) in powers of $\q$ around $\q = 0$ results in an expansion of the integral in powers of $(q_1 \xi)^{-1}$. The results are:

\begin{align}
    &\frac{4\pi}{\alpha} \frac{\delta w_0 - \delta w_0(\xi=\infty)}{w_0} \approx - 2\pi + \frac{\pi^2}{8} q_1d \ &&\text{double gate}, \ {q_1\xi \ll 1} \label{dwxi1} \\
    &\frac{4\pi}{\alpha} \frac{\delta w_0 - \delta w_0(\xi=\infty)}{w_0} \approx \frac{-8\pi\log(2)}{q_1 \xi} \ &&\text{double gate}, \  {q_1\xi \gg 1} \\
    &\frac{4\pi}{\alpha} \frac{\delta w_0 - \delta w_0(\xi=\infty)}{w_0} \approx {-2\pi + \frac{\pi^2}{4} q_1\xi} \ &&\text{single gate}, \ {q_1\xi \ll 1}  \\
    &\frac{4\pi}{\alpha} \frac{\delta w_0 - \delta w_0(\xi=\infty)}{w_0} \approx  \frac{-4\pi}{q_1\xi} \ &&\text{single gate}, \ {q_1\xi \gg 1} 
\end{align}
\begin{align}
    &\frac{4\pi}{\alpha} \frac{\delta w_1 - \delta w_1(\xi=\infty)}{w_0} \approx -\pi \log \bigg(\frac{4\pi e^{3/2 - \gamma - \pi}}{q_1 \xi}\bigg) - \frac{\pi^2}{8} q_1\xi \ &&\text{double gate}, \ q_1\xi \ll 1 \\
    &\frac{4\pi}{\alpha} \frac{\delta w_1 - \delta w_1(\xi=\infty)}{w_1} \approx \frac{-360\pi \zeta(5)}{(q_1\xi)^5} \ &&\text{double gate}, \ q_1\xi \gg 1 \\
    &\frac{4\pi}{\alpha} \frac{\delta w_1 - \delta w_1(\xi=\infty)}{w_0} \approx  -\pi \log \bigg(\frac{8 e^{3/2 - \gamma - \pi}}{q_1 \xi}\bigg) - \frac{\pi^2}{4} q_1\xi \ &&\text{single gate},\ q_1\xi \ll 1 \\
    &\frac{4\pi}{\alpha} \frac{\delta w_1 - \delta w_1(\xi=\infty)}{w_1} \approx \frac{-192\pi }{(q_1\xi)^5} \ &&\text{single gate},  \ q_1\xi \gg 1, \label{dwxi2} 
\end{align}
with $\zeta$ the Riemann zeta function. In Fig. \ref{renormalizedw01gate} we show the numerical corrections and the analytical formulas as a function of $\xi$ for $\theta=1.0^\circ$. Note that physical values of $\xi$, say $\xi > 5$ nm, lie in the large-$\xi$ region or the intermediate-$\xi$ region where no approximation is accurate.

\subsubsection{Non-local couplings.}

Now we study the non-local terms, in other words the $\k$ dependence of the renormalized interlayer couplings Eqs. (\ref{dw0intfull}), (\ref{dw1intfull}), (\ref{dw2intfull}), away from $\k=-\q_1/2$.

In order to make analytical progress, let us fix $|\k + \q_1/2| \gg q_1$. In this limit, $\delta w_0 (\k)$ can be approximated by the monopole term of Eq. (\ref{dw0intfull}), setting $V_{tb}(\k + \q_1/2 - \q) \approx V_{tb}(\k+\q_1/2)$,
\begin{align}
    &\delta w_0(\k) \approx \frac{1}{2} V_{tb}(\k + \q_1/2) \int \frac{d\q}{(2\pi)^2} \frac{w_0}{v_F\big( |\q+\q_1/2| + |\q-\q_1/2| \big)} \bigg( 1 - \frac{q^2 -q_1^2/4}{|\q+\q_1/2||\q-\q_1/2|}\bigg). \\
\end{align}
Evaluating the integral and assuming $\xi \gg d$ gives
\begin{align}
    \boxed{
    \frac{4\pi}{\alpha} \frac{\delta w_0(\k)}{\delta w_0} \approx \frac{\pi^2}{4} \frac{q_1e^{-|\k + \q_1/2|d}}{|\k + \q_1/2|}.
    }
    \label{dw0k}
\end{align}

For $\delta w_1(\k)$, again let us fix $|\k + \q_1/2| \gg q_1$ and $\xi=\infty$. From Eq. (\ref{dw1intfull}) we have to compute 
\begin{align}
    &\int {d\q} \frac{1}{ |\q+\q_1/2| + |\q-\q_1/2|}\frac{e^{-|\k + \q_1/2 -\q|d}}{ |\k + \q_1/2 -\q|} \nonumber \\
    &=  \int {d\q} \frac{1}{ |\q+\q_1/2| + |\q-\q_1/2|} \bigg( \frac{e^{-|\k + \q_1/2 -\q|d}}{ |\k + \q_1/2 -\q|} - \frac{e^{-qd}}{q}  \bigg) +  \int {d\q} \frac{1}{ |\q+\q_1/2| + |\q-\q_1/2|} \frac{e^{-qd}}{q}.  
\end{align}
The  second integral above was already computed for $\delta w_1(\xi=\infty)$, with the logarithmic term $\pi \log(8e^{1 - \gamma - \pi/2}/(q_1d))$ for $q_1d \ll 1$. We set $d=0$ in the first integral, which is justified for $|\k + \q_1/2| \ll d^{-1}$, giving
\begin{align}
    &\int {d\q} \frac{1}{ |\q+\q_1/2| + |\q-\q_1/2|} \bigg( \frac{1}{ |\k + \q_1/2 -\q|} - \frac{1}{q}  \bigg) \nonumber \\ &=  \int {d\q} \frac{1}{2q} \frac{1}{|\k + \q_1/2-\q|} +  \int {d\q} \bigg( \frac{1}{ |\q+\q_1/2| + |\q-\q_1/2|} - \frac{1}{2q} \bigg) \frac{1}{|\k + \q_1/2-\q|} - \int {d\q} \frac{1}{ |\q+\q_1/2| + |\q-\q_1/2|}\frac{1}{q} \nonumber \\
    &\approx \pi \log\bigg( \frac{4\Lambda}{|\k + \q_1/2|} \bigg) - \frac{\pi^2}{8} \frac{q_1}{|\k + \q_1/2|} - \pi \log \bigg( \frac{4e^{0.1227}\Lambda }{q_1} \bigg), 
    \end{align}
where we extracted the logarithmic and constant terms of the first and third integrals and the monopole term by exact evaluation of the second integral. The coefficient $e^{0.1227}$ is found numerically, and an analysis by Claude reveals its closed form $e^{1 + \log(2)-\pi/2}$. Rearranging the pieces yields
\begin{align}
    \int {d\q} \frac{1}{ |\q+\q_1/2| + |\q-\q_1/2|}\frac{e^{-|\k + \q_1/2 -\q|d}}{ |\k + \q_1/2 -\q|} \approx \pi \log \bigg( \frac{4e^{-\gamma}}{|\k + \q_1/2|d} \bigg) - \frac{\pi^2}{8} \frac{q_1}{|\k + \q_1/2|}.
\end{align}
The remaining integral from Eq. (\ref{dw1intfull}) reads
\begin{align}
 \int d\q \frac{1}{ |\q+\q_1/2| + |\q-\q_1/2|}  \frac{q_x^2 - q_y^2 + q_1^2/4}{|\q+\q_1/2||\q-\q_1/2|} \frac{e^{-|\k + \q_1/2 - \q|d}}{|\k + \q_1/2 - \q|}.
\end{align}
The limit $|\k + \q_1/2| \gg q_1$ amounts to setting $\q_1=0$ above. Setting also $d=0$ allows us to evaluate the expression exactly,
\begin{align}
     \int d\q \frac{1}{2q}  \frac{q_x^2 - q_y^2}{q^2} \frac{1}{|k \hat{\boldsymbol{x}} - \q|} &= \frac{1}{2} \int_0^{2\pi} d\nu \int_0^{\infty} d\mu\frac{\big( \cosh(\mu)\cos(\nu)+1 \big)^2 - \big( \sinh(\mu)\sin(\nu) \big)^2}{(\cosh(\mu) + \cos(\nu))^2} \nonumber \\ &=\frac{\pi}{2},
\end{align}
where we chose $\k + \q_1/2 = k \hat{\boldsymbol{x}}$ and used the change of variables $\Big\{q_x = \frac{k}{2}\big( \cosh(\mu)\cos(\nu) +1 \big),\ q_y = \frac{k}{2}\big( \sinh(\mu)\sin(\nu) \big) \Big\}$. The solution for generic $\k + \q_1/2$ follows from the structure of the integrand. In summary,
\begin{align}
    \int d\q \frac{1}{ |\q+\q_1/2| + |\q-\q_1/2|}  \frac{q_x^2 - q_y^2 + q_1^2/4}{|\q+\q_1/2||\q-\q_1/2|} \frac{e^{-|\k + \q_1/2 - \q|d}}{|\k + \q_1/2 - \q|} \approx \frac{\pi}{2}\cos(2 \varphi_{\k + \q_1/2}),
\end{align}
with $\varphi_{\k}$ the angle between $\k$ and the x axis. We have thus obtained an approximation for $\delta w_1(\k)$ when $d^{-1} \gg |\k + \q_1/2| \gg q_1$ and $\xi=\infty$,
\begin{align}
    \boxed{
    \frac{4\pi}{\alpha} \frac{\delta w_1(\k)}{w_1} \approx \pi \log \bigg( \frac{4e^{-\gamma}}{|\k + \q_1/2|d} \bigg) -\frac{\pi^2}{8} \frac{q_1}{|\k + \q_1/2|} - \frac{\pi}{2}\cos \big( 2\varphi_{\k + \q_1/2} \big).
    }
    \label{dw1k} 
\end{align}

Finally, for the calculation of $\delta w_2(\k)$, notice that Eq. (\ref{dw2intfull}) is the analogous to the last integral that we have just obtained for $\delta w_1(\k)$. Thus,  when $d^{-1} \gg |\k + \q_1/2| \gg q_1$ and $\xi=\infty$,
\begin{align}
    \boxed{
    \frac{4\pi}{\alpha} \frac{\delta w_2(\k)}{w_1} \approx \frac{\pi}{2} \sin \big( 2\varphi_{\k + \q_1/2} \big).
    }
    \label{dw2k}
\end{align}

The analytical features of Eqs. (\ref{dw0k}), (\ref{dw1k}) and (\ref{dw2k}) reproduce those from the numerically extracted $\delta w_0(\k)$, $\delta w_1(\k)$ and $\delta w_2(\k)$ in Fig. 3(c) of the main text. Indeed, $\delta w_0(\k)$ is approximately rotationally symmetric and decays with $|\k + \q_1/2|$, $\delta w_2(\k)$ oscillates like $\sin(2\varphi_{\k + \q_1/2})$ (notice also a non-negligible background making the points related by $\k + \q_1/2 \to -(\k + \q_1/2)$ have different values, for example $4$ and $2$ for the 'maxima' at the edges of the plot), and  in $\delta w_1(\k)$ we can readily infer a rotationally symmetric, decaying component plus an oscillating component like $-\cos(2 \varphi_{\k + \q_1/2})$. These results confirm the expectation that the dominant corrections to $w_0$, $w_1$ and $w_2$ are of first order in $w_0$ and $w_1$.
 
In the main text we have argued that for $w_3$ the first order correction is not sufficient, as higher perturbative orders of parametrically larger couplings should contribute with terms comparable in magnitude. The first order expression can be derived straightforwardly, and reads
\begin{align}
\delta w_3(\k) =& \frac{1}{2} \int \frac{d\q}{(2\pi)^2} \frac{w_3}{v_F\big( |\q+\q_1/2| + |\q-\q_1/2| \big)} \bigg( 1 + \frac{q^2 -q_1^2/4}{|\q+\q_1/2||\q-\q_1/2|}\bigg)V_{tb}(\k + \q_1/2 - \q). \label{dw3intfull} 
\end{align}
With $1 + \frac{q^2 -q_1^2/4}{|\q+\q_1/2||\q-\q_1/2|} = 2 - \Big(1 - \frac{q^2 -q_1^2/4}{|\q+\q_1/2||\q-\q_1/2|} \Big)$ the integral splits into two that we have studied above. Specifically, for $d^{-1} \gg |\k + \q_1/2| \gg q_1$ and $\xi=\infty$ we have
\begin{align}
    \frac{4\pi}{\alpha}\frac{\delta w_3(\k)}{w_3} \approx  2\pi \log \bigg( \frac{4e^{-\gamma}}{|\k + \q_1/2|d} \bigg) - \frac{\pi^2}{2} \frac{q_1}{|\k + \q_1/2|}.
\end{align}
We thus expect a rotationally symmetric, decaying behavior. This is in contrast to the numerical $\delta w_3(\k)$ in Fig. 3(c) of the main text that displays a clear angular dependence, demonstrating the emergence of higher-order contributions.

\section{Results for an alternative tight-binding model}

\begin{figure}[H]
    \raggedright
    (a) \hspace{9.3cm} (b) \\
    \centering
    \includegraphics[width=0.54\linewidth]{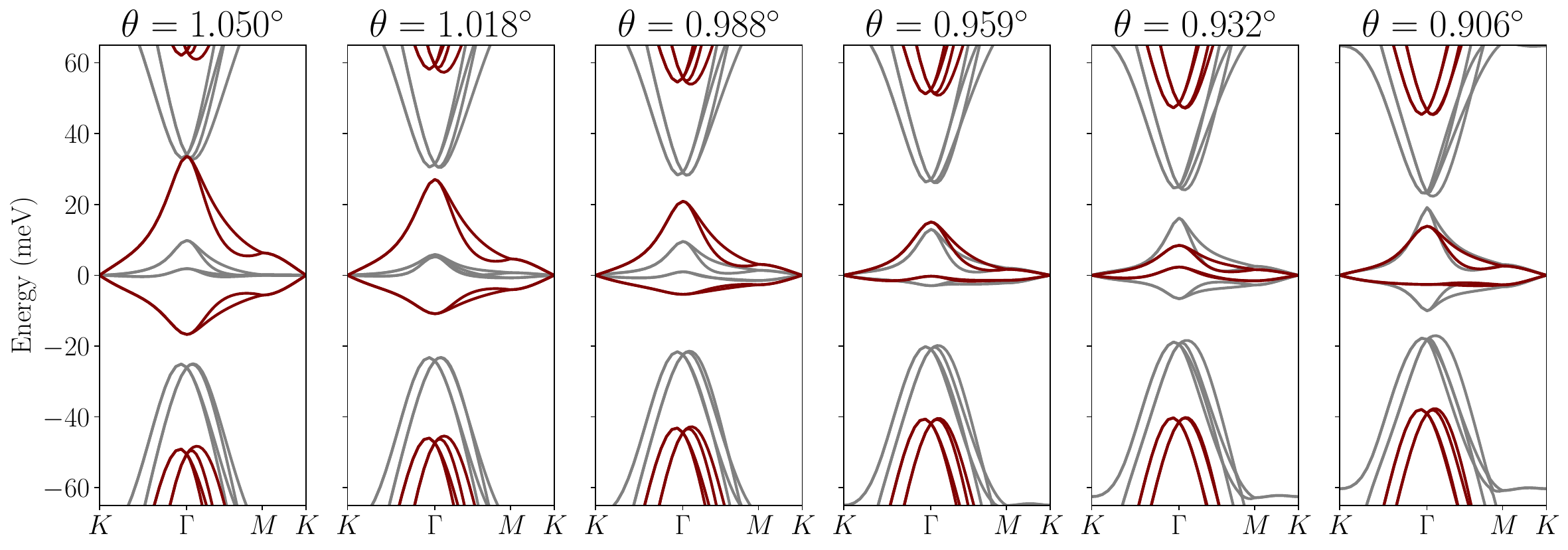}
    \includegraphics[width=0.44\linewidth]{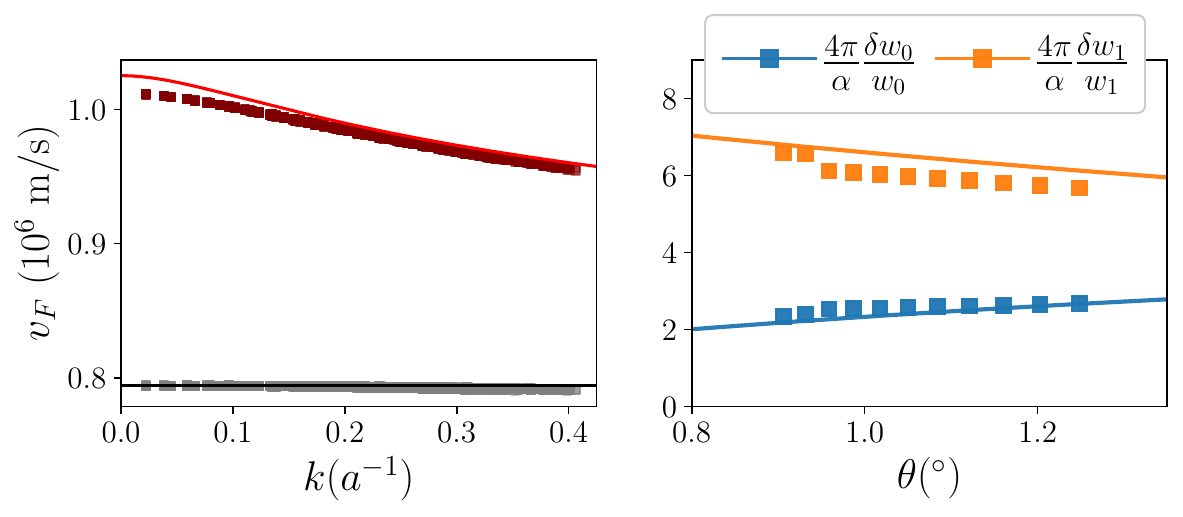}
    \caption{Results using the hopping function and atomic relaxation from Refs. \cite{koshino20_1}, \cite{koshino20_2}. The momentum cutoff of this model is fitted as $\Lambda = 4.02 a^{-1}$. (a) Same as Fig. 1 of the main text. (b) Same as Fig. 3(a) and (b) of the main text. In this model the magic angle gets shifted from $\approx 1.018^\circ$ to $\approx 0.932 ^\circ$. We set $\epsilon=10$ and $\xi=10$ nm, and $\theta = 1.050^\circ$ in (b) left panel.}
    \label{koshinomodel}
\end{figure}

\end{document}